\documentclass[
    aps,
    pre,
    reprint,
    amssymb,
    twocolumn,
    preprintnumbers,
    nofootinbib
]{revtex4-1}
\usepackage{graphicx}
\usepackage[utf8]{inputenc}
\DeclareUnicodeCharacter{2010}{-}
\usepackage{amsmath}
\usepackage[]{todonotes}
\usepackage{siunitx}
\usepackage{todonotes}
\newcommand{\de}[2]{\frac{\text{d} {#1}}{\text{d}{#2}}}

\newcommand{\pd}[2]{\frac{\partial #1}{\partial #2}}

\newcommand{\pdsec}[2]{\frac{\partial^2 #1}{\partial {#2}^2}}

\newcommand{\ignore}[1]{}
\newcommand{\intv}[1]{\text{d}{#1}}

\newcommand{\precite}{\,}
\newcommand{\preonlinecite}{Ref.\:}
\newcommand{\prefigNoWord}{\:}
\newcommand{\prefig}{Fig.\prefigNoWord}
\newcommand{\prefigs}{Figs.\prefigNoWord}
\newcommand{\Prefig}{Figure\prefigNoWord}
\newcommand{\Prefigs}{Figures\ }
\newcommand{\preequNoWord}{\,}
\newcommand{\preequ}{Eq.\preequNoWord}
\newcommand{\preequs}{Eqs.\preequNoWord}
\newcommand{\Preequ}{Equation\preequNoWord}
\newcommand{\preapp}{App.\,}
\newcommand{\prefootnote}{\,}

\newcommand{\prerefs}{Refs.~}
\newcommand{\presec}{Sec.\,}
\newcommand{\pretab}{Tab.\,}
\newcommand{\pretabs}{Tabs.\,}
\newcommand{\Presec}{Section\,}
\newcommand{\Presecs}{Sections~}
\newcommand{\figtop}{Top:~}
\newcommand{\figbottom}{Bottom:~}

\newcommand{\figbottomleft}{Bottom left:~}
\newcommand{\figbottomright}{Bottom right:~}
\newcommand{\figinset}{Inset:~}

\newcommand{\Dint}{D}

\newcommand{\inttraj}{\Lambda}
\newcommand{\inttrajpoints}{\ell}
\newcommand{\scalelength}{\sigma}
\newcommand{\scalelengthod}{\scalelength_\text{1d}}
\newcommand{\scalelengthtdexp}{\sigma_\text{e}}
\newcommand{\scalelengthtdg}{\sigma_\text{g}}

\renewcommand{\vec}[1]{\mathbf{#1}}

\begin{document}

\preprint{DRAFT - \today}

\title{On the weak scaling of the contact distance between two fluctuating interfaces with system size}


\author{Clemens Moritz}
\affiliation{Faculty of Physics, University of Vienna, Boltzmanngasse 5, 1090 Vienna, Austria}
\author{Marcello Sega}
\affiliation{Forschungszentrum Jülich GmbH, Helmholtz Institute Erlangen-Nürnberg for Renewable Energy (IEK-11), Fürther Straße 248, 90429 Nürnberg}
\author{Max Innerbichler}
\affiliation{Faculty of Physics, University of Vienna, Boltzmanngasse 5, 1090 Vienna, Austria}
\author{Phillip L. Geissler}
\affiliation{Department of Chemistry, University of California, Berkeley, California 94720}
\author{Christoph Dellago}
\email[]{christoph.dellago@univie.ac.at}
\affiliation{Faculty of Physics, University of Vienna, Boltzmanngasse 5, 1090 Vienna, Austria}
\affiliation{Erwin Schrödinger Institute for Mathematics and Physics, Boltzmanngasse 9, 1090, Vienna, Austria}

\date{\today}

\begin{abstract}
    A pair of flat parallel surfaces, each freely diffusing along the direction of their separation, will eventually come into contact. If the shapes of these surfaces also fluctuate, then contact will occur when their centers of mass remain separated by a nonzero distance $\ell$.  Here we examine the statistics of $\ell$ at the time of first contact for surfaces that evolve in time according to the Edwards-Wilkinson equation. We present a general approach to calculate its probability distribution and determine how its most likely value $\ell^*$ depends on the surfaces' lateral size $L$. We are motivated by an interest in the motion of interfaces between two phases at conditions of thermodynamic coexistence, and in particular the annihilation of domain wall pairs under periodic boundary conditions. Computer simulations of this scenario verify the predicted scaling behavior in two and three dimensions. In the latter case, slow growth where $\ell^\ast$ is an algebraic function of $\log L$ implies that slab-shaped domains remain topologically intact until $\ell$ becomes very small, contradicting expectations from equilibrium thermodynamics.
\end{abstract}

%\pacs{}
%\keywords{}

\maketitle

\section{Introduction\label{sec:intro}}
In the initial stages of many large-scale changes in the shape of interfaces, microscopic thermal fluctuations play a crucial role. The rupture of a thin liquid film\precite\cite{Craster2009a,Vrij1968,Vrij1966a}, the breakup of a liquid jet\precite\cite{Eggers2008,Hennequin2006,Eggers2002,Moseler2000a,Shi1994}, and the fusion of two liquid droplets\precite\cite{Perumanath2019,Aarts2005,Aarts2004,Eggers1999,Bradley1978} are examples of processes that on the meso- and macroscopic scale are governed by surface tension and hydrodynamics, while at their earliest stages they are controlled by thermal undulations.
\begin{figure}[tb]
    \centering
    \includegraphics[width=0.8\linewidth]{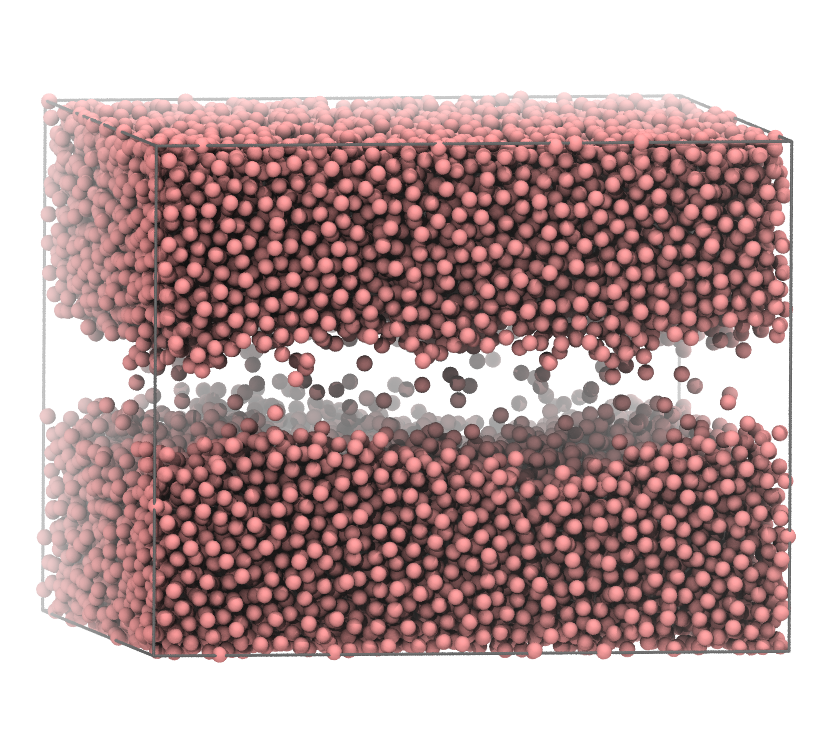}
    \caption{Example configuration of a Lennard-Jones gas slab enclosed by liquid. The configuration has been obtained from a molecular dynamics simulation at liquid-gas coexistence using periodic boundary conditions. This snapshot has been rendered using VMD\precite\cite{Stone1998,Humphrey1996}.}
    \label{fig:LJ_slab}
\end{figure}

\begin{figure}[tb]
\centering
\includegraphics[width=1.0\linewidth]{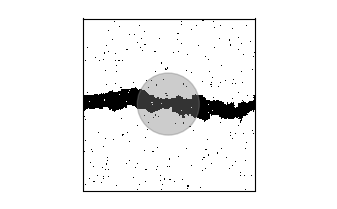}
\caption{Example configuration obtained from a simulation of the $200 \times 200$ Ising model with vanishing external field. The gray circle indicates a disk with the same area as the slab composed of black spins. This circle has a circumference that is roughly $45\%$ smaller than the surface of the slab. Matplotlib\precite\cite{Hunter2007} has been used to generate graphs and Ising model simulation snapshots throughout this work.}
\label{fig:example_config_slab}
\end{figure}

Thermal fluctuations also govern the distance up to which two planar interfaces will approach before they come into contact for the first time; a situation which can be viewed as an approximation to the droplet fusion problem for large radii. \Prefig\ref{fig:LJ_slab} shows a snapshot taken from a computer simulation of a Lennard-Jones fluid that is an example of this kind of interfacial system. The two interfaces are formed by a single slab of atoms in the liquid phase that is connected through the periodic boundaries of the simulation box. \Prefig\ref{fig:example_config_slab} shows a similar scenario for a two-dimensional Ising model at coexistence. The aim of this paper is to determine how close the centers-of-mass of two such interfaces approach before they come into contact for the first time.

Larger surfaces, whose topographical fluctuations are generally more extreme, should contact each other at larger average separation. The mean distance between the two interfaces at the time of first contact, $\ell^\ast$, indeed grows with the linear size of the system, $L$, in our analysis. But this growth will turn out to be very slow, especially in the case of a two-dimensional interface that is embedded in three-dimensional space (a (2+1)-dimensional interface). Here $L$ enters the result for $\ell^\ast$ as an algebraic function of $\log L$ only. This not only implies that the distance at which two freely diffusing, macroscopic interfaces come into contact for the first time is still likely microscopic, but also has interesting consequences for the situations shown in \prefigs\ref{fig:LJ_slab} and \ref{fig:example_config_slab}. In both cases, the slab is thermodynamically metastable at conditions that fix the relative proportions of coexisting phases. Alternate geometries---a spherical gas bubble and a disk formed by spins pointing in the same direction---would have a considerably smaller surface and, therefore, also a lower free energy\precite\cite{Troester2017,Moritz2017,Binder2012,Troster2005,Leung1990}. Nevertheless, these slab geometries are remarkably long-lived. In fact, we will show that the critical widths of these slabs, at which they transition into the thermodynamically preferred shape, grow with $L$ much more slowly than the scaling $\ell^\ast \sim L$ expected from equilibrium considerations (see \presec\ref{sec:slab_simulations} for details on how this scaling comes about). Hence, this two-interface system has the interesting property that with increasing system size the behavior of the system increasingly diverges from the behavior expected based on equilibrium principles.

The remainder of the article is structured as follows: in \presec\ref{sec:rough_argument} we present a rough estimate of the expected scaling of the most likely contact distance, $\ell^\ast$, with system size $L$. In \presec\ref{sec:framework} we explain our approach to calculating the distribution of contact distances $u(\ell)$. \Presec\ref{sec:application_ew} introduces the Edward-Wilkinson (EW) equation as a model for surface dynamics and goes through all the steps necessary to predict the scaling of the contact distance between two EW interfaces with system size. In \presec\ref{sec:slab_simulations} we report on simulations of slabs in the Ising model and in a Lennard-Jones system at liquid-gas coexistence. The predicted scaling behavior of the EW model is then compared with the critical width of slabs observed in simulation. \Presec\ref{sec:summary} provides a summary and a discussion of the results.

\section{A simple scaling argument}\label{sec:rough_argument}
One expects the meeting of initially distant interfaces to occur through a combination of two basic processes. First, the interfaces' mean positions (averaged over directions perpendicular to their separation) each execute a random walk with diffusion coefficient $D_{\rm int}\sim 1/L^d$, where $L$ is the surfaces' linear size and $d$ is their dimensionality. Second, as their shapes change in the course of natural fluctuations, a point on one interface can draw transiently nearer to a point on the other interface, even if the mean separation $\ell$ is fixed.  An extreme fluctuation in interfacial shapes can thus achieve contact while the mean positions remain separated by a considerable distance. Contact occurs when the rate $k(\ell)$ of contact-forming shape fluctuations is roughly comparable to a rate $k_D$ of further diffusion
\begin{equation}\label{equ:rough_condition}
    k(\ell)\sim k_{D}(\ell).
\end{equation}
This diffusion rate, $k_D \sim D/\lambda^2$, is determined by the diffusion constant $D = \sqrt{2} D_\text{int}$, that is associated with the time evolution of the mean distance between the two interfaces, together with a length scale $\lambda$ whose magnitude is not straightforward to anticipate. For the purpose of roughly calculating $\ell^*$, we assume simply that $\lambda$ is independent of $L$.

According to notions of transition state theory, the rate of an extreme shape fluctuation that establishes contact at fixed $\ell$ should be proportional to its equilibrium probability, i.e.~the probability of an undulation that closes the mean gap. The distribution $P(x)$ of the fluctuating height $x$ at a single point on an interface is simply estimated from Gaussian field theories as $P(x)\sim \exp[-x^2/(2 w^2)]$\precite\cite{Godreche1991}, where $w$ is the \emph{roughness} of the interface and $w^2\sim L$ for $d=1$ and $w^2\sim \log{L}$ for $d=2$. Assuming the gap-closing probability to follow these simple statistics, we estimate $k(\ell) \sim \exp[-\ell^2/(2\sigma^2)]$.

Assembling these arguments, we expect the typical separation $\ell^*$ at contact to scale with $L$ as
\begin{equation}
  \label{equ:rough-scaling}
\ell_{1d}^* \sim \sqrt{L \log{L}} \qquad \textrm{and} \qquad
\ell_{2d}^* \sim \log{L}
\end{equation}
for one- and two-dimensional interfaces, respectively. According to this rough prediction, growth of $\ell^\ast$ with system size is so gradual that $\ell^\ast$ can remain microscopic even for macroscopic systems.

This line of reasoning is loose, imprecise, and incomplete. First, as mentioned, the balance between $k(\ell)$ and $k_D$ involves a length scale $\lambda$, which is not specified here. Secondly, the gap-closing probability should reflect extreme value statistics of the \emph{closest} points on the two interfaces, not that of arbitrarily chosen points. Finally, since long-wavelength undulations of an interface evolve very slowly in time, the assumptions of transition state theory are difficult to justify in this context.

\Presecs\ref{sec:framework} and \ref{sec:application_ew} present a much more careful treatment of the dynamics that lead to interfacial contact. By making controlled approximations for well-defined models, they reveal and clarify subtleties associated with all of the issues listed above. The theory developed yields improved scaling predictions for $\ell^\ast$ that confirm an extremely slow divergence of $\ell^*$ with system size.

\section{A framework for calculating contact distance distributions}\label{sec:framework}
\subsection{Reaction-diffusion equation for the contact distance distribution}\label{sec:reaction_diffusion}
As a first step towards the estimation of the typical contact distance $\ell^\ast$, we devise a simple reaction-diffusion equation that governs the time evolution of the distribution of distances between fluctuating interfaces. Consider two planar interfaces, in a 2- or 3-dimensional simulation box with periodic boundary conditions, that do not interact with each other until they are brought in contact by thermal fluctuations. These interfaces are released at an initial distance $\ell_0$ that is much larger than the typical size of interface fluctuations.
\begin{figure}[tb]
    \centering
    \includegraphics[width=1.0\linewidth]{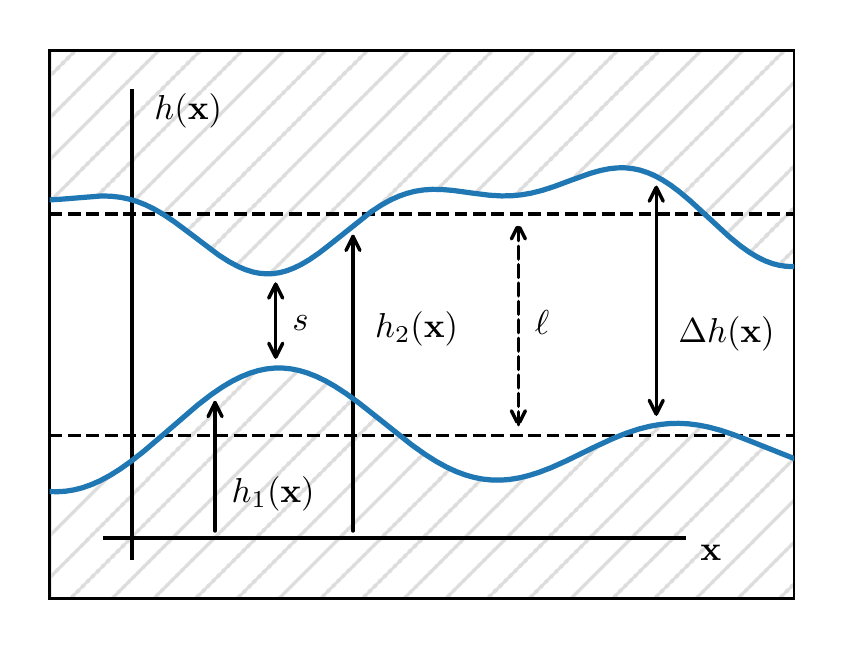}
    \caption{Two surfaces are described by functions $h_1(\vec{x})$ and $h_2(\vec{x})$ where $\vec{x}$ is a vector that collects the lateral coordinates (those orthogonal to the surface normal). The dashed lines indicate the average positions $\bar{h}_1$ and $\bar{h}_2$ of the interfaces, $\ell$ is the average distance between the interfaces, and $s$ is the minimum distance between the two interfaces, i.e.~$s = \min_\vec{x} \left[ \Delta h(\vec{x})\right] = \min_\vec{x} \left[h_2(\vec{x}) - h_1(\vec{x})\right]$.}
    \label{fig:theory_sketch}
\end{figure}
As shown in Fig.~\ref{fig:theory_sketch} we describe the geometry of the surfaces by two continuous functions $h_1(\vec{x})$ and $h_2(\vec{x})$, where the vector $\vec{x}$ collects the lateral coordinates (i.e., those orthogonal to the $h$-direction). The distance between the two interfaces at a given position $\vec{x}$ is denoted by $\Delta h(\vec{x}) = h_2(\vec{x})-h_1(\vec{x})$ and the mean distance of the two interfaces is given by $\ell = \bar{h}_2 - \bar{h}_1$, where $\bar{h}_1$ and $\bar{h}_2$ are the spatially averaged means of $h_1$ and $h_2$, respectively. The distance between the two interfaces at the point where they are closest is denoted by
\begin{equation}
  s = \min_{\vec{x}} \left[\Delta h(\vec{x})\right].
\end{equation}
As long as the two interfaces are not in contact with each other, $\bar{h}_1$ and $\bar{h}_2$ freely diffuse relative to each other along the $h$-direction governed by the diffusion coefficient $D_\text{int}$. In the case of a liquid slab, the interfaces diffuse due to mass transport in the slab, due to evaporation and condensation on the surface, and, in the case of simulations at constant pressure, due to fluctuations of the box size.

The rate $k(\ell)$ with which fluctuations of the interfaces are formed that are large enough to bridge the gap between the two interfaces depends on $\ell$, i.e.~$k = k(\ell)$. This situation is formally equivalent to a particle that diffuses along a single direction $\ell$ and undergoes a reaction with a position dependent reaction rate $k(\ell)$. Using this analogy we write down the following partial differential equation that governs the evolution of the probability density, $\rho(\ell,t)$, that two interfaces have not yet touched, and have mean distance $\ell$ at time $t$:
\begin{equation}\label{equ:reaction_diffusion}
    \pd{\rho (\ell,t)}{t} = \Dint\pdsec{\rho (\ell,t)}{\ell} - k(\ell) \rho(\ell,t)
\end{equation}
The initial condition is given by
\begin{equation}\label{equ:reaction_diffusion_initial}
    \rho (\ell,0) = \delta(\ell - \ell_0),
\end{equation}
where $\delta(x)$ is the Dirac $\delta$-function and $\ell_0$ is the initial separation of the interfaces.

This equation is reminiscent of the theory of diffusion-controlled reactions\precite\cite{Wilemski1973,Mattis1998,Prustel2017}, however, instead of reactions that occur at a boundary, the reaction in our model is controlled by a position- and concentration dependent loss-term $k(\ell) \rho (\ell,t)$. The time integral over the loss term yields the distribution of distances at which the reaction occurs,
\begin{equation}\label{equ:reaction_probability}
    u(\ell) = \int_0^{\infty} \text{d}t ~ k(\ell) \rho (\ell,t).
\end{equation}
In the next section we introduce a path-integral formalism that allows us to calculate $u(\ell)$.

\subsection{A path-integral formulation of $u(\ell)$}\label{sec:pde_solution}
In order to derive an expression for the probability distribution $u(\ell)$, we use a well known correspondence between reaction-diffusion equations and the Schrödinger equation that has been pointed out by various authors before; see Refs. \onlinecite{wiegel1986introduction,Risken1984,parisi1988statistical,kleinert2009path} for comprehensive treatments. Furthermore, our derivation is closely related to the Feynman-Kac path integral formula\precite\cite{Kac1949}.
\begin{figure}[tb]
  \centering
  \includegraphics[width=1.0\linewidth]{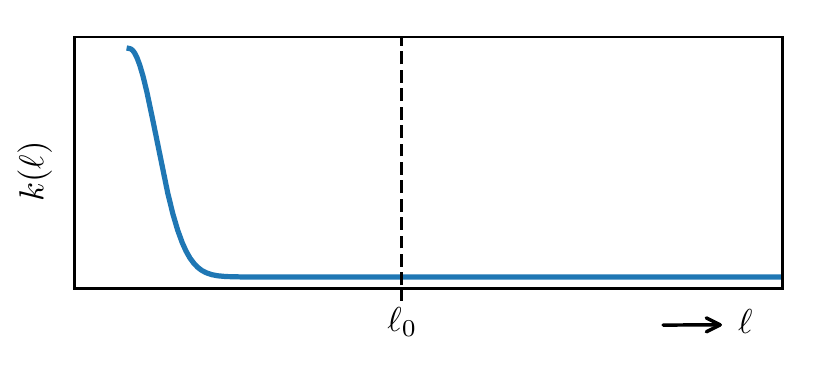}
  \caption{Sketch of a Gaussian rate function $k(\ell)$. The initial position is marked by $\ell_0$.}
  \label{fig:k_gaussian_sketch}
\end{figure}

We start by considering a discrete diffusive trajectory $\inttraj$ that consists of $N$ steps and that has been obtained by sampling a continuous trajectory $\inttrajpoints(t)$ at times $i \Delta t$. We write
\begin{equation}
    \inttraj = \left\{ \inttrajpoints_0, \inttrajpoints_1, \ldots, \inttrajpoints_n, \inttrajpoints_{n+1}, \ldots \inttrajpoints_{N}\right\},
\end{equation}
where $\inttrajpoints_i = \inttrajpoints(i \Delta t)$. The probability of observing $\Lambda$ for given initial position $\inttrajpoints_0$ is given by
\begin{equation}
    P_\text{diff}\left[\inttraj\right] = \prod_{i=0}^{N-1} p(\inttrajpoints_i \rightarrow \inttrajpoints_{i+1}),
\end{equation}
where $p(\inttrajpoints_i \rightarrow \inttrajpoints_{i+1})$ is the transition probability of a freely diffusing random walker:
\begin{equation}
    p(\inttrajpoints_i \rightarrow \inttrajpoints_{i+1}) = \frac{1}{\sqrt{4 \pi \Dint \Delta t}} \exp \left[ -\frac{\left(\inttrajpoints_{i+1} - \inttrajpoints_i\right)^2}{4 D \Delta t}\right]
\end{equation}

Now let us assume that the first contact between the interfaces occurs in the time interval between steps $n$ and $n+1$. The probability of observing such a trajectory is then given by the product of $P_\text{diff}$ with the conditional probability $P_\text{nr}\left[n | \inttraj\right]$ that no reaction occurs until time step $n$ given that we are following trajectory $\inttraj$, and the probability $P_\text{r}[n | \inttraj]$ that the reaction occurs between $n \Delta t$ and $(n+1)\Delta t$, %$P_\text{r}[n | \inttraj]$:
\begin{equation}\label{equ:path_prob}
    P\left[\inttraj\right] =
    P_\text{diff}\left[\inttraj\right]
    P_\text{nr}\left[n | \inttraj\right]
    P_\text{r}[n | \inttraj]
\end{equation}
We approximate $P_\text{r}[n | \inttraj]$ by
\begin{equation}\label{equ:rate_to_prob}
    P_\text{r}\left[\inttrajpoints_n,\inttrajpoints_{n+1}\right] \approx 1 - e^{-k\left(\inttrajpoints_n\right) \Delta t} \approx k\left(\inttrajpoints_n\right) \Delta t,
\end{equation}
i.e.~we assume that $k(\inttrajpoints)$ is approximately constant in the region the particle visits between $\inttrajpoints_n$ and $\inttrajpoints_{n+1}$. In the same manner we also approximate $P_\text{nr}[n | \inttraj]$ by
\begin{equation}
    P_\text{nr}[n | \inttraj] \approx \prod_{i=0}^{n-1}~e^{-k(\inttrajpoints_i) \Delta t}
    = \exp\left[-\Delta t \sum_{i=0}^{n-1} k(\inttrajpoints_i)\right].
\end{equation}
Note that for ease of notation we define here that the product in this equation is equal to one if its end index is smaller than its starting index and, by extension, that the sum on the right hand side is 0 in the same situation.
Substituting these expressions into~\eqref{equ:path_prob} and expanding $P_r$ for small $\Delta t$ yields the probability of a specific trajectory $\inttraj$ that reacts at time $n \Delta t$,
\begin{equation}\label{equ:single_path_probability}
	 \begin{aligned}
	 P\left[\inttraj\right] &= \frac{k(\inttrajpoints_n) \Delta t}{\left(4 \pi \Dint \Delta t\right)^{N/2}} \exp \left[-\Delta t \sum_{i=0}^{n-1} k(\inttrajpoints_i)  - \right. \\
	 &\qquad\left. - \sum_{i=0}^{N-1}\frac{\left(\inttrajpoints_{i+1} - \inttrajpoints_i\right)^2}{4 \Dint \Delta t} \right].
	 \end{aligned}
\end{equation}

The probability $u(\ell)$ of observing a reaction at a specific position $\ell$ is now the sum of $P[\inttraj]$ over all possible paths $\inttraj$ and path lengths $n$ where the initial point $\inttrajpoints_0$ and the final point $\inttrajpoints_n = \ell$ are kept fixed:
\begin{widetext}
\begin{align}
    u\left(\ell\right) &= \lim_{N \to \infty} \sum_{n=0}^{N-1} \int \frac{\intv{\inttrajpoints'_1} \ldots \intv{\inttrajpoints'_{N-1}}}{\left(4 \pi \Dint \Delta t\right)^{N/2}} \delta(\inttrajpoints'_n - \ell) \exp \left[ -\sum_{i=0}^{N-1}\frac{\left(\inttrajpoints'_{i+1} - \inttrajpoints'_i\right)^2}{4 \Dint \Delta t} -\Delta t \sum_{i=0}^{n-1} k(\inttrajpoints'_i)\right] k(\inttrajpoints'_n) \Delta t \nonumber \\
    &= k(\ell) ~ \sum_{n=0}^{\infty} \Delta t \int \frac{\intv{\inttrajpoints'_1} \ldots \intv{\inttrajpoints'_{n-1}}}{\left(4 \pi \Dint \Delta t\right)^{n/2}} \exp \left[ -\sum_{i=0}^{n-1}\Delta t\left(\frac{\left(\inttrajpoints'_{i+1} - \inttrajpoints'_i\right)^2}{4 \Dint (\Delta t)^2} + k(\inttrajpoints'_i)\right)\right]\label{equ:xR_probability}
\end{align}
\end{widetext}
In the second line we have carried out the integrations over all $\inttrajpoints'_i$ with $i \geq n$ and have taken the limit $N \to \infty$. We now take the limit $\Delta t \to 0$ while keeping the trajectory lengths $T = n \Delta t$ fixed. With the abbreviations
\begin{equation}
     \int_0^{\infty} \intv{T} \int \mathcal{D}[\ell_T] \equiv \lim_{\Delta t \to 0} \sum_{n=0}^{\infty} \Delta t \int \frac{\intv{\inttrajpoints_1} \ldots \intv{\inttrajpoints_{n-1}}}{\left(4 \pi \Dint \Delta t\right)^{n/2}},
\end{equation}
and
\begin{equation}\label{equ:action}
    \begin{aligned}
    S[\inttrajpoints_T(t)] &= \int_0^T \intv{t'} \left\{ \frac{v_T(t')^2}{4 \Dint} + k\left[\inttrajpoints_T(t')\right] \right\} \\
    &\equiv \lim_{\Delta t \to 0} \sum_{i=0}^{n-1}\Delta t\left\{\frac{\left(\inttrajpoints_{i+1} - \inttrajpoints_i\right)^2}{4 \Dint (\Delta t)^2} + k(\inttrajpoints_i)\right\},
    \end{aligned}
\end{equation}
this yields
\begin{equation}\label{equ:path_integral}
    u(\ell) = k(\ell) \int_0^{\infty} \intv{T} \int \mathcal{D}[\inttrajpoints_T(t)] e^{-S[\inttrajpoints_T(t)]}.
\end{equation}
Here $\inttrajpoints_T(t)$ is a trajectory of length $T$ and $v_T(t) = \dot{\inttrajpoints}_T(t)$. As suggested by a comparison to \preequ\eqref{equ:reaction_probability}, the probability density $\rho(\ell,t)$ is given by
\begin{equation}\label{equ:path_integral_density}
    \rho(\ell,t) = \int \mathcal{D}[\inttrajpoints_T(t)] e^{-S[\inttrajpoints_T(t)]}.
\end{equation}

\Preequ\eqref{equ:path_integral_density} is known as a path-integral in quantum mechanics (QM) and in the following we will use results obtained there to calculate an approximation for $u(\ell)$. In particular, in analogy to the semi-classical approximation in QM, we can calculate an expression for the probability of the most likely path between two points $\ell_0$ and $\ell$. This can be done most easily by comparing \preequ\eqref{equ:action} to the classical action of a particle in a potential that follows a trajectory $x(t)$,
\begin{equation}
    S\left[x(t)\right] = \int_0^T \text{d}t \left[ \frac{m\dot{x}(t)^2}{2} - V(x(t)) \right],
\end{equation}
with the kinetic energy $K = m \dot{x}^2 / 2$ and the potential energy $V$. The paths with extremal action are the paths that are solutions of the classical equations of motion. Hence, the path $\tilde{\inttrajpoints}(t)$ that minimizes the action\preequNoWord\eqref{equ:action} is the Newtonian trajectory taken by a classical particle with mass $1/2\Dint$ in a potential $V(\inttrajpoints) = -k(\inttrajpoints)$\precite\cite{wiegel1986introduction}, i.e.~the inverted reaction rate function. Hence, the trajectories with maximum probability (or the \emph{classical paths}), $\tilde{\inttrajpoints}_T(t)$, are solutions of the equation
\begin{equation}\label{equ:velocity}
    \dot{\tilde{\inttrajpoints}}_T(t) = \pm \sqrt{4\Dint\left(E+k(\tilde{\inttrajpoints}_T(t))\right)}
\end{equation}
that fulfill the boundary conditions $\tilde{\inttrajpoints}_T(0) = \ell_0$ and $\tilde{\inttrajpoints}_T(T) = \ell$. The constant $E$ that is determined by these boundary conditions together with $T$ (i.e. $E = E(\ell,\ell_0,T)$), resembles the total energy of the classical particle $E = K + V$. Alternatively, \preequ\eqref{equ:velocity} can be derived by optimizing the $\inttrajpoints_i$ in  \preequ\eqref{equ:single_path_probability} for maximum $P[\inttraj]$ and subsequently taking the limit $\Delta t \to 0$.

It is reasonable to assume that the rate of gap-closing fluctuations shrinks with increasing mean distance $\ell$, so in the following we assume that the derivative of the rate function fulfills $k'(\ell) < 0$. This situation is sketched in \prefig\ref{fig:k_gaussian_sketch}. In addition, we place a reflecting boundary at the initial position $\ell_0$. This has two effects: first, the average length of trajectories is finite in this case (which is not necessarily the case if the rate function $k(\ell)$ goes to zero fast enough as $\ell \to \infty$); secondly, the solution  to \preequ\eqref{equ:velocity} is unique for any given set of boundary conditions ($\ell_0$, $\ell$, and $T$), i.e.~there is only a single classical path that leads from $\ell_0$ to $\ell$ in time $T$. Note, that we have chosen $\ell_0 > \ell$ and will keep this convention through the rest of the derivation.

Dynamics with such a reflecting boundary can be mapped onto an unbounded system with a symmetrized rate function $k_\text{sym}(x)$ such that $k_\text{sym}(\ell) = k_\text{sym}(2\ell_0 - \ell)$\cite{Bastianelli2007,kleinert2009path}, i.e.
\begin{equation}
  k_\text{sym}(\ell) = \Theta\left(\ell-\ell_0\right)k(\ell) + \Theta\left(-\ell+\ell_0\right)k(2\ell_0 - \ell),
\end{equation}
where $\Theta(x)$ is the Heaviside step function. The transition probability from $\ell_0$ to $\ell$ in time $T$, $p(\ell_0,\ell,T)$, is then given by
\begin{equation}
  \begin{aligned}
    p(\ell_0,\ell,T) &= p_\text{sym}(\ell_0,\ell,T) + p_\text{sym}(\ell_0,2\ell_0-\ell,T) \\
    &= 2 p_\text{sym}(\ell_0,\ell,T)
  \end{aligned}
\end{equation}
where $p_\text{sym}(\ell_0,\ell,T)$ is the transition probability calculated using the extended rate function $k_\text{sym}$ without boundaries and in the second line we have used the symmetry of $k_\text{sym}$ and the fact that $\ell=\ell_0$ at $t=0$. In other words, one has to sum the path probabilities that move from $\ell_0$ to $\ell$ and another set of paths that move from $\ell_0$ to the symmetrically equivalent $2\ell_0 - \ell$. We calculate the probabilities of these paths using a semiclassical approximation. The principal difficulty in this calculation lies in the fact that the extended rate function $k_\text{sym}(\ell)$ is not analytic at $\ell_0$ and, hence, can only be treated perturbatively\precite\cite{Bastianelli2007}. In the following calculation, we assume that, due to the shape of the rate function $k(\ell)$, the corrections to our semiclassical calculations are small. In \preapp\ref{app:contact_distance_dists} we then numerically check our results for exponential and Gaussian rate functions and find that our calculations are in excellent agreement in the limit $\Dint \to 0$.

We approximate the path integral in \preequ\eqref{equ:path_integral} by expanding the action $S$ up to leading order around the likeliest paths\precite\cite{wiegel1986introduction,Chaichian2001,kleinert2009path}. This approximation yields
\begin{equation}\label{equ:reaction_prob}
    u\left(\ell\right) = \frac{k(\ell)}{Z} \int_{0}^{T_0} \intv{T} ~ e^{-\tilde{S}_T(\ell,\ell_0,T)} \tilde{F}_T(\ell,\ell_0,T),
\end{equation}
where $T_0$ is the length of the longest possible path which has the minimum ``energy'' $E_0 = -k(\ell_0)$, $\tilde{S}_T(\ell,\ell_0,T) = S[\tilde{\inttrajpoints}_T(t)]$ is the action of the likeliest path and the factor $\tilde{F}_T(\ell,\ell_0,T)$ arises from the expansion around the likeliest path. It is given by\precite\cite{kleinert2009path}
\begin{equation}\label{equ:expansion_weight}
  \begin{aligned}
    \tilde{F}_T(\ell,\ell_0,T) &= \left\{ 2 \pi \sqrt{\Dint} \kappa(\ell,\ell_0,E(T)) \vphantom{\int_{\ell_0}^{\ell}}\right. \\ &\left.\qquad\times\int_{\ell}^{\ell_0} \intv{\inttrajpoints'} ~ \left[E(T) + k(\inttrajpoints')\right]^{-3/2}\right\}^{-1/2}
  \end{aligned}
\end{equation}
with
\begin{equation}\label{equ:kappa}
  \kappa(\ell,\ell_0,E) = \sqrt{\left(E + k(\ell_0)\right) \left(E + k(\ell)\right)}.
\end{equation}
The constant $Z$ is introduced in \preequ\eqref{equ:reaction_prob} to normalize the distribution $u(\ell)$ and includes contributions that are neglected due to the assumptions we made regarding the boundary condition at $\ell_0$ and the multiplicity of the classical paths; because the initial position $\ell_0$ is also the location of a reflecting boundary, there are two identical paths for each value of $T$ with the same energy $E$.

The action of the likeliest path of length $T$ can be expressed in terms of an integral over the velocity $\dot{\tilde{\inttrajpoints}}(t)$,
\begin{align*}
    &\tilde{S}_T(\ell,\ell_0,T) = \int_0^T \text{d}t ~ \left[ K - V \right] = \int_0^T \text{d}t ~ \left[-E + 2 K\right] \\
    & \qquad = -E(T) T + \int_0^T \text{d}t ~ \dot{\tilde{\inttrajpoints}}_T \frac{\dot{\tilde{\inttrajpoints}}_T}{2\Dint} \\
    & \qquad = -E(T)\int_{\ell_0}^{\ell} \frac{\intv{\inttrajpoints'}}{\dot{\tilde{\inttrajpoints}}_T(\inttrajpoints')} + \int_{\ell_0}^{\ell} \text{d}\inttrajpoints' ~ \frac{\dot{\tilde{\inttrajpoints}}_T(\inttrajpoints')}{2\Dint},
\end{align*}
where we have used the fact that
\begin{equation}\label{equ:path_T}
    T = \int_{\ell_0}^{\ell} \frac{\intv{\inttrajpoints'}}{\dot{\tilde{\inttrajpoints}}_T(\ell')}.
\end{equation}
By substituting \preequ\eqref{equ:velocity} (using the minus sign because we consider the path from $\ell_0 > \ell$ to $\ell$) we arrive at the expression
\begin{equation}\label{equ:likeliest_action}
    \tilde{S}_T(\ell,\ell_0,T) = \int_{\ell}^{\ell_0} \intv{\inttrajpoints'}~\frac{E(T)/2 + k(\inttrajpoints')}{\sqrt{\Dint(E(T)+k(\inttrajpoints'))}}.
\end{equation}

Due to the placement of the reflecting boundary at $\ell_0$ and the assumption that $k' < 0$, $E(T)$ is a monotonic function of $T$ so that we can rewrite the integral in \preequ\eqref{equ:reaction_prob} as an integral over $E$ by using the derivative
\begin{equation}
  \de{T}{E} = -\frac{1}{8 \pi D \kappa(\ell,\ell_0,E) \tilde{F}_T(\ell,\ell_0,E)^2},
\end{equation}
leading us to the expression
\begin{equation}\label{equ:prob_integral_E}
    u\left(\ell\right) = \frac{k(\ell)}{8 \pi D Z} \int_{E_0}^{\infty} \intv{E}~\frac{e^{-\tilde{S}_E(\ell,\ell_0,E)}}{\kappa(\ell,\ell_0,E) \tilde{F}_E(\ell,\ell_0,E)},
\end{equation}
where the notation $\tilde{S}_E$ and $\tilde{F}_E$ indicates that these functions are now evaluated using $E$ as the independent variable instead of $T$.

This integral can be evaluated using a saddle point approximation. A calculation of the derivatives
\begin{equation}
  \pd{\tilde{S}_E(\ell,\ell_0,E)}{E} = \int_{\ell}^{\ell_0}\intv{\inttrajpoints'}\,\frac{E}{4\sqrt{\Dint {\left(E + k(\inttrajpoints)\right)}^{3}}}.
\end{equation}
and
\begin{equation}
  \pdsec{\tilde{S}_E(\ell,\ell_0,E)}{E} = \int_{\ell}^{\ell_0}\intv{\inttrajpoints'}\,\frac{2 k(\inttrajpoints) - E}{\sqrt{64 \Dint {\left(E + k(\inttrajpoints)\right)}^5}},
\end{equation}
demonstrates that the minimum of $\tilde{S}_E$ can be found at $E=0$, as long as the rate function itself is larger than zero in the whole range between $\ell_0$ and $\ell$. $E=0$ corresponds to the paths where
\begin{equation}
  \dot{\tilde{\inttrajpoints}}(0) = -\sqrt{4 \Dint k{(\tilde{\inttrajpoints}{(0)})}}
\end{equation}
and their action is given by
\begin{equation}\label{equ:action_s0}
  \tilde{S}_0(\ell,\ell_0) = \int_{\ell}^{\ell_0} \intv{\inttrajpoints'}\, \sqrt{k(\inttrajpoints')/\Dint}.
\end{equation}
Rewriting \preequ\eqref{equ:prob_integral_E} by setting $\tilde{S}_E(\ell,\ell_0,E) = \tilde{s}_E(E)/\sqrt{\Dint}$ so that $\tilde{s}_E(E)$ is independent of $\Dint$, yields
\begin{equation}\label{equ:prob_integral_E_D_factored}
    u\left(\ell\right) = \frac{k(\ell)}{8 \pi D Z} \int_{E_0}^{\infty} \intv{E}~\frac{e^{-\tilde{s}_E(E)/\sqrt{\Dint}}}{\kappa(E,\ell) \tilde{F}_E(E)}.
\end{equation}
In the limit $\Dint \to 0$ we can now use Laplace's method\precite\cite{Mathews1970} to approximate the integral using the minimum of $\tilde{S}_E$ at $E = 0$, resulting in
\begin{equation}\label{equ:pi_solution_final}
  u(\ell) \approx \frac{e^{-\tilde{S}_0(\ell,\ell_0)} k(\ell)^{3/4}}{2\Dint^{1/2} k(\ell_0)^{1/4}Z}.
\end{equation}

It is instructive to calculate the location of maximum probability, $\ell^\ast$, by setting
\begin{equation}
  \left. \de{\log u(\ell)}{\ell}\right|_{\ell=\ell^\ast} = 0.
\end{equation}
Evaluating the derivative yields one of the main results of this paper: a simple condition that determines the likeliest reaction distance $\ell^\ast$ in terms of the rate function $k(\ell)$ and the diffusion coefficient $\Dint$:
\begin{equation}\label{equ:maximum_cond}
  \sqrt{\frac{k(\ell^\ast)}{\Dint}} = \frac{3}{4}\left|\frac{k'(\ell^\ast)}{k(\ell^\ast)}\right|
\end{equation}
The prime indicates a derivative with respect to $\ell^\ast$. Note, that this result hinges on the validity of the quadratic expansion\preequNoWord\eqref{equ:reaction_prob} which depends on the details of $k(\ell)$\prefootnote\footnote{The problem discussed in this paper is in structure very similar to the problem of diffusion in a potential as discussed e.g.~in \prerefs\onlinecite{Kampen1977,Caroli1981,Autieri2009}. In this context it has been pointed out that the second order expansion of the path integral breaks down in the case of effective potentials $V$ that feature degenerate, or quasi-degenerate minima. Similarly, for rate functions $k(\ell)$ that feature minima at either $\ell$ or $\ell_0$ a treatment along the lines of \prerefs\onlinecite{Caroli1981} or \onlinecite{Autieri2009} may be necessary. In the case of the Gaussian rate function, which is discussed in the following, $\ell_0$ is never located at a minimum of the effective potential and, hence, the expansion becomes exact in the limit $\Dint \to 0$.}. In \preapp\ref{app:contact_distance_dists} we calculate the distributions $u(\ell)$ for the rate functions $k(\ell)$ that are of interest to us later on and compare them to numeric results in order to test the theory developed so far. This comparison shows excellent agreement in the limit $\Dint \to 0$ which corresponds to the limit of inifinite system size when we consider the contact distance of two interfaces later on.

By squaring both sides and rearranging the factors we can rewrite \preequ\eqref{equ:maximum_cond} to read
\begin{equation}\label{equ:maximum_cond_app}
  k(\ell^\ast) = \frac{\Dint}{[(4/3) (k(\ell^\ast)/k'(\ell^\ast))]^2} =  \frac{\Dint}{[\lambda(\ell^\ast)]^2},
\end{equation}
where $\lambda(\ell^\ast) = (4/3) (k(\ell^\ast)/ k'(\ell^\ast))$. This expression recalls the reasoning of \presec\ref{sec:rough_argument}, equating reaction and diffusion rates at the critical separation $\ell^\ast$. The systematic approximation to the reaction-diffusion equation developed in this section allows us to identify the diffusive length scale ($\lambda$ in \preequ\eqref{equ:rough_condition}) that was previously left ambiguous. It is dictated by how quickly $\log [k(\ell)]$ changes with distance. Specifically, contact of the two interfaces occurs when the reaction time scale $1/k(\ell)$ becomes comparable to the time it takes to diffuse to a location where $k(\ell)$ has changed significantly.

By using a rate function $k(\ell)$, the ansatz\preequNoWord\eqref{equ:reaction_diffusion} tacitly assumes that the equilibration of the shape of the interface happens faster than the center-of-mass diffusion. Based on \preequ\eqref{equ:maximum_cond_app} we can check the internal consistency of this assumption using the pertinent timescale of diffusion, $\tau_D(\ell) = \lambda(\ell)^2 / D$. For the interfaces to be equilibrated up to the first contact between them, $\tau_D$ has to be larger than the timescale associated with the relaxation of the interface, $\tau_\text{rlx}$, for all values of $\ell$ that the system visits before first contact. This yields the condition
\begin{equation}\label{equ:relax_condition}
 1 \ll \left.\frac{\tau_D}{\tau_\text{rlx}}\right|_{\ell > \ell^\ast} = \left.\frac{1}{\tau_\text{rlx}}\frac{(k(\ell) / k'(\ell))^2}{D}\right|_{\ell > \ell^\ast},
\end{equation}

Coming back to our initial problem of determining the scaling of contact distance $\ell^\ast$ with system size, \preequ\eqref{equ:maximum_cond} demonstrates that this scaling is determined by the scaling of both $k(\ell)$ and $\Dint$. In the following section we investigate these scaling behaviors for a simple interface model as an example: the Edward-Wilkinson model.

\section{Application to the EW model}\label{sec:application_ew}
\subsection{The EW-equation as a microscopic model for interfaces}
The Edward-Wilkinson (EW) model\precite\cite{Edwards1982} was first introduced to model the statistics of surfaces that are produced by depositing a granular material onto a surface. It is given by the stochastic differential equation
\begin{equation}\label{equ:ew}
    \pd{h(\vec{x},t)}{t} = \beta \gamma c_{D} \pdsec{h(\vec{x},t)}{\vec{x}} + \tilde{\eta}(t),
\end{equation}
where $h(\vec{x},t)$ is the height of the surface, as shown in \prefig\ref{fig:theory_sketch}, $\gamma$ is the surface- or line tension, $c_D$ is an effective diffusion constant, and $\tilde{\eta}(t)$ represents random noise that is specified separately. Together with the Kardar-Parisi-Zhang equation\precite\cite{Kardar1986} it has since become one of the basic equations used to model surface growth\precite\cite{Barabasi1995}. In order to perform simulations we discretize \preequ\eqref{equ:ew} in space resulting, in the case of the (1+1)-dimensional EW interface, in the set of $N = L/\Delta x$ equations of motion
\begin{equation}\label{equ:dynamics-subst}
    \dot{h}_j = \beta \gamma \Dint_0 \Delta x \left(\frac{h_{j+1} - 2 h_j + h_{j-1}}{(\Delta x)^2}\right) + \sqrt{2 \Dint_0} \eta_j(t),
\end{equation}
where $\Dint_0 = c_D / \Delta x$. We choose the $\eta_j(t)$ to be independent, delta-correlated noise processes with $\left<\eta_j\right> = 0$ and $\left<\eta_i(t) \eta_j(t')\right> = \delta_{ij}\delta(t-t')$. The equations\preequNoWord\eqref{equ:dynamics-subst} then become a set of $N$ coupled overdamped Langevin equations where each random walker $h_j$ diffuses with a diffusion coefficient $\Dint_0$ and is coupled to its immediate neighbors by harmonic springs. These equations can also be derived from a discretization of the Hamiltonian
\begin{equation}\label{equ:surface_model}
  \mathcal{H} = \frac{\gamma}{2} \int \text{d}^{(d-1)}x\, \left(\nabla h(\vec{x})\right)^2,
\end{equation}
where $d$ is the dimensionality. This is the Hamiltonian of an interface with an energy that is proportional to its surface area provided that gradients $\nabla h(\vec{x})$ in the height function are small. The generalization of the above equations to $d$ dimensions is given in \preapp\ref{app:nd_hamiltonian}.

An expansion of $h(\vec{x},t)$ into a Fourier series, substituted into the equations of motion\preequNoWord\eqref{equ:dynamics-subst}, demonstrates that the relaxation timescale $\tau_\text{rlx}$ scales with system size like $\tau_\text{rlx} \sim L^2$\precite\cite{Godreche1991,Gross2018a,Gross2018b}. This result is independent of the dimensionality of the system.

Consider now a pair of identical EW interfaces which we assume to be non-interacting. We are ultimately interested in the rate $k(\ell)$ of interfacial fluctuations that result in the two interfaces coming into contact with each other. As before, $\ell = \bar{h}_2 - \bar{h}_1$ is the distance between the mean positions of the two interfaces. The position of the interface at each point $\vec{x}$ can be written as $h_i(\vec{x}) = \bar{h}_i + \delta h_i(\vec{x})$, where the $\delta h_i$ are the \emph{relative heights} of the two interfaces with respect to their mean position $\bar{h}_i$. We can then write the separation between the two interfaces as
\begin{equation}
  \begin{aligned}
    \Delta h(\vec{x}) &= h_2(\vec{x}) - h_1(\vec{x}) \\
    & = \ell - (\delta h_1(\vec{x}) -   \delta h_2(\vec{x})) = \ell - \Delta(\vec{x}),
  \end{aligned}
\end{equation}
where we have used the abbreviation $\Delta(\vec{x}) = \delta h_1(\vec{x}) - \delta h_2(\vec{x})$ in the last line. The two interfaces touch if there is a point where the separation between them is less than or equal to zero, i.e.~if
\begin{equation}
  \min_\vec{x}\left[\Delta h(\vec{x})\right] = \ell - \max_\vec{x}\left[\Delta (\vec{x})\right] \leq 0.
\end{equation}
Rearranging these terms results in the condition
\begin{equation}\label{equ:cond_gc}
  \max_\vec{x}\left[\Delta (\vec{x})\right] \geq \ell
\end{equation}
that all fluctuations that close the gap between the interfaces must fulfill. These are the events that contribute to the rate $k(\ell)$.

Note, that we can infer the equilibrium statistics of $\Delta(\vec{x})$ from the equilibrium statistics of a solitary interface $h(\vec{x})$ by using the fact that the equilibrium fluctuations of the EW interface are Gaussian\precite\cite{Godreche1991}. Hence, also the $\Delta(\vec{x})$ are distributed according to a Gaussian distribution. In addition, the spatial correlations of $\Delta(\vec{x})$ are simply given by\precite\cite{Majumdar2004,Majumdar2020}
\begin{equation}
  \left< \Delta(0) \Delta(\vec{x}) \right> = 2 \left< \delta h_i(0) \delta h_i(\vec{x}) \right>,
\end{equation}
i.e.~the correlations are the same, except their variance $\left< \Delta(0) \Delta(0) \right>$ is doubled. This variance is also known as the square of the \emph{roughness}, $w$, of the EW interface which is given by\precite\cite{Majumdar2004}
\begin{equation}
  w^2 = \frac{L}{12\beta\gamma}
\end{equation}
for the (1+1) dimensional EW interface and by\precite\cite{Rcz1994}
\begin{equation}
  w^2 = \frac{f(L)}{2 \pi^2 \beta \gamma}
\end{equation}
for the (2+1)-dimensional EW interface. Here, $f(L)$ is a system size dependent, dimensionless factor that approaches $f(L)\sim \log(L)$ as $L \to \infty$\precite\cite{Rcz1994}. Hence, the correlations of $\Delta(\vec{x})$ are the same as the correlations of a solitary EW-interface where the surface tension $\gamma$ has been halved. In the following we use this property to make contact with results that are available in literature as well as to reduce the computational effort required to perform simulations. In particular, instead of calculating the rate of fluctuations of two interfaces that fulfill condition\preequNoWord\eqref{equ:cond_gc}, we instead calculate the rate $K(\zeta)$ at which the \emph{maximum relative height} (MRH) of a single EW-interface, $\max_\vec{x}\left[\delta h(\vec{x})\right]$, reaches a threshold value $\zeta$.

For $\gamma = \tilde{\gamma}/2$, where $\tilde{\gamma}$ is the surface tension of each of the two interfaces of interest, the rate $k(\ell)$ is then given by
\begin{equation}
  k(\ell) = K(\ell).
\end{equation}

\subsection{Simulation details and rate calculation method}\label{sec:rate_calc}
To calculate $K(\zeta)$ for the EW interface we numerically integrate the equations of motion\preequNoWord\eqref{equ:dynamics-subst} using the Euler forward-like scheme
\begin{equation}
  h_i(t+\Delta t) = h_i(t) + \beta D_0 \Delta t F_i(t) + \sqrt{2 D_0 \Delta t} \xi,
\end{equation}
where $\xi$ is a random number chosen from a Gaussian distribution with unit variance and $F_i(t)$ is the force on random walker $i$ that is given by
\begin{equation}
  F_i(t) = \gamma \left(\frac{h_{i+1}(t) - 2 h_i(t) + h_{i-1}(t)}{\Delta x}\right)
\end{equation}
in the case of a 1-dimensional interface. The straightforward generalization to higher dimensions is given in \preapp\ref{app:nd_hamiltonian}. The timestep $\Delta t$ is chosen such that $(\beta\gamma D_0 /\Delta x) \Delta t = 0.005$ and $\beta \gamma D_0 \Delta t = 0.01$ for (1+1) and (2+1) dimensional simulations, respectively. A calculation of the MRH at each timestep then yields trajectories $\zeta(t)$ that we use in the following to calculate the rate $K(\zeta)$.
\begin{figure}[tb]
    \centering
    \includegraphics[width=1.0\linewidth]{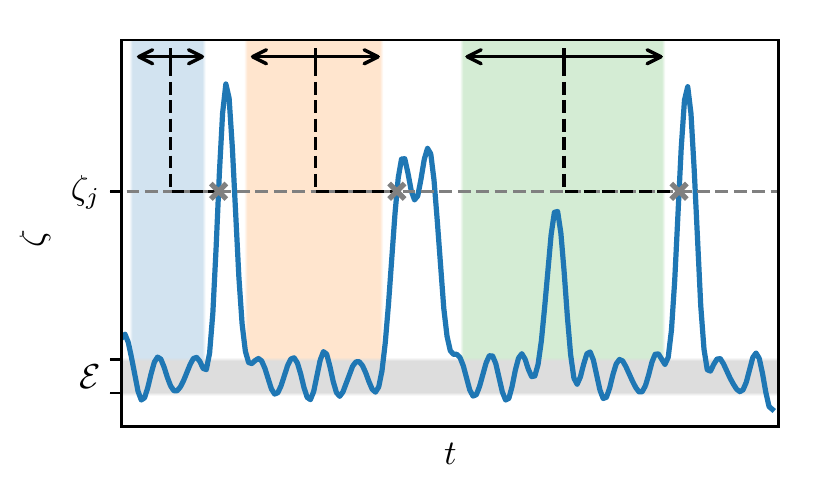}
    \caption{Sketch of the direct method used to calculate the mean first passage time at a checkpoint $\zeta_i$ starting from a region $\mathcal{E}$ close to the minimum of the free energy landscape. Each time the checkpoint $\zeta_i$ is crossed, the time difference to all points within the equilibrium region $\mathcal{E}$ that have been visited since the last time $\zeta_i$ has been crossed, are added to a running average.}\label{fig:mfpt_calc_sketch}
\end{figure}

We define the rate $K(\zeta)$ via the so-called mean first-passage time (MFPT), $\tau(\zeta)$, which is the average time it takes for a configuration to occur with an MRH that is larger than $\zeta$ given that one starts from a configuration drawn from an initial ensemble $\mathcal{E}$ (see \prefig\ref{fig:mfpt_calc_sketch}). If $t_j$ are the times where configurations from the ensemble $\mathcal{E}$ are found, then this average is given by
\begin{equation}\label{equ:mfpt_def}
    \tau(\zeta_i) = \lim_{n\to\infty} \frac{1}{n} \sum_{j=1}^{n} \left( T_i(t_j) - t_j\right),
\end{equation}
where $n$ is the number of samples $t_j$ and $T_i(t_j)$ is the next time checkpoint $\zeta_i$ is reached after time $t_j$. $K(\zeta_i)$ is then calculated via the relationship
\begin{equation}
  K(\zeta_i) = \frac{1}{\tau(\zeta_i)},
\end{equation}
that holds for the so called flux-over-population rate\precite\cite{Reimann1999}. In our case the ensemble $\mathcal{E}$ consists of configurations with an MRH close to the most likely MRH (see \preapp\ref{app:MFPT_calculation} for details).

\begin{figure*}[tb]
    \centering
    \includegraphics[width=1.0\linewidth]{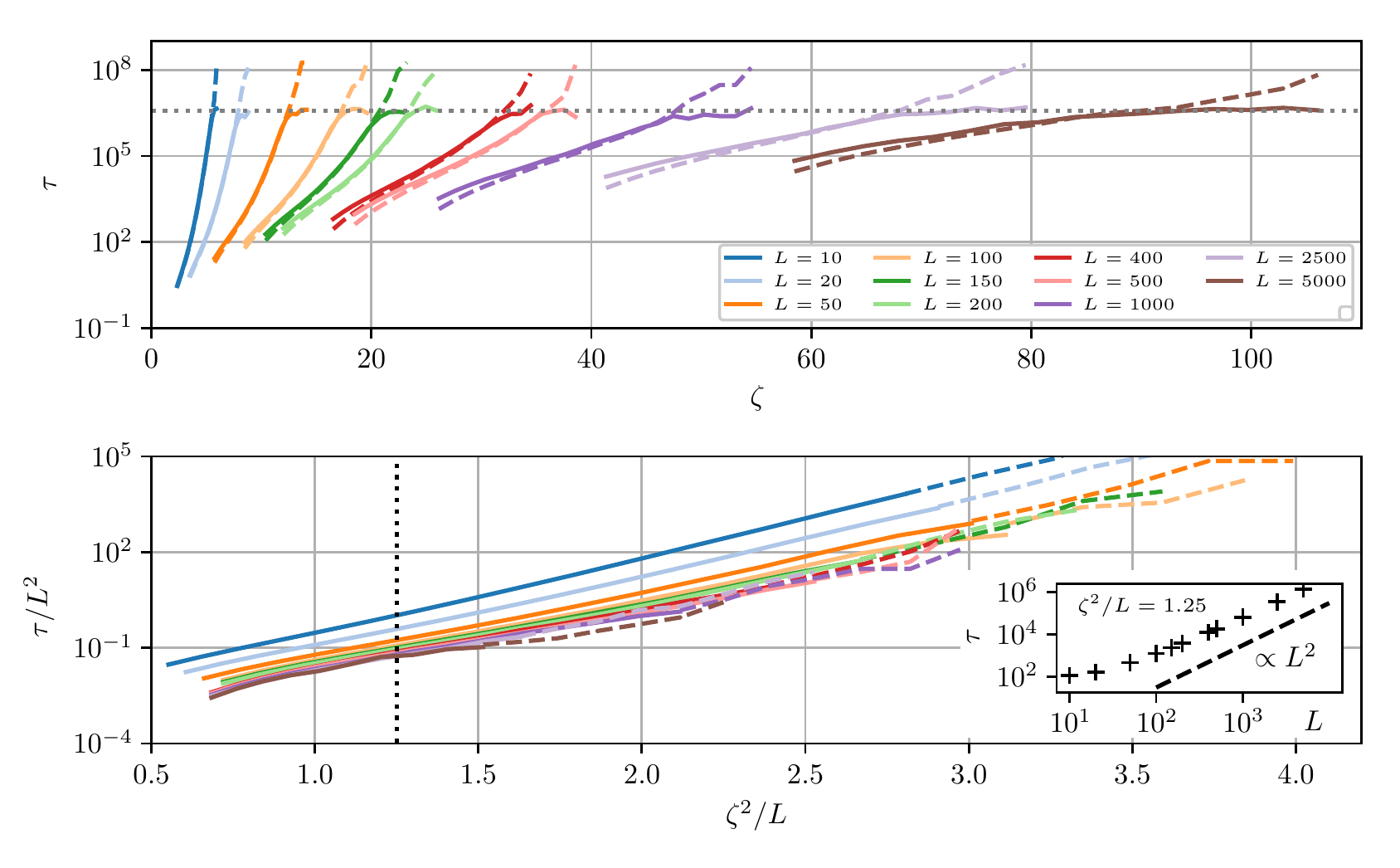}
    \caption{\figtop Mean first-passage times $\tau$ as functions of the MRH $\zeta$ calculated by integrating the Langevin equations of motion\preequNoWord\eqref{equ:dynamics-subst} for (1+1) dimensional interfaces for several system sizes $L$. The MFPTs are calculated from initial configurations where $\zeta$ is close to its average value (see \preapp\ref{app:MFPT_calculation} for details of the calculation). Solid and dashed lines indicate results obtained using the direct and the Poisson method, respectively. The dotted gray line indicates the value of $\tau = \mathcal{T}/4 = 3.75 \times 10^6$ (in dimensionless units) that we expect to be the result of the direct method in the limit $\tau \to \infty$ (see \preapp\ref{app:additional_results} for details). \figbottom Same data with scaled axes as suggested by \preequ\eqref{equ:mfpt_1d_gross}. For clarity, the results of the direct and the Poisson method are stitched together so that the results of each method are shown where we expect them to be accurate. \figinset $\tau$ as a function of $L$ at a fixed value of $\zeta^2 / L = 1.25$. See \preapp\ref{app:scaling} for the same plot at other values of $\zeta^2 / L$.}
    \label{fig:mfpt_histo_analysis_1d_mfpt}
\end{figure*}

The definition\preequNoWord\eqref{equ:mfpt_def} can be used to estimate the MFPT directly by evaluating it based on a finite set of samples, e.g.~those generated by propagating a long unbiased trajectory. This has the advantage that no assumptions are made about the dynamics of $\zeta(t)$ and that, in principle, the method is exact in the limit $n\to\infty$. However, in practice this \emph{direct method} is strongly limited by the length of trajectories available. If the MFPT that one wants to estimate is on the same order as or larger than the length of the trajectory used, $\mathcal{T}$, then the resulting estimates of $\tau(\zeta)$ are systematically smaller than the true MFPT because not all waiting times can be observed with an equal prior probability (consider, for example, waiting times longer than $\mathcal{T}$ which can not be observed at all). In \preapp\ref{app:additional_results} it is shown that for a flat distribution of waiting times the expected observed waiting time is $\mathcal{T}/4$, which is where the MFPTs measured using the direct method level off in \prefigs\ref{fig:mfpt_histo_analysis_1d_mfpt} and \prefigNoWord\ref{fig:mfpt_histo_analysis_2d_mfpt}.

Despite this limitation, in the following we use a slight variation of this direct method (see \preapp\ref{app:MFPT_calculation} for details) to estimate MFPTs without having to make prior assumptions about the statistics of crossing events. We then compare these results to a second set of calculations, where we assume that the statistics of first crossing events---i.e.~the crossings of a checkpoint for the first time after visiting $\mathcal{E}$---are those of a Poisson process. With this approximation the rate can simply be estimated by calculating
\begin{equation}
    K(\zeta_i) = \frac{n_i}{\mathcal{T}},
\end{equation}
where $n_i$ is the number of first-crossing events observed at checkpoint $\zeta_i$ over time $\mathcal{T}$. For this method (the \emph{Poisson method}) the results from different simulations can be combined by simply summing over the respective $n_i$s and the simulation lengths of the individual simulations. The Poisson method significantly extends the range of MFPTs that can be estimated and, additionally, allows us to gauge the influence of correlations between crossing events on their statistics.

\subsection{The (1+1)-dimensional EW interface}
For the (1+1)-dimensional EW interface a wide array of results is available. It has been shown that a tagged random walker within the interface behaves according to fractional Brownian motion\precite\cite{Kolmogorov1940,Mandelbrot1968} with a Hurst exponent of $H=1/4$\precite\cite{Krug1997,Gross2018b} (for conventional diffusion $H=1/2$). This suggests that also the MRH shows some non-Markovianity. Nevertheless, a calculation of first-passage times (FPTs) by \textcite{Gross2018b}, from an initially flat interface to a given value of the MRH, found that they follow a Kramers-like form\precite\cite{Kramers1940} of
\begin{equation}\label{equ:mfpt_1d_gross}
  \tau(\zeta) = \tau_0 \exp\left[\frac{\zeta^2}{2 \scalelengthod^2}\right]
\end{equation}
where $\scalelengthod^2 \sim L / (\beta \gamma)$ and $\zeta \gg w$. A dimensional analysis in \preonlinecite\onlinecite{Gross2018b} shows that $\tau_0 \sim L^{\alpha}$. No theoretical argument has yet predicted the value of the exponent $\alpha$, but numerical results in \preonlinecite\onlinecite{Gross2018b} suggest that $\alpha = 2$.

\Prefig\ref{fig:mfpt_histo_analysis_1d_mfpt} presents a similar analysis, however, we calculate the MFPT not from an initially flat configuration but rather from configurations observed along long trajectories. For small $\tau$ the Poisson method underestimates the MFPTs, as judged by results from the direct method that should be accurate in this regime. In the intermediate range of $\tau$-values we find excellent agreement between the two methods, indicating that different crossing events are indeed largely uncorrelated. For large values of $\tau$ we expect the MFPTs measured using the direct method to approach $\mathcal{T}/4$ (see \preapp\ref{app:additional_results}) and, indeed, this is the behavior we observe.

To estimate the scaling behavior of $\tau(\zeta)$, we show in the lower panel of \prefig\ref{fig:mfpt_histo_analysis_1d_mfpt} the MFPTs as a function of $\zeta^2 / L$ as suggested by \preequ\eqref{equ:mfpt_1d_gross}. We observe that for values $\zeta^2/L$ larger than 1, $\tau(\zeta)$ is approximately proportional to $e^{a \zeta^2/L}$, where $a$ is a constant (as was previously observed for the FPTs). The scaling of the prefactor $\tau_0$ approaches a behavior proportional to $L^2$ in the limit $L \to \infty$.

In summary, we observe the MFPTs to approach the form of \preequ\eqref{equ:mfpt_1d_gross} as the system size and the value of $\zeta$ increase. Assuming, that \preequ\eqref{equ:mfpt_1d_gross} holds, we substitute $k(\ell) = K(\ell) = 1/\tau(\ell)$ into \preequ\eqref{equ:maximum_cond_app}. Solving for $\ell^\ast$ then yields a prediction for the most likely contact distance between two interfaces,
\begin{equation}\label{equ:l_star_gauss}
  \left(\ell^\ast\right)^2 = 2 \scalelengthod^2 W \left( \frac{8 \scalelengthod^2}{9 D \tau_0}\right),
\end{equation}
where $W(x)$ is the inverse of $f(x) = x \exp(x)$, otherwise known as the Lambert-W function. Substituting $\tau_0 \sim L^{\alpha}$, $\scalelengthod^2 \sim L$, $D \sim L^{-1}$ and using the fact that for large arguments $W$ becomes logarithmic\precite\cite{Hassani2007}\prefootnote\footnote{In particular
\begin{equation*}
    \log x - \log \log x < W(x) < \log x.
\end{equation*}}, we get
\begin{equation}\label{equ:scaling_l_star_1p1_final}
  \ell^\ast \sim \sqrt{L \log \left(c L\right)},
\end{equation}
where $c$ is an $L$-independent constant, just as anticipated from the simple reasoning of \presec\ref{sec:rough_argument}. The exponent $\alpha$ ultimately only appears as an argument to a logarithm and, hence, becomes a multiplicative factor in front of the scaling function. Notice the key result here, that the most likely contact distance between two EW interfaces scales sublinearly with the system size.

The equilibrium condition\preequNoWord\eqref{equ:relax_condition} can now be assessed by substituting the rate function\preequNoWord\eqref{equ:mfpt_1d_gross}. Since the diffusion timescale increases with increasing $\ell$, it is sufficient to show that the equilibrium condition is fulfilled at $\ell = \ell^\ast$. Hence, we substitute the solution\preequNoWord\eqref{equ:l_star_gauss}, approximate $W$ by the logarithm and simplify the expression to
\begin{equation}\label{equ:equ_cond_gauss}
  \left.\frac{\tau_D}{\tau_\text{rlx}}\right|_{\ell=\ell^\ast} \approx \frac{1}{2}\frac{\scalelengthod^2}{\tau_\text{rlx} D} \sim \frac{L}{L^2 (1/L)} \sim \mathrm{const.}
\end{equation}
This implies that, if the interfaces have time to relax at one system size around $\ell^\ast$, the relaxation condition\preequNoWord\eqref{equ:relax_condition} is fulfilled for all system sizes, allowing us to extrapolate our results to large system sizes.

In \presec\ref{sec:slab_simulations} we test these results using simulations of the
two-dimensional Ising model as an example. First, however, we present a similar analysis of the (2+1)-dimensional interface.

\subsection{The (2+1)-dimensional EW interface}
\begin{figure}[tb]
    \centering
    \includegraphics[width=1.0\linewidth]{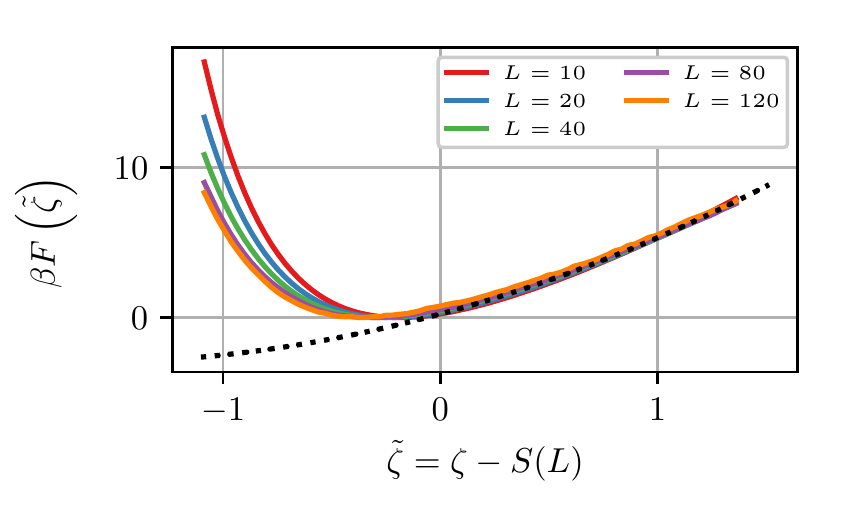}
    \caption{Free energy $\beta F(\tilde{\zeta}) = -\log P(\tilde{\zeta})$ as a function of the shifted maximum relative height, $\tilde{\zeta}$, for the (2+1)-dimensional Edward-Wilkinson interface and several system size $L$. The dotted line is a fit of \preequ\ref{equ:quadr_fit} with parameters  $\tilde{\zeta}_0 = -1.66$ and width $\scalelengthtdg = 0.65$. The data has been obtained using umbrella sampling simulations with harmonic biases that were subsequently matched using the WHAM method\precite\cite{Ferrenberg1989,Kumar1992,Grossfield2013}. For details on these simulations see \preapp\ref{app:mc_simulations}.}
    \label{fig:surface_mc_p_fe_2d_collapsed}
\end{figure}

Less is known about the properties of the (2+1)-dimensional EW interface than the (1+1)-dimensional interface. The equilibrium probability distribution $P(\zeta)$ is expected to be Gaussian for large values of $\zeta$\precite\cite{Gyorgyi2003,Lee2005,Oliveira2008} and in the limit $N = L/\Delta x \to \infty$ the distributions for different system sizes (but the same value of $\beta\gamma$) can be collapsed by applying the transformation\precite\cite{Lee2005}
\begin{equation}\label{equ:sfc_2d_transform}
    \tilde{\zeta} = \zeta - \sqrt{\frac{2}{\pi\beta\gamma}} \log \left(\frac{L}{\Delta x} \right) = \zeta - S(L),
\end{equation}
as shown in \prefig\ref{fig:surface_mc_p_fe_2d_collapsed}. Here, we have defined the shift $S(L) = \sqrt{2/(\pi\beta\gamma)}\log (L/\Delta x)$. Notice, that this expression explicitly references the discretization length $\Delta x$ even in the continuous limit. This is a direct consequence of the fact that the width of the (2+1)-dimensional interface diverges in the same limit\precite\cite{Edwards1982,Rcz1994} and, therefore, one has to choose a fixed discretization length in order to predict quantities related to the width of the interface. \Prefig\ref{fig:surface_mc_p_fe_2d_collapsed} shows the free energies $\beta F(\tilde{\zeta}) = - \log P(\tilde{\zeta})$ for a number of system sizes together with a fit of
\begin{equation}\label{equ:quadr_fit}
    \beta F_\text{gauss}(\tilde{\zeta}) = \frac{\left(\tilde{\zeta} - \tilde{\zeta}_0\right)^2}{2 \scalelengthtdg^2}
\end{equation}
to the large $\tilde{\zeta}$ tail of $\beta F$. While the fact that the tail of these distributions is Gaussian can be shown from first principles\precite\cite{Lee2005}, no theoretical argument that determines the constant $\tilde{\zeta}_0$ exists so far. We determined $\sigma_g$ and $\tilde{\zeta}_0$ from the fit shown in \prefig\ref{fig:surface_mc_p_fe_2d_collapsed}.

The functional form of $\tau(\zeta)$ is not known analytically. As was done in the (1+1) dimensional case, also here one can propose a Kramers-like form for the MFPT based on \preequ\eqref{equ:quadr_fit}, given by
\begin{equation}\label{equ:mfpt_2-d-kramers}
    \tau(\tilde{\zeta}) = \tau_0 \exp \left[\frac{(\tilde{\zeta} -\tilde{\zeta}_0)^2}{2\scalelengthtdg^2}\right].
\end{equation}
\begin{figure*}[tb]
    \centering
    \includegraphics[width=1.0\linewidth]{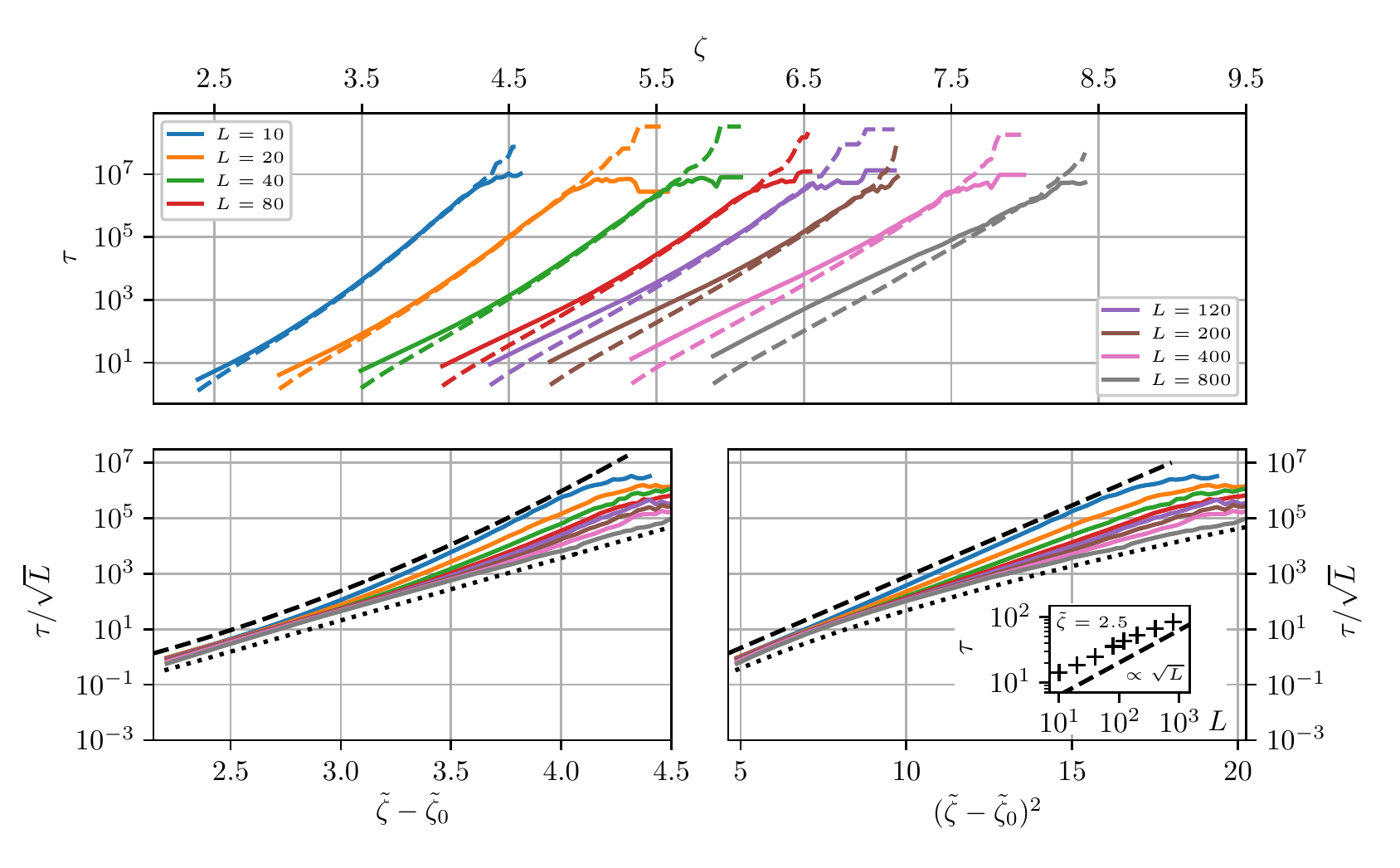}
    \caption{Mean first-passage times of the maximum relative height at checkpoints $\zeta$, $\tau(\zeta)$ calculated from long equilibrium simulations of (2+1)-dimensional EW interfaces of several sizes $L$. \figtop shown are the results obtained using the direct method (solid lines), and using the Poisson method (dashed lines). \figbottomleft same data scaled by $\sqrt{L}$ and as a function of the shifted values $\tilde{\zeta} - \tilde{\zeta}_0$. \figbottomright same data as a function of $(\tilde{\zeta} - \tilde{\zeta}_0)^2$. The dashed black lines represent \preequ\eqref{equ:mfpt_2-d-kramers} where $\tau_0$ has been used to shift the curve. The values of $\scalelengthtdg$ and $\tilde{\zeta}_0$ were taken from the fit shown in \prefig\ref{fig:surface_mc_p_fe_2d_collapsed}. The dotted lines show the exponential behavior of \preequ\eqref{equ:mfpt_2-d-exp} where the parameters $\tau_0$ and $\scalelengthtdexp$ were chosen by fitting to the data and the curve was subsequently shifted towards smaller $\tau$ for clarity. As in the (1+1)-dimensional case, $\tau_\text{direct}$ approaches $\mathcal{T}/{4}$ in the limit of large $\tau$. \figinset $\tau$ as a function of $L$ at a fixed value of $\tilde{\zeta} = \zeta - S(L)= 1.25$. See \preapp\ref{app:scaling} for the same plot at other values of $\tilde{\zeta}$.}\label{fig:mfpt_histo_analysis_2d_mfpt}
\end{figure*}
\Prefig\ref{fig:mfpt_histo_analysis_2d_mfpt} shows the MFPTs calculated from simulations. For the smallest system size of $L=10$ excellent agreement with \preequ\eqref{equ:mfpt_2-d-kramers} is found by fitting $\tau_0$ while using the parameters $\tilde{\zeta_0}$ and $\scalelengthtdg$ determined from the free energy landscape.

As $L$ increases, however, two changes can be observed: first, the $\zeta$-ranges where direct- and Poisson method yield different results become larger and, secondly, as $L$ becomes larger the $\tau(\zeta)$ is not well described by \preequ\eqref{equ:mfpt_2-d-kramers}. Rather, the MFPTs depend exponentially on $\zeta$, like
\begin{equation}\label{equ:mfpt_2-d-exp}
    \tau(\tilde{\zeta}) = \tau_0 \exp\left[\frac{\tilde{\zeta} - \tilde{\zeta}_0}{\scalelengthtdexp}\right],
\end{equation}
where $\tau_0 \sim \text{const}$. We are not aware of any theoretical prediction of this behavior, and in the following we will explore the consequences of both the behavior of \preequ\eqref{equ:mfpt_2-d-kramers} and the exponential behavior of \preequ\eqref{equ:mfpt_2-d-exp} on the most likely contact distance, $\ell^\ast$.

We start with the exponential rate function implied by \preequ\eqref{equ:mfpt_2-d-exp}, which appears to hold for large $L$ in the EW model:
\begin{equation}
    k(\ell) = \tau_0^{-1} \exp\left[-\frac{\tilde{\zeta} - \tilde{\zeta}_0}{\scalelengthtdexp}\right]
\end{equation}
Substituting into \preequ\eqref{equ:maximum_cond} and solving for $\ell^\ast$ yields
\begin{equation}\label{equ:l_star_2d_exp}
  \ell^\ast  = \scalelengthtdexp\log\left(\frac{16 \scalelengthtdexp^2}{9 D \tau_0}\right) + S(L) + \tilde{\zeta}_0.
\end{equation}
With the scalings $\scalelengthtdexp \sim \text{const.}$, $D \sim L^{-2}$, and $\tau_0 \sim \sqrt{L}$ the scaling of $\ell^\ast$ with system size becomes
\begin{equation}
    \ell^\ast \sim \log(c L),
\end{equation}
where $c$ gathers all $L$-independent factors. With the same scalings the equilibrium condition\preequNoWord\eqref{equ:relax_condition} yields
\begin{equation}
  \frac{\tau_D}{\tau_\text{rlx}} = \frac{\scalelengthtdexp^2}{\tau_\text{rlx} D} \sim \frac{\mathrm{const.}}{L^2 L^{-2}} \sim \mathrm{const.},
\end{equation}
indicating that, if the system fulfills the equilibrium condition at one system size, it also fulfills it at larger system sizes and that we can extrapolate our results to large $L$.

The above scaling is exactly what was expected based on the simple derivation sketched in \presec\ref{sec:rough_argument}. However, the path that leads to this result is unexpected: it requires a careful treatment of the contact distance distribution and is based on the unexpected rate expression\preequNoWord\eqref{equ:mfpt_2-d-exp} that likely is the result of strong correlations in the dynamics of the interface.

The Kramers rate expression\preequNoWord\eqref{equ:mfpt_2-d-kramers}, that was the basis of the argument in \presec\ref{sec:rough_argument}, leads to the slightly different result
\begin{equation}
  \left(\ell^\ast - S(L) - \tilde{\zeta}_0\right)^2 = 2 \lambda^2 W \left( \frac{8 \scalelengthtdg^2}{9 D \tau_0}\right),
\end{equation}
where, again, $W(x)$ is the Lambert W-function. Substituting the above scalings then leads to
\begin{equation}\label{equ:l_star_2d_gauss}
    \ell^\ast \sim \sqrt{\log\left(c L\right)} + \sqrt{\frac{2}{\pi\beta\gamma}} \log \left(\frac{L}{\Delta x}\right)  + \tilde{\zeta}_0,
\end{equation}
where $c$ is a constant with regard to $L$ and we have approximated $W(x)$ by $\log(x)$\precite\cite{Hassani2007} as before. The equilibrium condition\preequNoWord\eqref{equ:relax_condition} reads as
\begin{equation}
  \frac{\tau_D}{\tau_\text{rlx}} = \frac{1}{2}\frac{\scalelengthtdg^2}{D \tau_\text{rlx}} \sim \frac{\mathrm{const.}}{L^2 (1/L^2)} \sim \mathrm{const.},
\end{equation}
once again indicating that, if the interface has time to equilibrate at a given system size, it also has sufficient time to do so at larger system sizes. The above result is somewhat more involved since it involves the discretization length $\Delta x$. But also here, the growth of $\ell^\ast$ with system size $L$ is expected to be slow.

In the next section we compare these predictions derived using the EW model to simulations of more realistic systems: the Ising model and a system that contains Lennard-Jones liquid-vapor interfaces.

\section{The stability of slabs in simulations}\label{sec:slab_simulations}
As an example for systems where the interactions of two interfaces play an important role, we examine the stability of slabs formed in molecular simulations of phase-separated systems where periodic boundary conditions are used (see \prefigs\ref{fig:LJ_slab} and \ref{fig:example_config_slab} for examples). Here, the slab shape is the thermodynamically stable shape for clusters within a certain range of sizes, because the overall surface is smaller than the surface that would be formed by other geometries (spheres, cylinders, or disks). The sizes where the slab is thermodynamically stable can be determined by comparing the surface area of the slab to the competing geometries---a disk in (1+1) dimensions and a cylinder in (2+1) dimensions. Denoting with $\Omega$ the area or volume of the system simulated and with $\Omega_\text{c}$ the area or volume of the critical cluster where the slab geometry becomes stable with respect to the competing geometries, the result is $\Omega_\text{c} / \Omega = 1/\pi$ in both (1+1)- and (2+1) dimensions. In other words: the slab geometry becomes thermodynamically stable when the cluster makes up more than a certain area/volume fraction of the system. As the cluster becomes larger than half the size of the system, the two phases switch their roles and the sequence of geometries inverts; the phase that previously formed the cluster now envelopes a cluster of the other phase that takes on the different geometries.

In order for a slab to change its shape to another geometry the two interfaces have to come into contact first. Assuming that this is the rate controlling step of the process, the width of slabs that are observed immediately prior to their transition to a disk- or a spherical shape is therefore expected to scale according to \preequ\eqref{equ:scaling_l_star_1p1_final} and \preequ\eqref{equ:l_star_2d_exp}, in (1+1) and (2+1) dimensions, respectively.

In the following we investigate these widths in simulations of the ferromagnetic Ising model for zero field in (1+1) and (2+1) dimensions as well as in simulations of a Lennard-Jones liquid at coexistence with the gas phase. All quantities are presented in terms of reduced units of the respective system.

\subsection{Simulation details}
Ising model simulations are performed on square and cubic lattices with ferromagnetic nearest-neighbor interactions vanishing external field and periodic boundary conditions in all directions. We set up a simulation box that contains a slab of spins pointing in the up direction, which includes $50\%$ of the available spins. Trajectories are run until the first hole appears in the slab and the configuration observed just prior is analyzed. The detection of holes in the slabs is detailed in \preapp\ref{sec:geometry_detection}. Roughly $500$ trajectories have been generated for each system size.

\begin{figure}[tb]
    \centering
    \includegraphics[width=1.0\linewidth]{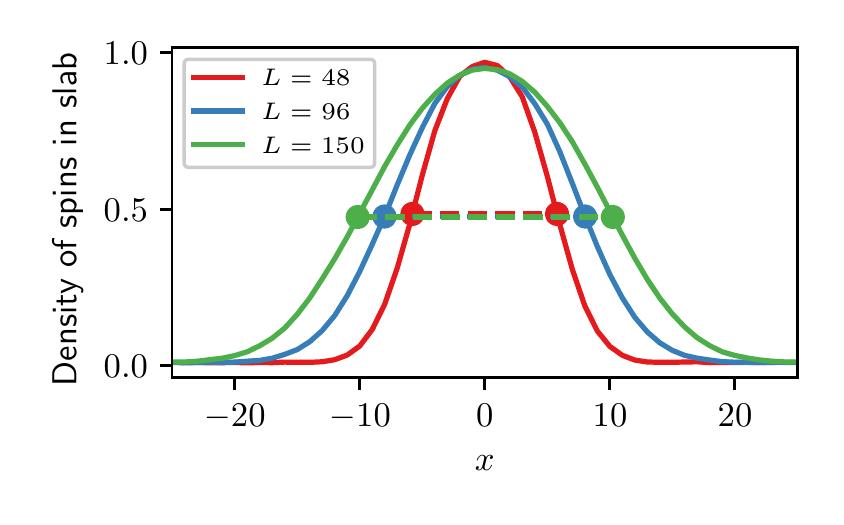}
    \caption{Density of spins that belong to slabs that are observed just prior to a hole forming in trajectories of the 2d-Ising model. Prior to averaging, the slabs have been aligned with respect to their centers-of-mass at $x = 0$. The distance $x$ is measured in units of the lattice spacing between spins. Different colors correspond to different linear system sizes $L$. The widths indicated are the full width at half maximum.}
    \label{fig:slab_density}
\end{figure}

In order to find the average width of the slabs in these configurations, the configurations are then aligned according to each slab's center-of-mass and the spin values are averaged over the in-plane directions of the slab. From the resulting distribution (see Fig.~\ref{fig:slab_density} for examples) we calculate the full-width-half-maximum distance (FWHM) which is plotted as $l^{\ast}$ in Figs.\,\ref{fig:l_star_L_2d} and \ref{fig:l_star_L_3d}. Alternatively, the width of the slab can be calculated from the magnetization of the configurations, taking into account the bulk magnetization of the all-up and the all-down phase at the given temperature (see App. \ref{sec:slab_width_from_magn} for details). Both approaches yield similar results.

In the case of the 3-dimensional LJ system, we first calculate the vapor pressure of a system of 20133 atoms in a slab configuration (periodic boundary conditions applied in all directions), in the canonical ensemble at $T = 0.63$, using a simulation box of $35\times{}35\times{}35$. After averaging over \SI{500000}{} timesteps, sampling every 10 steps, we obtain a pressure of $p = \SI[separate-uncertainty = true]{5.9(5)e-3}{}$, which we use as the reference pressure for the subsequent sets of $NpT$ simulations. All molecular dynamics simulations were performed using the GROMACS simulation package\precite\cite{Abraham2015} (version 5.1.2) with an interaction cut-off of $2.05$. We integrated the equations of motion using the leap-frog integrator with a time step of $4.65 \times 10^{-4} $ which is equivalent to $\SI{1}{\femto\second}$ using parameters for argon. To simulate the $NVT$ and $NpT$ ensembles, we used the Nosè--Hoover\precite\cite{nose84a,Hoover1985a} thermostat and the Parrinello--Rahman barostat\precite\cite{parrinello1981}, respectively.

Next, we prepare the systems for the $NpT$ simulations with different cross-sectional areas by cutting out boxes of  $4.4\times{}4.4$, $8.8\times{}8.8$, $17\times{}17$, and $26\times{}26$ from the original one. For each size, we run a short $NVT$ relaxation and generate 20 different initial states by choosing random initial velocities taken from a Maxwell distribution at $T = 0.63$. We integrate the equations of motion at constant temperature ($T = 0.63$) and normal pressure ($p = 5.9$) for \SI{200000}{} timesteps for each of the randomized starting configurations.  For this particular setup that sets only the pressure normal to the interfaces, we use three uncoupled Parrinello-Rahman barostats, each in one of the three spatial directions, but we vary the box length only along the direction of the interface normal. In this way, we impose a constant length for the simulation box edges along the two directions perpendicular to the surface normal. Because the conditions chosen are close to coexistence, during these runs some of the simulation boxes increase their volume, while some others decrease it, bringing the two interfaces eventually in contact.

Unlike in the Ising model, in an off-lattice system like the Lennard-Jones fluid it is necessary to introduce an {\textsl ad-hoc} characteristic distance $d_c$ to determine which molecules belong to the vapor phase, and when two interfaces can be considered to be in contact. We have chosen the value $d_c = 1.32$, roughly corresponding to the distance at which the pair distribution function crosses 1.0 after the first maximum. We used the distance $d_c$ to perform a cluster analysis of the frames, where we determined the liquid fraction of the system to be the largest connected cluster in the system. Next, we performed an analysis of the interfacial atoms using the ITIM algorithm\precite\cite{partay2008new} in the \texttt{pytim} analysis package\precite\cite{Sega2018}\footnote{Available at \texttt{https://github.com/Marcello-Sega/pytim}.} using a probe sphere radius of $0.5$\precite\cite{sega2016role}.  In a last step, we use an additional clustering analysis with the same cutoff parameter $d_c$,  this time performed only on the set of interfacial atoms, and consider the two interfaces to be in contact as soon as the two interfacial layers become a single cluster.

Since the interfacial analysis is computationally expensive, a bisection approach can be helpful in the analysis of trajectories, where one checks only the frame in the middle of the search interval, updating the search interval to the right half if no contact is found.  This strategy reduces the number of frames to be analyzed from the order of the total number  $\mathcal{O} (N_t)$ to $\mathcal{O}(\log_2 N_t)$, where $N_t$ is the number of frames obtained from the simulation over time $t$.  Once the frame at which the contact takes place is identified, we compute the root mean square width of the slab as
\begin{equation}\label{equ:delta_lj}
    \delta = \sqrt{\frac{1}{N_\textrm{s}}\sum_i^{N_\textrm{s}} (z_i - \bar{z})^2},
\end{equation}
where the sum extends over the $N_\textrm{s}$ surface atoms, $z_i$ is the position along the surface normal of an interfacial atom, and $\bar{z} = \sum{z_i}/ N_s$. Figure~\ref{fig:l_star_L_LJ} shows the averages of $\delta$ over all trajectories observed for a given linear system size $L$ as $l^\ast(L)$.

\subsection{Results}
\begin{figure}[tb]
    \centering
    \includegraphics[width=1.0\linewidth]{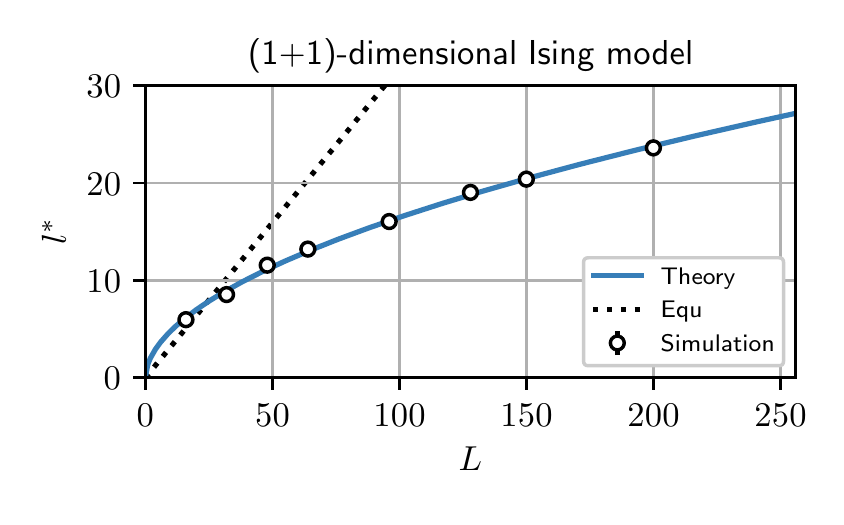}
    \caption{$l^{\ast}(L)$ for the (1+1)-dimensional Ising model. The widths $\ell^\ast$ are the average of the full-width at half-maximum (FWHM) of the spatial distribution of spins that belong to a cluster that is observed just prior to a hole forming. Error bars indicating an error level of one standard deviation are smaller than the symbol size. The blue line is a fit to $C_1 \sqrt{L W(C_2 L)}$ in line with \preequ\eqref{equ:l_star_gauss}. The resulting parameters are shown in \pretab\ref{tab:l_star_L_fit_pars}. The dotted black line indicates the slab width below which a disk-shaped cluster becomes the stable configuration in equilibrium.}
    \label{fig:l_star_L_2d}
\end{figure}

\Prefig\ref{fig:l_star_L_2d} shows the data obtained using the (1+1)-dimensional Ising model together with a fit to the predicted scaling\preequNoWord\eqref{equ:l_star_gauss}. The data is in excellent agreement with the predicted scaling.

\begin{figure}[tb]
    \centering
    \includegraphics[width=1.0\linewidth]{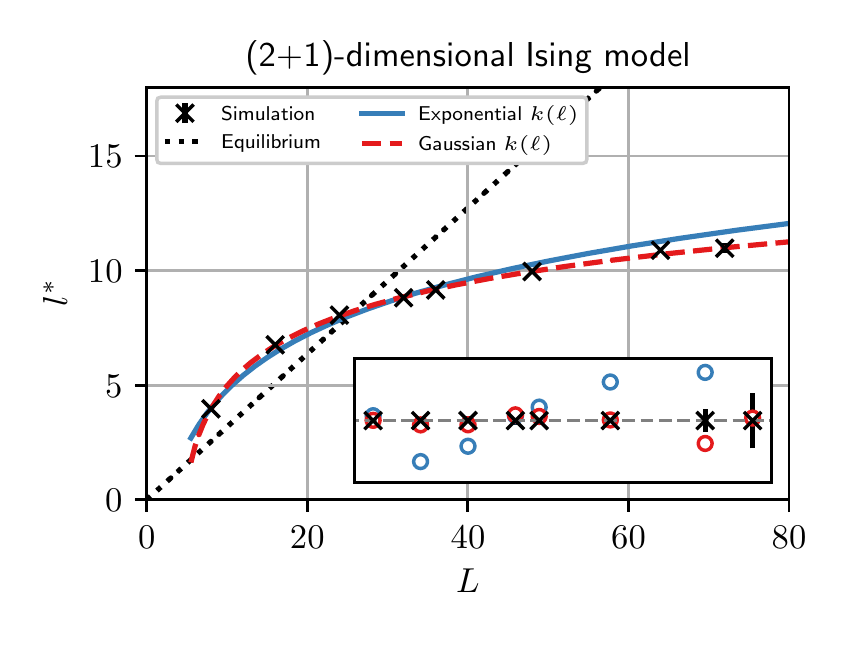}
    \caption{$l^{\ast}(L)$ for the (2+1)-dimensional Ising model obtained by calculating the full-width at half-maximum (FWHM) of the spatial distribution of spins that belong to a cluster that is observed just prior to a hole forming. Error bars indicate an error level of one standard deviation. The lines are fits to $\alpha_1 \log(\alpha_2 L)$ and $\beta_1 \sqrt{\log(\beta_2)} + S(L) + \tilde{\zeta}_0$, corresponding to choosing $k(\ell)$ as an exponential function and a Gaussian, respectively (see \preequs\eqref{equ:l_star_2d_exp} and\preequNoWord\eqref{equ:l_star_2d_gauss}). The resulting parameters are shown in \pretab\ref{tab:l_star_L_fit_pars}. The dotted black line indicates the slab width below which a cylindrical cluster becomes the stable configuration in equilibrium. The inset shows the deviations of the fitted functions from the observed data.}\label{fig:l_star_L_3d}
\end{figure}

\begin{figure}[htpb]
    \centering
    \includegraphics[width=1.0\linewidth]{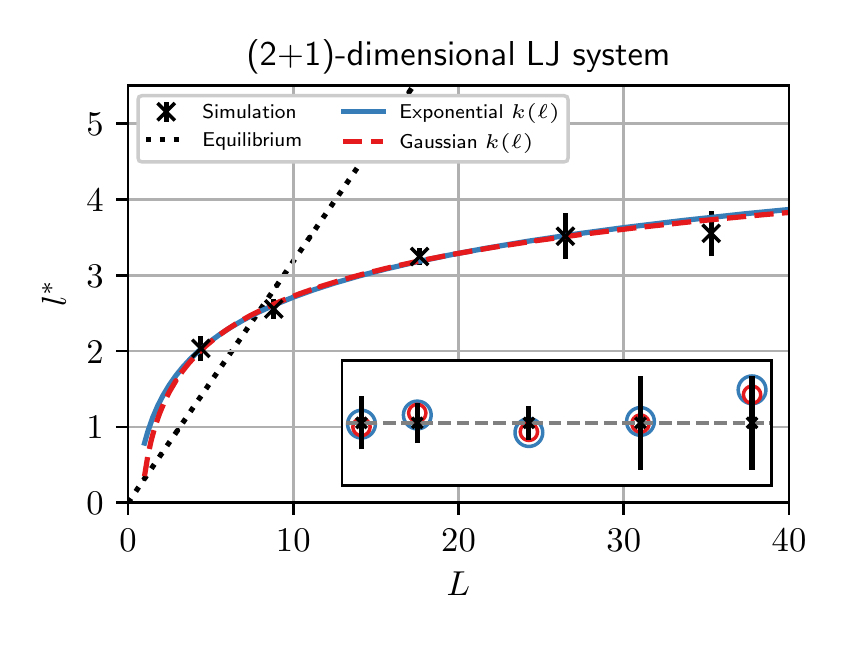}
    \caption{$l^{\ast}(L)$ for the (2+1)-dimensional Lennard-Jones system where a slab of liquid coexists with a slab of gas. The widths are determined by averaging the values $\delta$ in \preequ\eqref{equ:delta_lj} over all observed trajectories. The error bars indicate an error level of one standard deviation. The lines are fits to $\alpha_1 \log(\alpha_2 L)$ and $\beta_1 \sqrt{\log(\beta_2)} + S(\beta_3 L)$, corresponding to choosing $k(\ell)$ as an exponential function and a Gaussian, respectively (see \preequs\eqref{equ:l_star_2d_exp} and\preequNoWord\eqref{equ:l_star_2d_gauss}). The resulting parameters are shown in \pretab\ref{tab:l_star_L_fit_pars}. The dashed black line indicates the slab width below which a cylindrical cluster becomes the stable configuration in equilibrium. The inset shows the deviations of the fitted functions from the observed data.}\label{fig:l_star_L_LJ}
\end{figure}

For (2+1)-dimensional interfaces two predictions have been made in \presec\ref{sec:application_ew} that are derived from assuming different dynamics of the maximum relative height of the interface expressed by the rate function $k(\ell)$. In \prefigs\ref{fig:l_star_L_3d} and\prefigNoWord\ref{fig:l_star_L_LJ} fits to both predicted scalings are shown. Note, that the second term in the scaling law for the (2+1)-dimensional case, \preequ\eqref{equ:l_star_2d_gauss}, requires knowledge of the surface tension $\gamma$. In case of the (2+1)-dimensional Ising model we use a value of $\gamma = 1.15095$ as obtained in \preonlinecite\onlinecite{Bittner2009}. We use the surface tension extrapolated to $L = \infty$ here, since finite size effects are already included in the EW model. In the case of the LJ liquid-vapor interface we use a value of $\gamma = 1.97$, which is obtained from a linear extrapolation of the data provided in \preonlinecite\onlinecite{Neyt2011} to the temperature used in the simulations presented here.

\begin{table}[tb]
    \centering
    \caption{Parameters obtained by the fits shown in \prefigs\ref{fig:l_star_L_2d} through \prefigNoWord\ref{fig:l_star_L_LJ}. All parameters are given in reduced units.}\label{tab:l_star_L_fit_pars}
    \begin{tabular}{|c|c|c|}
      \hline
        Ising (1+1)-dim. & Ising (2+1)-dim. & LJ (2+1)-dim. \\
      \hline
        $C_1 = 0.43 $             & $\alpha_1 = 3.5$    & $\alpha_1 = 0.84$\\
        $C_2 = 2.6 \times 10^5 $ & $\alpha_2 = 0.39$   & $\alpha_2 = 2.6$\\
                                            & $\beta_1  = 6.1$    & $\beta_1  = 1.6 $\\
                                            & $\beta_2  = 0.21 $   & $\beta_2  = 7.4$\\
                                            &                               & $\beta_3  = 0.49$\\
      \hline
    \end{tabular}
\end{table}

For the (2+1)-dimensional Ising model the best fit is obtained by assuming a Gaussian form for $k(\ell)$, while the prediction for exponential $k(\ell)$ shows systematic deviations. The agreement between data and the prediction for a Gaussian rate function is excellent, indicating that the fluctuations of the interfaces in the system are, indeed, well described by Gaussian rate functions.

In the (2+1)-dimensional LJ system such a clear distinction can not be made. Both assumptions for $k(\ell)$ yield fits of similar quality. Note, that since the value of $\Delta x$ is unknown in the case of the LJ model, we have used its value as an additional fit parameter. Nevertheless, the slow growth of the contact distance is confirmed.

\Prefigs\ref{fig:l_star_L_2d} through \ref{fig:l_star_L_LJ} also show equilibrium expectations for the stability of slabs. For values of the distance between the two interfaces $\ell$ below the dashed black lines, slabs are thermodynamically unstable with respect to other geometries. From both theory and simulations we find that slabs persist to much smaller interface separations. For large enough system sizes, the transition from slab to the other geometries is suppressed in all cases by the slow dynamics of extreme interfacial fluctuations. Also, the discrepancy with equilibrium expectations grows with increasing system size. Conversely, for small system sizes, we see that slabs are unstable even though they are predicted to be stable based on their volume alone, which is consistent with our fits and follows from the modulation of the surfaces by capillary waves. Also this regime is well described by the theory developed in this paper.

\section{Summary and Conclusions}\label{sec:summary}
We have presented a general approach to calculate the distribution of contact distances between two surfaces that are modulated by fluctuations and at the same time diffuse freely relative to each other. The corresponding stochastic PDE (\preequ\eqref{equ:reaction_diffusion_initial}), which can be understood as a generalized reaction-diffusion equation, is solved approximately for different rate functions $k(\ell)$ using a path-integral method. This calculation yields the distribution of contact distances as well as a condition for the most likely contact distance, $\ell^\ast$ (\preequ\eqref{equ:maximum_cond}).

The rate function $k(\ell)$, which represents the rate of forming a closing fluctuation between two interfaces that are on average $\ell$ apart, is in principle system specific. We have presented simulations using a basic interface model---the Edward-Wilkinson\precite\cite{Edwards1982} (EW) model in both (1+1)- and (2+1) dimensions---where we calculated the rate functions $k(\ell)$ numerically. Our results on the (1+1)-dimensional interface confirmed previous results\precite\cite{Gross2018b} that the rate function $k(\ell)$ is a Gaussian function of $\ell$. In the (2+1)-dimensional system a more complex behavior was observed where $k(\ell)$ for small system sizes is a Gaussian, whose mean and variance are consistent with expectations from the known free energy landscape. For larger system sizes, $k(\ell)$ instead decays exponentially.

Based on these functional forms of $k(\ell)$, we calculated expressions for the distribution of contact distances, $u(\ell)$, and the most likely contact distance $\ell^\ast$. A comparison to the results of numerical calculations of $u(\ell)$ show excellent agreement with our theory in the limit of slow diffusion of the interface, $\Dint \to 0$. Furthermore, a scaling analysis based on the results achieved using the EW model, yielded a scaling law for $\ell^\ast$ that shows remarkably slow growth of the most likely contact distance with system size. In particular, in the (1+1)-dimensional case the growth is given by $\ell^\ast \sim \sqrt{L \log(c L)}$ and in the (2+1)-dimensional case $\ell^\ast$ grows like $\ell^\ast \sim \log(c L)$ if $k(\ell)$ is exponential and like $\ell^\ast \sim \sqrt{\log\left(c L\right)} + \sqrt{2/(\pi\beta\gamma)} \log \left(L/\Delta x\right)  + \tilde{\zeta}_0$ for Gaussian $k(\ell)$. Here, $\beta\gamma$ is the reduced interface tension, $\Delta x$ is the chosen discretization length, and $\tilde{\zeta}_0$ is a constant. In either case the growth of $\ell^\ast$ is slow in the sense that the $L$-dependence enters into the fastest growing terms through $\log L$ only.

As an application of these results, we investigated the stability of slabs of particles in molecular simulations, in particular, the widths $\ell$ at which they are found to collapse into more compact cluster shapes. The expectation from equilibrium considerations alone, is that $\ell^\ast \sim L$ in all cases. However, in order for this collapse to occur, the two interfaces of a given slab have to come into contact with each other first. Assuming that the dynamics of the interfaces formed in these simulations are similar to the ones exhibited by the EW model, it follows that the scaling laws derived above also apply to the most likely collapse distance; a result that is in stark contrast to equilibrium expectations. Indeed, our simulations of the (1+1)-dimensional Ising model show that $\ell^\ast$ scales proportional to $\sqrt{L \log(c L)}$, as stated above. The results obtained with the (2+1)-dimensional Ising model are best fit by a model that assumes $k(\ell)$ to be Gaussian in shape and also here the fit is excellent. Another set of simulations was performed using a slab of a Lennard-Jones liquid that forms two liquid-gas interfaces. In this case, based on our results we can not distinguish, whether a better fit can be achieved using a $k(\ell)$ that is exponential or one that is Gaussian. However, both theories fit well to the data obtained, once again confirming the extremely slow growth of contact distance with system size.

These results imply that, even in the macroscopic limit, the most likely contact distance $\ell^\ast$ is microscopic, an observation that has been made in experiments using colloidal dispersions\precite\cite{Aarts2004}, where colloidal droplets coalesce under the influence of gravity when there is a gap on the order of \SI{1}{\micro\meter} between them. This distance has previously been argued to come about based on the roughness of the interface. Our analysis---which pertains to a system where the systematic drift of the droplets is negligible compared to the diffusion of their centers of mass---suggests that in addition to the roughness of the interface, also the diffusion coefficient that describes the motion of the center of mass of the droplets, determines the size of the gap observed when two droplets coalesce.

The results presented in this work have been achieved by using results about the maximum relative height (MRH) observed in the Edward-Wilkinson model and extending them to other systems. This yielded an excellent match between simulation results and theory for the systems that were investigated. These results should be transferable to other interfacial systems provided that the dynamics of the MRH are qualitatively similar to the EW model. An extension to other situations can be made using the remarkably simple condition\preequNoWord\eqref{equ:maximum_cond} that allows a prediction of the likely contact distance under some general assumptions, should interfaces exhibit different dynamics.

\begin{acknowledgments}
    C.M.~has been supported by an uni:docs fellowship of the University of Vienna and acknowledges support from the Austrian Science Fund (FWF) Project No. P27738-N28. C.M.~and C.D.~acknowledge support from FWF Project No. I3163-N36. P.G.~acknowledges the generous support of the Erwin Schrödinger Institute for Mathematics and Physics (ESI). The computational results presented have been achieved in part using the Vienna Scientific Cluster (VSC).
\end{acknowledgments}

\bibliography{paper_surface_fluctuations}

\begin{thebibliography}{70}%
\makeatletter
\providecommand \@ifxundefined [1]{%
 \@ifx{#1\undefined}
}%
\providecommand \@ifnum [1]{%
 \ifnum #1\expandafter \@firstoftwo
 \else \expandafter \@secondoftwo
 \fi
}%
\providecommand \@ifx [1]{%
 \ifx #1\expandafter \@firstoftwo
 \else \expandafter \@secondoftwo
 \fi
}%
\providecommand \natexlab [1]{#1}%
\providecommand \enquote  [1]{``#1''}%
\providecommand \bibnamefont  [1]{#1}%
\providecommand \bibfnamefont [1]{#1}%
\providecommand \citenamefont [1]{#1}%
\providecommand \href@noop [0]{\@secondoftwo}%
\providecommand \href [0]{\begingroup \@sanitize@url \@href}%
\providecommand \@href[1]{\@@startlink{#1}\@@href}%
\providecommand \@@href[1]{\endgroup#1\@@endlink}%
\providecommand \@sanitize@url [0]{\catcode `\\12\catcode `\$12\catcode
  `\&12\catcode `\#12\catcode `\^12\catcode `\_12\catcode `\%12\relax}%
\providecommand \@@startlink[1]{}%
\providecommand \@@endlink[0]{}%
\providecommand \url  [0]{\begingroup\@sanitize@url \@url }%
\providecommand \@url [1]{\endgroup\@href {#1}{\urlprefix }}%
\providecommand \urlprefix  [0]{URL }%
\providecommand \Eprint [0]{\href }%
\providecommand \doibase [0]{http://dx.doi.org/}%
\providecommand \selectlanguage [0]{\@gobble}%
\providecommand \bibinfo  [0]{\@secondoftwo}%
\providecommand \bibfield  [0]{\@secondoftwo}%
\providecommand \translation [1]{[#1]}%
\providecommand \BibitemOpen [0]{}%
\providecommand \bibitemStop [0]{}%
\providecommand \bibitemNoStop [0]{.\EOS\space}%
\providecommand \EOS [0]{\spacefactor3000\relax}%
\providecommand \BibitemShut  [1]{\csname bibitem#1\endcsname}%
\let\auto@bib@innerbib\@empty
%</preamble>
\bibitem [{\citenamefont {Craster}\ and\ \citenamefont
  {Matar}(2009)}]{Craster2009a}%
  \BibitemOpen
  \bibfield  {author} {\bibinfo {author} {\bibfnamefont {R.~V.}\ \bibnamefont
  {Craster}}\ and\ \bibinfo {author} {\bibfnamefont {O.~K.}\ \bibnamefont
  {Matar}},\ }\href {\doibase 10.1103/RevModPhys.81.1131} {\bibfield  {journal}
  {\bibinfo  {journal} {Rev. Mod. Phys.}\ }\textbf {\bibinfo {volume} {81}},\
  \bibinfo {pages} {1131} (\bibinfo {year} {2009})}\BibitemShut {NoStop}%
\bibitem [{\citenamefont {Vrij}\ and\ \citenamefont
  {Overbeek}(1968)}]{Vrij1968}%
  \BibitemOpen
  \bibfield  {author} {\bibinfo {author} {\bibfnamefont {A.}~\bibnamefont
  {Vrij}}\ and\ \bibinfo {author} {\bibfnamefont {J.~T.}\ \bibnamefont
  {Overbeek}},\ }\href {\doibase 10.1021/ja01014a015} {\bibfield  {journal}
  {\bibinfo  {journal} {J. Am. Chem. Soc.}\ }\textbf {\bibinfo {volume} {90}},\
  \bibinfo {pages} {3074} (\bibinfo {year} {1968})}\BibitemShut {NoStop}%
\bibitem [{\citenamefont {Vrij}(1966)}]{Vrij1966a}%
  \BibitemOpen
  \bibfield  {author} {\bibinfo {author} {\bibfnamefont {A.}~\bibnamefont
  {Vrij}},\ }\href {\doibase 10.1039/DF9664200023} {\bibfield  {journal}
  {\bibinfo  {journal} {Discuss. Faraday Soc.}\ }\textbf {\bibinfo {volume}
  {42}},\ \bibinfo {pages} {23} (\bibinfo {year} {1966})}\BibitemShut {NoStop}%
\bibitem [{\citenamefont {Eggers}\ and\ \citenamefont
  {Villermaux}(2008)}]{Eggers2008}%
  \BibitemOpen
  \bibfield  {author} {\bibinfo {author} {\bibfnamefont {J.}~\bibnamefont
  {Eggers}}\ and\ \bibinfo {author} {\bibfnamefont {E.}~\bibnamefont
  {Villermaux}},\ }\href {\doibase 10.1088/0034-4885/71/3/036601} {\bibfield
  {journal} {\bibinfo  {journal} {Reports Prog. Phys.}\ }\textbf {\bibinfo
  {volume} {71}},\ \bibinfo {pages} {036601} (\bibinfo {year}
  {2008})}\BibitemShut {NoStop}%
\bibitem [{\citenamefont {Hennequin}\ \emph {et~al.}(2006)\citenamefont
  {Hennequin}, \citenamefont {Aarts}, \citenamefont {van~der Wiel},
  \citenamefont {Wegdam}, \citenamefont {Eggers}, \citenamefont
  {Lekkerkerker},\ and\ \citenamefont {Bonn}}]{Hennequin2006}%
  \BibitemOpen
  \bibfield  {author} {\bibinfo {author} {\bibfnamefont {Y.}~\bibnamefont
  {Hennequin}}, \bibinfo {author} {\bibfnamefont {D.~G. A.~L.}\ \bibnamefont
  {Aarts}}, \bibinfo {author} {\bibfnamefont {J.~H.}\ \bibnamefont {van~der
  Wiel}}, \bibinfo {author} {\bibfnamefont {G.}~\bibnamefont {Wegdam}},
  \bibinfo {author} {\bibfnamefont {J.}~\bibnamefont {Eggers}}, \bibinfo
  {author} {\bibfnamefont {H.~N.~W.}\ \bibnamefont {Lekkerkerker}}, \ and\
  \bibinfo {author} {\bibfnamefont {D.}~\bibnamefont {Bonn}},\ }\href {\doibase
  10.1103/PhysRevLett.97.244502} {\bibfield  {journal} {\bibinfo  {journal}
  {Phys. Rev. Lett.}\ }\textbf {\bibinfo {volume} {97}},\ \bibinfo {pages}
  {244502} (\bibinfo {year} {2006})}\BibitemShut {NoStop}%
\bibitem [{\citenamefont {Eggers}(2002)}]{Eggers2002}%
  \BibitemOpen
  \bibfield  {author} {\bibinfo {author} {\bibfnamefont {J.}~\bibnamefont
  {Eggers}},\ }\href {\doibase 10.1103/PhysRevLett.89.084502} {\bibfield
  {journal} {\bibinfo  {journal} {Phys. Rev. Lett.}\ }\textbf {\bibinfo
  {volume} {89}},\ \bibinfo {pages} {084502} (\bibinfo {year}
  {2002})}\BibitemShut {NoStop}%
\bibitem [{\citenamefont {Moseler}\ and\ \citenamefont
  {Landman}(2000)}]{Moseler2000a}%
  \BibitemOpen
  \bibfield  {author} {\bibinfo {author} {\bibfnamefont {M.}~\bibnamefont
  {Moseler}}\ and\ \bibinfo {author} {\bibfnamefont {U.}~\bibnamefont
  {Landman}},\ }\href {\doibase 10.1126/science.289.5482.1165} {\bibfield
  {journal} {\bibinfo  {journal} {Science}\ }\textbf {\bibinfo {volume}
  {289}},\ \bibinfo {pages} {1165} (\bibinfo {year} {2000})}\BibitemShut
  {NoStop}%
\bibitem [{\citenamefont {Shi}\ \emph {et~al.}(1994)\citenamefont {Shi},
  \citenamefont {Brenner},\ and\ \citenamefont {Nagel}}]{Shi1994}%
  \BibitemOpen
  \bibfield  {author} {\bibinfo {author} {\bibfnamefont {X.~D.}\ \bibnamefont
  {Shi}}, \bibinfo {author} {\bibfnamefont {M.~P.}\ \bibnamefont {Brenner}}, \
  and\ \bibinfo {author} {\bibfnamefont {S.~R.}\ \bibnamefont {Nagel}},\ }\href
  {\doibase 10.1126/science.265.5169.219} {\bibfield  {journal} {\bibinfo
  {journal} {Science}\ }\textbf {\bibinfo {volume} {265}},\ \bibinfo {pages}
  {219} (\bibinfo {year} {1994})}\BibitemShut {NoStop}%
\bibitem [{\citenamefont {Perumanath}\ \emph {et~al.}(2019)\citenamefont
  {Perumanath}, \citenamefont {Borg}, \citenamefont {Chubynsky}, \citenamefont
  {Sprittles},\ and\ \citenamefont {Reese}}]{Perumanath2019}%
  \BibitemOpen
  \bibfield  {author} {\bibinfo {author} {\bibfnamefont {S.}~\bibnamefont
  {Perumanath}}, \bibinfo {author} {\bibfnamefont {M.~K.}\ \bibnamefont
  {Borg}}, \bibinfo {author} {\bibfnamefont {M.~V.}\ \bibnamefont {Chubynsky}},
  \bibinfo {author} {\bibfnamefont {J.~E.}\ \bibnamefont {Sprittles}}, \ and\
  \bibinfo {author} {\bibfnamefont {J.~M.}\ \bibnamefont {Reese}},\ }\href
  {\doibase 10.1103/PhysRevLett.122.104501} {\bibfield  {journal} {\bibinfo
  {journal} {Phys. Rev. Lett.}\ }\textbf {\bibinfo {volume} {122}},\ \bibinfo
  {pages} {104501} (\bibinfo {year} {2019})}\BibitemShut {NoStop}%
\bibitem [{\citenamefont {Aarts}\ \emph {et~al.}(2005)\citenamefont {Aarts},
  \citenamefont {Lekkerkerker}, \citenamefont {Guo}, \citenamefont {Wegdam},\
  and\ \citenamefont {Bonn}}]{Aarts2005}%
  \BibitemOpen
  \bibfield  {author} {\bibinfo {author} {\bibfnamefont {D.~G. A.~L.}\
  \bibnamefont {Aarts}}, \bibinfo {author} {\bibfnamefont {H.~N.~W.}\
  \bibnamefont {Lekkerkerker}}, \bibinfo {author} {\bibfnamefont
  {H.}~\bibnamefont {Guo}}, \bibinfo {author} {\bibfnamefont {G.~H.}\
  \bibnamefont {Wegdam}}, \ and\ \bibinfo {author} {\bibfnamefont
  {D.}~\bibnamefont {Bonn}},\ }\href {\doibase 10.1103/PhysRevLett.95.164503}
  {\bibfield  {journal} {\bibinfo  {journal} {Phys. Rev. Lett.}\ }\textbf
  {\bibinfo {volume} {95}},\ \bibinfo {pages} {164503} (\bibinfo {year}
  {2005})}\BibitemShut {NoStop}%
\bibitem [{\citenamefont {Aarts}\ \emph {et~al.}(2004)\citenamefont {Aarts},
  \citenamefont {Schmidt},\ and\ \citenamefont {Lekkerkerker}}]{Aarts2004}%
  \BibitemOpen
  \bibfield  {author} {\bibinfo {author} {\bibfnamefont {D.~G. A.~L.}\
  \bibnamefont {Aarts}}, \bibinfo {author} {\bibfnamefont {M.}~\bibnamefont
  {Schmidt}}, \ and\ \bibinfo {author} {\bibfnamefont {H.~N.~W.}\ \bibnamefont
  {Lekkerkerker}},\ }\href {\doibase 10.1126/science.1097116} {\bibfield
  {journal} {\bibinfo  {journal} {Science}\ }\textbf {\bibinfo {volume}
  {304}},\ \bibinfo {pages} {847} (\bibinfo {year} {2004})}\BibitemShut
  {NoStop}%
\bibitem [{\citenamefont {Eggers}\ \emph {et~al.}(1999)\citenamefont {Eggers},
  \citenamefont {Lister},\ and\ \citenamefont {Stone}}]{Eggers1999}%
  \BibitemOpen
  \bibfield  {author} {\bibinfo {author} {\bibfnamefont {J.}~\bibnamefont
  {Eggers}}, \bibinfo {author} {\bibfnamefont {J.~R.}\ \bibnamefont {Lister}},
  \ and\ \bibinfo {author} {\bibfnamefont {H.~A.}\ \bibnamefont {Stone}},\
  }\href {\doibase 10.1017/S002211209900662X} {\bibfield  {journal} {\bibinfo
  {journal} {J. Fluid Mech.}\ }\textbf {\bibinfo {volume} {401}},\ \bibinfo
  {pages} {293} (\bibinfo {year} {1999})}\BibitemShut {NoStop}%
\bibitem [{\citenamefont {Bradley}\ and\ \citenamefont
  {Stow}(1978)}]{Bradley1978}%
  \BibitemOpen
  \bibfield  {author} {\bibinfo {author} {\bibfnamefont {S.~G.}\ \bibnamefont
  {Bradley}}\ and\ \bibinfo {author} {\bibfnamefont {C.~D.}\ \bibnamefont
  {Stow}},\ }\href {\doibase 10.1098/rsta.1978.0001} {\bibfield  {journal}
  {\bibinfo  {journal} {Philos. Trans. R. Soc. A Math. Phys. Eng. Sci.}\
  }\textbf {\bibinfo {volume} {287}},\ \bibinfo {pages} {635} (\bibinfo {year}
  {1978})}\BibitemShut {NoStop}%
\bibitem [{\citenamefont {Stone}(1998)}]{Stone1998}%
  \BibitemOpen
  \bibfield  {author} {\bibinfo {author} {\bibfnamefont {J.}~\bibnamefont
  {Stone}},\ }\emph {\bibinfo {title} {{An Efficient Library for Parallel Ray
  Tracing and Animation}}},\ \href@noop {} {Ph.D. thesis},\ \bibinfo  {school}
  {Computer Science Department, University of Missouri-Rolla} (\bibinfo {year}
  {1998})\BibitemShut {NoStop}%
\bibitem [{\citenamefont {Humphrey}\ \emph {et~al.}(1996)\citenamefont
  {Humphrey}, \citenamefont {Dalke},\ and\ \citenamefont
  {Schulten}}]{Humphrey1996}%
  \BibitemOpen
  \bibfield  {author} {\bibinfo {author} {\bibfnamefont {W.}~\bibnamefont
  {Humphrey}}, \bibinfo {author} {\bibfnamefont {A.}~\bibnamefont {Dalke}}, \
  and\ \bibinfo {author} {\bibfnamefont {K.}~\bibnamefont {Schulten}},\
  }\href@noop {} {\bibfield  {journal} {\bibinfo  {journal} {J. Mol. Graph.}\
  }\textbf {\bibinfo {volume} {14}},\ \bibinfo {pages} {33} (\bibinfo {year}
  {1996})}\BibitemShut {NoStop}%
\bibitem [{\citenamefont {Tr{\"{o}}ster}\ \emph {et~al.}(2018)\citenamefont
  {Tr{\"{o}}ster}, \citenamefont {Schmitz}, \citenamefont {Virnau},\ and\
  \citenamefont {Binder}}]{Troester2017}%
  \BibitemOpen
  \bibfield  {author} {\bibinfo {author} {\bibfnamefont {A.}~\bibnamefont
  {Tr{\"{o}}ster}}, \bibinfo {author} {\bibfnamefont {F.}~\bibnamefont
  {Schmitz}}, \bibinfo {author} {\bibfnamefont {P.}~\bibnamefont {Virnau}}, \
  and\ \bibinfo {author} {\bibfnamefont {K.}~\bibnamefont {Binder}},\ }\href
  {\doibase 10.1021/acs.jpcb.7b10392} {\bibfield  {journal} {\bibinfo
  {journal} {J. Phys. Chem. B}\ }\textbf {\bibinfo {volume} {122}},\ \bibinfo
  {pages} {3407} (\bibinfo {year} {2018})}\BibitemShut {NoStop}%
\bibitem [{\citenamefont {Moritz}\ \emph {et~al.}(2017)\citenamefont {Moritz},
  \citenamefont {Tr{\"{o}}ster},\ and\ \citenamefont {Dellago}}]{Moritz2017}%
  \BibitemOpen
  \bibfield  {author} {\bibinfo {author} {\bibfnamefont {C.}~\bibnamefont
  {Moritz}}, \bibinfo {author} {\bibfnamefont {A.}~\bibnamefont
  {Tr{\"{o}}ster}}, \ and\ \bibinfo {author} {\bibfnamefont {C.}~\bibnamefont
  {Dellago}},\ }\href {\doibase 10.1063/1.4997479} {\bibfield  {journal}
  {\bibinfo  {journal} {J. Chem. Phys.}\ }\textbf {\bibinfo {volume} {147}},\
  \bibinfo {pages} {152714} (\bibinfo {year} {2017})}\BibitemShut {NoStop}%
\bibitem [{\citenamefont {Binder}\ \emph {et~al.}(2012)\citenamefont {Binder},
  \citenamefont {Block}, \citenamefont {Virnau},\ and\ \citenamefont
  {Tr{\"{o}}ster}}]{Binder2012}%
  \BibitemOpen
  \bibfield  {author} {\bibinfo {author} {\bibfnamefont {K.}~\bibnamefont
  {Binder}}, \bibinfo {author} {\bibfnamefont {B.~J.}\ \bibnamefont {Block}},
  \bibinfo {author} {\bibfnamefont {P.}~\bibnamefont {Virnau}}, \ and\ \bibinfo
  {author} {\bibfnamefont {A.}~\bibnamefont {Tr{\"{o}}ster}},\ }\href {\doibase
  10.1119/1.4754020} {\bibfield  {journal} {\bibinfo  {journal} {Am. J. Phys.}\
  }\textbf {\bibinfo {volume} {80}},\ \bibinfo {pages} {1099} (\bibinfo {year}
  {2012})}\BibitemShut {NoStop}%
\bibitem [{\citenamefont {Tr{\"{o}}ster}\ \emph {et~al.}(2005)\citenamefont
  {Tr{\"{o}}ster}, \citenamefont {Dellago},\ and\ \citenamefont
  {Schranz}}]{Troster2005}%
  \BibitemOpen
  \bibfield  {author} {\bibinfo {author} {\bibfnamefont {A.}~\bibnamefont
  {Tr{\"{o}}ster}}, \bibinfo {author} {\bibfnamefont {C.}~\bibnamefont
  {Dellago}}, \ and\ \bibinfo {author} {\bibfnamefont {W.}~\bibnamefont
  {Schranz}},\ }\href {\doibase 10.1103/PhysRevB.72.094103} {\bibfield
  {journal} {\bibinfo  {journal} {Phys. Rev. B}\ }\textbf {\bibinfo {volume}
  {72}},\ \bibinfo {pages} {094103} (\bibinfo {year} {2005})}\BibitemShut
  {NoStop}%
\bibitem [{\citenamefont {Leung}\ and\ \citenamefont {Zia}(1990)}]{Leung1990}%
  \BibitemOpen
  \bibfield  {author} {\bibinfo {author} {\bibfnamefont {K.}~\bibnamefont
  {Leung}}\ and\ \bibinfo {author} {\bibfnamefont {R.~K.~P.}\ \bibnamefont
  {Zia}},\ }\href {\doibase 10.1088/0305-4470/23/20/021} {\bibfield  {journal}
  {\bibinfo  {journal} {J. Phys. A. Math. Gen.}\ }\textbf {\bibinfo {volume}
  {23}},\ \bibinfo {pages} {4593} (\bibinfo {year} {1990})}\BibitemShut
  {NoStop}%
\bibitem [{\citenamefont {Hunter}(2007)}]{Hunter2007}%
  \BibitemOpen
  \bibfield  {author} {\bibinfo {author} {\bibfnamefont {J.~D.}\ \bibnamefont
  {Hunter}},\ }\href {\doibase 10.1109/MCSE.2007.55} {\bibfield  {journal}
  {\bibinfo  {journal} {Comput. Sci. Eng.}\ }\textbf {\bibinfo {volume} {9}},\
  \bibinfo {pages} {90} (\bibinfo {year} {2007})}\BibitemShut {NoStop}%
\bibitem [{\citenamefont {Godr{\`{e}}che}(1991)}]{Godreche1991}%
  \BibitemOpen
  \bibinfo {editor} {\bibfnamefont {C.}~\bibnamefont {Godr{\`{e}}che}},\ ed.,\
  \href@noop {} {\emph {\bibinfo {title} {{Solids far from equilibrium}}}}\
  (\bibinfo  {publisher} {Cambridge Univ. Press},\ \bibinfo {address}
  {Cambridge},\ \bibinfo {year} {1991})\BibitemShut {NoStop}%
\bibitem [{\citenamefont {Wilemski}\ and\ \citenamefont
  {Fixman}(1973)}]{Wilemski1973}%
  \BibitemOpen
  \bibfield  {author} {\bibinfo {author} {\bibfnamefont {G.}~\bibnamefont
  {Wilemski}}\ and\ \bibinfo {author} {\bibfnamefont {M.}~\bibnamefont
  {Fixman}},\ }\href {\doibase 10.1063/1.1679757} {\bibfield  {journal}
  {\bibinfo  {journal} {J. Chem. Phys.}\ }\textbf {\bibinfo {volume} {58}},\
  \bibinfo {pages} {4009} (\bibinfo {year} {1973})}\BibitemShut {NoStop}%
\bibitem [{\citenamefont {Mattis}\ and\ \citenamefont
  {Glasser}(1998)}]{Mattis1998}%
  \BibitemOpen
  \bibfield  {author} {\bibinfo {author} {\bibfnamefont {D.~C.}\ \bibnamefont
  {Mattis}}\ and\ \bibinfo {author} {\bibfnamefont {M.~L.}\ \bibnamefont
  {Glasser}},\ }\href {\doibase 10.1103/RevModPhys.70.979} {\bibfield
  {journal} {\bibinfo  {journal} {Rev. Mod. Phys.}\ }\textbf {\bibinfo {volume}
  {70}},\ \bibinfo {pages} {979} (\bibinfo {year} {1998})}\BibitemShut
  {NoStop}%
\bibitem [{\citenamefont {Pr{\"{u}}stel}\ and\ \citenamefont
  {Meier-Schellersheim}(2017)}]{Prustel2017}%
  \BibitemOpen
  \bibfield  {author} {\bibinfo {author} {\bibfnamefont {T.}~\bibnamefont
  {Pr{\"{u}}stel}}\ and\ \bibinfo {author} {\bibfnamefont {M.}~\bibnamefont
  {Meier-Schellersheim}},\ }\href {\doibase 10.1103/PhysRevE.96.022151}
  {\bibfield  {journal} {\bibinfo  {journal} {Phys. Rev. E}\ }\textbf {\bibinfo
  {volume} {96}},\ \bibinfo {pages} {022151} (\bibinfo {year}
  {2017})}\BibitemShut {NoStop}%
\bibitem [{\citenamefont {Wiegel}(1986)}]{wiegel1986introduction}%
  \BibitemOpen
  \bibfield  {author} {\bibinfo {author} {\bibfnamefont {F.~W.}\ \bibnamefont
  {Wiegel}},\ }\href {\doibase 10.1142/0178} {\emph {\bibinfo {title}
  {{Introduction to Path-integral Methods in Physics and Polymer Science}}}}\
  (\bibinfo  {publisher} {World Scientific},\ \bibinfo {year}
  {1986})\BibitemShut {NoStop}%
\bibitem [{\citenamefont {Risken}(1984)}]{Risken1984}%
  \BibitemOpen
  \bibfield  {author} {\bibinfo {author} {\bibfnamefont {H.}~\bibnamefont
  {Risken}},\ }\href {\doibase 10.1007/978-3-642-96807-5_4} {\emph {\bibinfo
  {title} {{The Fokker-Planck Equation: Methods of Solution and
  Applications}}}}\ (\bibinfo  {publisher} {Springer Berlin Heidelberg},\
  \bibinfo {address} {Berlin, Heidelberg},\ \bibinfo {year} {1984})\BibitemShut
  {NoStop}%
\bibitem [{\citenamefont {Parisi}(1988)}]{parisi1988statistical}%
  \BibitemOpen
  \bibfield  {author} {\bibinfo {author} {\bibfnamefont {G.}~\bibnamefont
  {Parisi}},\ }\href@noop {} {\emph {\bibinfo {title} {{Statistical field
  theory}}}},\ edited by\ \bibinfo {editor} {\bibfnamefont {D.}~\bibnamefont
  {Pines}}\ (\bibinfo  {publisher} {Addison-Wesley},\ \bibinfo {year}
  {1988})\BibitemShut {NoStop}%
\bibitem [{\citenamefont {Kleinert}(2006)}]{kleinert2009path}%
  \BibitemOpen
  \bibfield  {author} {\bibinfo {author} {\bibfnamefont {H.}~\bibnamefont
  {Kleinert}},\ }\href@noop {} {\emph {\bibinfo {title} {{Path integrals in
  quantum mechanics, statistics, polymer physics, and financial markets}}}},\
  \bibinfo {edition} {4th}\ ed.\ (\bibinfo  {publisher} {World scientific},\
  \bibinfo {address} {New Jersey},\ \bibinfo {year} {2006})\BibitemShut
  {NoStop}%
\bibitem [{\citenamefont {Kac}(1949)}]{Kac1949}%
  \BibitemOpen
  \bibfield  {author} {\bibinfo {author} {\bibfnamefont {M.}~\bibnamefont
  {Kac}},\ }\href {\doibase 10.1090/S0002-9947-1949-0027960-X} {\bibfield
  {journal} {\bibinfo  {journal} {Trans. Am. Math. Soc.}\ }\textbf {\bibinfo
  {volume} {65}},\ \bibinfo {pages} {1} (\bibinfo {year} {1949})}\BibitemShut
  {NoStop}%
\bibitem [{\citenamefont {Bastianelli}\ \emph {et~al.}(2007)\citenamefont
  {Bastianelli}, \citenamefont {Corradini},\ and\ \citenamefont
  {Pisani}}]{Bastianelli2007}%
  \BibitemOpen
  \bibfield  {author} {\bibinfo {author} {\bibfnamefont {F.}~\bibnamefont
  {Bastianelli}}, \bibinfo {author} {\bibfnamefont {O.}~\bibnamefont
  {Corradini}}, \ and\ \bibinfo {author} {\bibfnamefont {P.~A.}\ \bibnamefont
  {Pisani}},\ }\href {\doibase 10.1088/1126-6708/2007/02/059} {\bibfield
  {journal} {\bibinfo  {journal} {J. High Energy Phys.}\ }\textbf {\bibinfo
  {volume} {2007}} (\bibinfo {year} {2007})}\BibitemShut {NoStop}%
\bibitem [{\citenamefont {Chaichian}\ and\ \citenamefont
  {Demichev}(2001)}]{Chaichian2001}%
  \BibitemOpen
  \bibfield  {author} {\bibinfo {author} {\bibfnamefont {M.}~\bibnamefont
  {Chaichian}}\ and\ \bibinfo {author} {\bibfnamefont {A.}~\bibnamefont
  {Demichev}},\ }\href@noop {} {\emph {\bibinfo {title} {{Path integrals in
  physics: Volume I stochastic processes and quantum mechanics}}}}\ (\bibinfo
  {publisher} {IOP Publishing},\ \bibinfo {year} {2001})\BibitemShut {NoStop}%
\bibitem [{\citenamefont {Mathews}(1970)}]{Mathews1970}%
  \BibitemOpen
  \bibfield  {author} {\bibinfo {author} {\bibfnamefont {J.}~\bibnamefont
  {Mathews}},\ }\href@noop {} {\emph {\bibinfo {title} {{Mathematical methods
  of physics}}}},\ \bibinfo {edition} {2nd}\ ed.\ (\bibinfo  {publisher}
  {Benjamin},\ \bibinfo {address} {New York, NY},\ \bibinfo {year}
  {1970})\BibitemShut {NoStop}%
\bibitem [{\citenamefont {Kampen}(1977)}]{Kampen1977}%
  \BibitemOpen
  \bibfield  {author} {\bibinfo {author} {\bibfnamefont {N.~V.}\ \bibnamefont
  {Kampen}},\ }\href {http://www.springerlink.com/index/q618780412046144.pdf}
  {\bibfield  {journal} {\bibinfo  {journal} {J. Stat. Phys.}\ }\textbf
  {\bibinfo {volume} {17}},\ \bibinfo {pages} {71} (\bibinfo {year}
  {1977})}\BibitemShut {NoStop}%
\bibitem [{\citenamefont {Caroli}\ \emph {et~al.}(1981)\citenamefont {Caroli},
  \citenamefont {Caroli},\ and\ \citenamefont {Roulet}}]{Caroli1981}%
  \BibitemOpen
  \bibfield  {author} {\bibinfo {author} {\bibfnamefont {B.}~\bibnamefont
  {Caroli}}, \bibinfo {author} {\bibfnamefont {C.}~\bibnamefont {Caroli}}, \
  and\ \bibinfo {author} {\bibfnamefont {B.}~\bibnamefont {Roulet}},\ }\href
  {\doibase 10.1007/BF01106788} {\bibfield  {journal} {\bibinfo  {journal} {J.
  Stat. Phys.}\ }\textbf {\bibinfo {volume} {26}},\ \bibinfo {pages} {83}
  (\bibinfo {year} {1981})}\BibitemShut {NoStop}%
\bibitem [{\citenamefont {Autieri}\ \emph {et~al.}(2009)\citenamefont
  {Autieri}, \citenamefont {Faccioli}, \citenamefont {Sega}, \citenamefont
  {Pederiva},\ and\ \citenamefont {Orland}}]{Autieri2009}%
  \BibitemOpen
  \bibfield  {author} {\bibinfo {author} {\bibfnamefont {E.}~\bibnamefont
  {Autieri}}, \bibinfo {author} {\bibfnamefont {P.}~\bibnamefont {Faccioli}},
  \bibinfo {author} {\bibfnamefont {M.}~\bibnamefont {Sega}}, \bibinfo {author}
  {\bibfnamefont {F.}~\bibnamefont {Pederiva}}, \ and\ \bibinfo {author}
  {\bibfnamefont {H.}~\bibnamefont {Orland}},\ }\href {\doibase
  10.1063/1.3074271} {\bibfield  {journal} {\bibinfo  {journal} {J. Chem.
  Phys.}\ }\textbf {\bibinfo {volume} {130}},\ \bibinfo {pages} {064106}
  (\bibinfo {year} {2009})}\BibitemShut {NoStop}%
\bibitem [{\citenamefont {Edwards}\ and\ \citenamefont
  {Wilkinson}(1982)}]{Edwards1982}%
  \BibitemOpen
  \bibfield  {author} {\bibinfo {author} {\bibfnamefont {S.~F.}\ \bibnamefont
  {Edwards}}\ and\ \bibinfo {author} {\bibfnamefont {D.~R.}\ \bibnamefont
  {Wilkinson}},\ }\href {\doibase 10.1098/rspa.1982.0056} {\bibfield  {journal}
  {\bibinfo  {journal} {Proc. R. Soc. A Math. Phys. Eng. Sci.}\ }\textbf
  {\bibinfo {volume} {381}},\ \bibinfo {pages} {17} (\bibinfo {year}
  {1982})}\BibitemShut {NoStop}%
\bibitem [{\citenamefont {Kardar}\ \emph {et~al.}(1986)\citenamefont {Kardar},
  \citenamefont {Parisi},\ and\ \citenamefont {Zhang}}]{Kardar1986}%
  \BibitemOpen
  \bibfield  {author} {\bibinfo {author} {\bibfnamefont {M.}~\bibnamefont
  {Kardar}}, \bibinfo {author} {\bibfnamefont {G.}~\bibnamefont {Parisi}}, \
  and\ \bibinfo {author} {\bibfnamefont {Y.-C.}\ \bibnamefont {Zhang}},\ }\href
  {\doibase 10.1103/PhysRevLett.56.889} {\bibfield  {journal} {\bibinfo
  {journal} {Phys. Rev. Lett.}\ }\textbf {\bibinfo {volume} {56}},\ \bibinfo
  {pages} {889} (\bibinfo {year} {1986})}\BibitemShut {NoStop}%
\bibitem [{\citenamefont {Barab{\'{a}}si}\ and\ \citenamefont
  {Stanley}(1995)}]{Barabasi1995}%
  \BibitemOpen
  \bibfield  {author} {\bibinfo {author} {\bibfnamefont {A.-L.}\ \bibnamefont
  {Barab{\'{a}}si}}\ and\ \bibinfo {author} {\bibfnamefont {H.~E.}\
  \bibnamefont {Stanley}},\ }\href@noop {} {\emph {\bibinfo {title} {{Fractal
  concepts in surface growth}}}}\ (\bibinfo  {publisher} {Cambridge university
  press},\ \bibinfo {year} {1995})\BibitemShut {NoStop}%
\bibitem [{\citenamefont {Gross}(2018{\natexlab{a}})}]{Gross2018a}%
  \BibitemOpen
  \bibfield  {author} {\bibinfo {author} {\bibfnamefont {M.}~\bibnamefont
  {Gross}},\ }\href {\doibase 10.1088/1742-5468/aaa386} {\bibfield  {journal}
  {\bibinfo  {journal} {J. Stat. Mech. Theory Exp.}\ }\textbf {\bibinfo
  {volume} {2018}},\ \bibinfo {pages} {033213} (\bibinfo {year}
  {2018}{\natexlab{a}})}\BibitemShut {NoStop}%
\bibitem [{\citenamefont {Gross}(2018{\natexlab{b}})}]{Gross2018b}%
  \BibitemOpen
  \bibfield  {author} {\bibinfo {author} {\bibfnamefont {M.}~\bibnamefont
  {Gross}},\ }\href {\doibase 10.1088/1742-5468/aaa792} {\bibfield  {journal}
  {\bibinfo  {journal} {J. Stat. Mech. Theory Exp.}\ }\textbf {\bibinfo
  {volume} {2018}},\ \bibinfo {pages} {033212} (\bibinfo {year}
  {2018}{\natexlab{b}})}\BibitemShut {NoStop}%
\bibitem [{\citenamefont {Majumdar}\ and\ \citenamefont
  {Comtet}(2004)}]{Majumdar2004}%
  \BibitemOpen
  \bibfield  {author} {\bibinfo {author} {\bibfnamefont {S.~N.}\ \bibnamefont
  {Majumdar}}\ and\ \bibinfo {author} {\bibfnamefont {A.}~\bibnamefont
  {Comtet}},\ }\href {\doibase 10.1103/PhysRevLett.92.225501} {\bibfield
  {journal} {\bibinfo  {journal} {Phys. Rev. Lett.}\ }\textbf {\bibinfo
  {volume} {92}},\ \bibinfo {pages} {225501} (\bibinfo {year}
  {2004})}\BibitemShut {NoStop}%
\bibitem [{\citenamefont {Majumdar}\ \emph {et~al.}(2020)\citenamefont
  {Majumdar}, \citenamefont {Pal},\ and\ \citenamefont
  {Schehr}}]{Majumdar2020}%
  \BibitemOpen
  \bibfield  {author} {\bibinfo {author} {\bibfnamefont {S.~N.}\ \bibnamefont
  {Majumdar}}, \bibinfo {author} {\bibfnamefont {A.}~\bibnamefont {Pal}}, \
  and\ \bibinfo {author} {\bibfnamefont {G.}~\bibnamefont {Schehr}},\ }\href
  {\doibase 10.1016/j.physrep.2019.10.005} {\bibfield  {journal} {\bibinfo
  {journal} {Phys. Rep.}\ }\textbf {\bibinfo {volume} {840}},\ \bibinfo {pages}
  {1} (\bibinfo {year} {2020})}\BibitemShut {NoStop}%
\bibitem [{\citenamefont {R{\'{a}}cz}\ and\ \citenamefont
  {Plischke}(1994)}]{Rcz1994}%
  \BibitemOpen
  \bibfield  {author} {\bibinfo {author} {\bibfnamefont {Z.}~\bibnamefont
  {R{\'{a}}cz}}\ and\ \bibinfo {author} {\bibfnamefont {M.}~\bibnamefont
  {Plischke}},\ }\href {\doibase 10.1103/PhysRevE.50.3530} {\bibfield
  {journal} {\bibinfo  {journal} {Phys. Rev. E}\ }\textbf {\bibinfo {volume}
  {50}},\ \bibinfo {pages} {3530} (\bibinfo {year} {1994})}\BibitemShut
  {NoStop}%
\bibitem [{\citenamefont {Reimann}\ \emph {et~al.}(1999)\citenamefont
  {Reimann}, \citenamefont {Schmid},\ and\ \citenamefont
  {H{\"{a}}nggi}}]{Reimann1999}%
  \BibitemOpen
  \bibfield  {author} {\bibinfo {author} {\bibfnamefont {P.}~\bibnamefont
  {Reimann}}, \bibinfo {author} {\bibfnamefont {G.~J.}\ \bibnamefont {Schmid}},
  \ and\ \bibinfo {author} {\bibfnamefont {P.}~\bibnamefont {H{\"{a}}nggi}},\
  }\href {\doibase 10.1103/PhysRevE.60.R1} {\bibfield  {journal} {\bibinfo
  {journal} {Phys. Rev. E}\ }\textbf {\bibinfo {volume} {60}},\ \bibinfo
  {pages} {R1} (\bibinfo {year} {1999})}\BibitemShut {NoStop}%
\bibitem [{\citenamefont {Kolmogorov}(1940)}]{Kolmogorov1940}%
  \BibitemOpen
  \bibfield  {author} {\bibinfo {author} {\bibfnamefont {A.~N.}\ \bibnamefont
  {Kolmogorov}},\ }in\ \href@noop {} {\emph {\bibinfo {booktitle} {Dokl. Akad.
  Nauk SSSR}}},\ Vol.~\bibinfo {volume} {26}\ (\bibinfo {year} {1940})\ pp.\
  \bibinfo {pages} {115--118}\BibitemShut {NoStop}%
\bibitem [{\citenamefont {Mandelbrot}\ and\ \citenamefont {{Van
  Ness}}(1968)}]{Mandelbrot1968}%
  \BibitemOpen
  \bibfield  {author} {\bibinfo {author} {\bibfnamefont {B.~B.}\ \bibnamefont
  {Mandelbrot}}\ and\ \bibinfo {author} {\bibfnamefont {J.~W.}\ \bibnamefont
  {{Van Ness}}},\ }\href {\doibase 10.1137/1010093} {\bibfield  {journal}
  {\bibinfo  {journal} {SIAM Rev.}\ }\textbf {\bibinfo {volume} {10}},\
  \bibinfo {pages} {422} (\bibinfo {year} {1968})}\BibitemShut {NoStop}%
\bibitem [{\citenamefont {Krug}\ \emph {et~al.}(1997)\citenamefont {Krug},
  \citenamefont {Kallabis}, \citenamefont {Majumdar}, \citenamefont {Cornell},
  \citenamefont {Bray},\ and\ \citenamefont {Sire}}]{Krug1997}%
  \BibitemOpen
  \bibfield  {author} {\bibinfo {author} {\bibfnamefont {J.}~\bibnamefont
  {Krug}}, \bibinfo {author} {\bibfnamefont {H.}~\bibnamefont {Kallabis}},
  \bibinfo {author} {\bibfnamefont {S.~N.}\ \bibnamefont {Majumdar}}, \bibinfo
  {author} {\bibfnamefont {S.~J.}\ \bibnamefont {Cornell}}, \bibinfo {author}
  {\bibfnamefont {A.~J.}\ \bibnamefont {Bray}}, \ and\ \bibinfo {author}
  {\bibfnamefont {C.}~\bibnamefont {Sire}},\ }\href {\doibase
  10.1103/PhysRevE.56.2702} {\bibfield  {journal} {\bibinfo  {journal} {Phys.
  Rev. E - Stat. Physics, Plasmas, Fluids, Relat. Interdiscip. Top.}\ }\textbf
  {\bibinfo {volume} {56}},\ \bibinfo {pages} {2702} (\bibinfo {year}
  {1997})}\BibitemShut {NoStop}%
\bibitem [{\citenamefont {Kramers}(1940)}]{Kramers1940}%
  \BibitemOpen
  \bibfield  {author} {\bibinfo {author} {\bibfnamefont {H.}~\bibnamefont
  {Kramers}},\ }\href {\doibase 10.1016/S0031-8914(40)90098-2} {\bibfield
  {journal} {\bibinfo  {journal} {Physica}\ }\textbf {\bibinfo {volume} {7}},\
  \bibinfo {pages} {284} (\bibinfo {year} {1940})}\BibitemShut {NoStop}%
\bibitem [{\citenamefont {Hassani}(2005)}]{Hassani2007}%
  \BibitemOpen
  \bibfield  {author} {\bibinfo {author} {\bibfnamefont {M.}~\bibnamefont
  {Hassani}},\ }\href {http://vuir.vu.edu.au/id/eprint/18365} {\emph {\bibinfo
  {title} {Res. Rep. Collect.}}},\ \bibinfo {type} {Tech. Rep.}\ \bibinfo
  {number} {4}\ (\bibinfo  {institution} {School of Communications and
  Informatics, Faculty of Engineering and Science, Victoria University of
  Technology},\ \bibinfo {year} {2005})\BibitemShut {NoStop}%
\bibitem [{\citenamefont {Gy{\"{o}}rgyi}\ \emph {et~al.}(2003)\citenamefont
  {Gy{\"{o}}rgyi}, \citenamefont {Holdsworth}, \citenamefont {Portelli},\ and\
  \citenamefont {R{\'{a}}cz}}]{Gyorgyi2003}%
  \BibitemOpen
  \bibfield  {author} {\bibinfo {author} {\bibfnamefont {G.}~\bibnamefont
  {Gy{\"{o}}rgyi}}, \bibinfo {author} {\bibfnamefont {P.~C.~W.}\ \bibnamefont
  {Holdsworth}}, \bibinfo {author} {\bibfnamefont {B.}~\bibnamefont
  {Portelli}}, \ and\ \bibinfo {author} {\bibfnamefont {Z.}~\bibnamefont
  {R{\'{a}}cz}},\ }\href {\doibase 10.1103/PhysRevE.68.056116} {\bibfield
  {journal} {\bibinfo  {journal} {Phys. Rev. E}\ }\textbf {\bibinfo {volume}
  {68}},\ \bibinfo {pages} {056116} (\bibinfo {year} {2003})}\BibitemShut
  {NoStop}%
\bibitem [{\citenamefont {Lee}(2005)}]{Lee2005}%
  \BibitemOpen
  \bibfield  {author} {\bibinfo {author} {\bibfnamefont {D.-S.}\ \bibnamefont
  {Lee}},\ }\href {\doibase 10.1103/PhysRevLett.95.150601} {\bibfield
  {journal} {\bibinfo  {journal} {Phys. Rev. Lett.}\ }\textbf {\bibinfo
  {volume} {95}},\ \bibinfo {pages} {150601} (\bibinfo {year}
  {2005})}\BibitemShut {NoStop}%
\bibitem [{\citenamefont {Oliveira}\ and\ \citenamefont {{Aar{\~{a}}o
  Reis}}(2008)}]{Oliveira2008}%
  \BibitemOpen
  \bibfield  {author} {\bibinfo {author} {\bibfnamefont {T.~J.}\ \bibnamefont
  {Oliveira}}\ and\ \bibinfo {author} {\bibfnamefont {F.~D.~A.}\ \bibnamefont
  {{Aar{\~{a}}o Reis}}},\ }\href {\doibase 10.1103/PhysRevE.77.041605}
  {\bibfield  {journal} {\bibinfo  {journal} {Phys. Rev. E}\ }\textbf {\bibinfo
  {volume} {77}},\ \bibinfo {pages} {041605} (\bibinfo {year}
  {2008})}\BibitemShut {NoStop}%
\bibitem [{\citenamefont {Ferrenberg}\ and\ \citenamefont
  {Swendsen}(1989)}]{Ferrenberg1989}%
  \BibitemOpen
  \bibfield  {author} {\bibinfo {author} {\bibfnamefont {A.}~\bibnamefont
  {Ferrenberg}}\ and\ \bibinfo {author} {\bibfnamefont {R.~H.}\ \bibnamefont
  {Swendsen}},\ }\href {http://www.ncbi.nlm.nih.gov/pubmed/10040500} {\bibfield
   {journal} {\bibinfo  {journal} {Phys. Rev. Lett.}\ }\textbf {\bibinfo
  {volume} {63}},\ \bibinfo {pages} {1195} (\bibinfo {year}
  {1989})}\BibitemShut {NoStop}%
\bibitem [{\citenamefont {Kumar}\ \emph {et~al.}(1992)\citenamefont {Kumar},
  \citenamefont {Rosenberg}, \citenamefont {Bouzida}, \citenamefont
  {Swendsen},\ and\ \citenamefont {Kollman}}]{Kumar1992}%
  \BibitemOpen
  \bibfield  {author} {\bibinfo {author} {\bibfnamefont {S.}~\bibnamefont
  {Kumar}}, \bibinfo {author} {\bibfnamefont {J.~M.}\ \bibnamefont
  {Rosenberg}}, \bibinfo {author} {\bibfnamefont {D.}~\bibnamefont {Bouzida}},
  \bibinfo {author} {\bibfnamefont {R.~H.}\ \bibnamefont {Swendsen}}, \ and\
  \bibinfo {author} {\bibfnamefont {P.~A.}\ \bibnamefont {Kollman}},\ }\href
  {\doibase 10.1002/jcc.540130812} {\bibfield  {journal} {\bibinfo  {journal}
  {J. Comput. Chem.}\ }\textbf {\bibinfo {volume} {13}},\ \bibinfo {pages}
  {1011} (\bibinfo {year} {1992})}\BibitemShut {NoStop}%
\bibitem [{\citenamefont {Grossfield}(2013)}]{Grossfield2013}%
  \BibitemOpen
  \bibfield  {author} {\bibinfo {author} {\bibfnamefont {A.}~\bibnamefont
  {Grossfield}},\ }\href {http://membrane.urmc.rochester.edu/wordpress/?page_id=126}
  {\enquote {\bibinfo {title} {{WHAM: the weighted histogram analysis
  method}}}\ } (\bibinfo {year} {2013})\BibitemShut {NoStop}%
\bibitem [{\citenamefont {Abraham}\ \emph {et~al.}(2015)\citenamefont
  {Abraham}, \citenamefont {Murtola}, \citenamefont {Schulz}, \citenamefont
  {P{\'{a}}ll}, \citenamefont {Smith}, \citenamefont {Hess},\ and\
  \citenamefont {Lindahl}}]{Abraham2015}%
  \BibitemOpen
  \bibfield  {author} {\bibinfo {author} {\bibfnamefont {M.~J.}\ \bibnamefont
  {Abraham}}, \bibinfo {author} {\bibfnamefont {T.}~\bibnamefont {Murtola}},
  \bibinfo {author} {\bibfnamefont {R.}~\bibnamefont {Schulz}}, \bibinfo
  {author} {\bibfnamefont {S.}~\bibnamefont {P{\'{a}}ll}}, \bibinfo {author}
  {\bibfnamefont {J.~C.}\ \bibnamefont {Smith}}, \bibinfo {author}
  {\bibfnamefont {B.}~\bibnamefont {Hess}}, \ and\ \bibinfo {author}
  {\bibfnamefont {E.}~\bibnamefont {Lindahl}},\ }\href {\doibase
  10.1016/j.softx.2015.06.001} {\bibfield  {journal} {\bibinfo  {journal}
  {SoftwareX}\ }\textbf {\bibinfo {volume} {1}},\ \bibinfo {pages} {19}
  (\bibinfo {year} {2015})}\BibitemShut {NoStop}%
\bibitem [{\citenamefont {Nos{\'{e}}}(1984)}]{nose84a}%
  \BibitemOpen
  \bibfield  {author} {\bibinfo {author} {\bibfnamefont {S.}~\bibnamefont
  {Nos{\'{e}}}},\ }\href
  {http://www.informaworld.com/10.1080/00268978400101201} {\bibfield  {journal}
  {\bibinfo  {journal} {Mol. Phys.}\ }\textbf {\bibinfo {volume} {52}},\
  \bibinfo {pages} {255} (\bibinfo {year} {1984})}\BibitemShut {NoStop}%
\bibitem [{\citenamefont {Hoover}(1985)}]{Hoover1985a}%
  \BibitemOpen
  \bibfield  {author} {\bibinfo {author} {\bibfnamefont {W.~G.}\ \bibnamefont
  {Hoover}},\ }\href {\doibase 10.1103/PhysRevA.31.1695} {\bibfield  {journal}
  {\bibinfo  {journal} {Phys. Rev. A}\ }\textbf {\bibinfo {volume} {31}},\
  \bibinfo {pages} {1695} (\bibinfo {year} {1985})}\BibitemShut {NoStop}%
\bibitem [{\citenamefont {Parrinello}\ and\ \citenamefont
  {Rahman}(1981)}]{parrinello1981}%
  \BibitemOpen
  \bibfield  {author} {\bibinfo {author} {\bibfnamefont {M.}~\bibnamefont
  {Parrinello}}\ and\ \bibinfo {author} {\bibfnamefont {A.}~\bibnamefont
  {Rahman}},\ }\href {\doibase 10.1063/1.328693} {\bibfield  {journal}
  {\bibinfo  {journal} {J. Appl. Phys.}\ }\textbf {\bibinfo {volume} {52}},\
  \bibinfo {pages} {7182} (\bibinfo {year} {1981})}\BibitemShut {NoStop}%
\bibitem [{\citenamefont {P{\'{a}}rtay}\ \emph {et~al.}(2008)\citenamefont
  {P{\'{a}}rtay}, \citenamefont {Hantal}, \citenamefont {Jedlovszky},
  \citenamefont {Vincze},\ and\ \citenamefont {Horvai}}]{partay2008new}%
  \BibitemOpen
  \bibfield  {author} {\bibinfo {author} {\bibfnamefont {L.~B.}\ \bibnamefont
  {P{\'{a}}rtay}}, \bibinfo {author} {\bibfnamefont {G.}~\bibnamefont
  {Hantal}}, \bibinfo {author} {\bibfnamefont {P.}~\bibnamefont {Jedlovszky}},
  \bibinfo {author} {\bibfnamefont {{\'{A}}.}~\bibnamefont {Vincze}}, \ and\
  \bibinfo {author} {\bibfnamefont {G.}~\bibnamefont {Horvai}},\ }\href
  {\doibase 10.1002/jcc.20852} {\bibfield  {journal} {\bibinfo  {journal} {J.
  Comput. Chem.}\ }\textbf {\bibinfo {volume} {29}},\ \bibinfo {pages} {945}
  (\bibinfo {year} {2008})}\BibitemShut {NoStop}%
\bibitem [{\citenamefont {Sega}\ \emph
  {et~al.}(2018{\natexlab{a}})\citenamefont {Sega}, \citenamefont {Hantal},
  \citenamefont {F{\'{a}}bi{\'{a}}n},\ and\ \citenamefont
  {Jedlovszky}}]{Sega2018}%
  \BibitemOpen
  \bibfield  {author} {\bibinfo {author} {\bibfnamefont {M.}~\bibnamefont
  {Sega}}, \bibinfo {author} {\bibfnamefont {G.}~\bibnamefont {Hantal}},
  \bibinfo {author} {\bibfnamefont {B.}~\bibnamefont {F{\'{a}}bi{\'{a}}n}}, \
  and\ \bibinfo {author} {\bibfnamefont {P.}~\bibnamefont {Jedlovszky}},\
  }\href {\doibase 10.1002/jcc.25384} {\bibfield  {journal} {\bibinfo
  {journal} {J. Comput. Chem.}\ }\textbf {\bibinfo {volume} {39}},\ \bibinfo
  {pages} {2118} (\bibinfo {year} {2018}{\natexlab{a}})}\BibitemShut {NoStop}%
\bibitem [{\citenamefont {Sega}(2016)}]{sega2016role}%
  \BibitemOpen
  \bibfield  {author} {\bibinfo {author} {\bibfnamefont {M.}~\bibnamefont
  {Sega}},\ }\href {\doibase 10.1039/C6CP04788B} {\bibfield  {journal}
  {\bibinfo  {journal} {Phys. Chem. Chem. Phys.}\ }\textbf {\bibinfo {volume}
  {18}},\ \bibinfo {pages} {23354} (\bibinfo {year} {2016})}\BibitemShut
  {NoStop}%
\bibitem [{\citenamefont {Bittner}\ \emph {et~al.}(2009)\citenamefont
  {Bittner}, \citenamefont {Nu{\ss}baumer},\ and\ \citenamefont
  {Janke}}]{Bittner2009}%
  \BibitemOpen
  \bibfield  {author} {\bibinfo {author} {\bibfnamefont {E.}~\bibnamefont
  {Bittner}}, \bibinfo {author} {\bibfnamefont {A.}~\bibnamefont
  {Nu{\ss}baumer}}, \ and\ \bibinfo {author} {\bibfnamefont {W.}~\bibnamefont
  {Janke}},\ }\href {\doibase 10.1016/j.nuclphysb.2009.05.009} {\bibfield
  {journal} {\bibinfo  {journal} {Nucl. Phys. B}\ }\textbf {\bibinfo {volume}
  {820}},\ \bibinfo {pages} {694} (\bibinfo {year} {2009})}\BibitemShut
  {NoStop}%
\bibitem [{\citenamefont {Neyt}\ \emph {et~al.}(2011)\citenamefont {Neyt},
  \citenamefont {Wender}, \citenamefont {Lachet},\ and\ \citenamefont
  {Malfreyt}}]{Neyt2011}%
  \BibitemOpen
  \bibfield  {author} {\bibinfo {author} {\bibfnamefont {J.-C.}\ \bibnamefont
  {Neyt}}, \bibinfo {author} {\bibfnamefont {A.}~\bibnamefont {Wender}},
  \bibinfo {author} {\bibfnamefont {V.}~\bibnamefont {Lachet}}, \ and\ \bibinfo
  {author} {\bibfnamefont {P.}~\bibnamefont {Malfreyt}},\ }\href {\doibase
  10.1021/jp204056d} {\bibfield  {journal} {\bibinfo  {journal} {J. Phys. Chem.
  B}\ }\textbf {\bibinfo {volume} {115}},\ \bibinfo {pages} {9421} (\bibinfo
  {year} {2011})}\BibitemShut {NoStop}%
\bibitem [{\citenamefont {Schmitz}\ \emph {et~al.}(2013)\citenamefont
  {Schmitz}, \citenamefont {Virnau},\ and\ \citenamefont
  {Binder}}]{Schmitz2013}%
  \BibitemOpen
  \bibfield  {author} {\bibinfo {author} {\bibfnamefont {F.}~\bibnamefont
  {Schmitz}}, \bibinfo {author} {\bibfnamefont {P.}~\bibnamefont {Virnau}}, \
  and\ \bibinfo {author} {\bibfnamefont {K.}~\bibnamefont {Binder}},\ }\href
  {\doibase 10.1103/PhysRevE.87.053302} {\bibfield  {journal} {\bibinfo
  {journal} {Phys. Rev. E}\ }\textbf {\bibinfo {volume} {87}},\ \bibinfo
  {pages} {053302} (\bibinfo {year} {2013})}\BibitemShut {NoStop}%
\bibitem [{\citenamefont {Binder}\ and\ \citenamefont
  {Virnau}(2016)}]{Binder2016}%
  \BibitemOpen
  \bibfield  {author} {\bibinfo {author} {\bibfnamefont {K.}~\bibnamefont
  {Binder}}\ and\ \bibinfo {author} {\bibfnamefont {P.}~\bibnamefont
  {Virnau}},\ }\href {\doibase 10.1063/1.4959235} {\bibfield  {journal}
  {\bibinfo  {journal} {J. Chem. Phys.}\ }\textbf {\bibinfo {volume} {145}},\
  \bibinfo {pages} {211701} (\bibinfo {year} {2016})}\BibitemShut {NoStop}%
\bibitem [{\citenamefont {Metropolis}\ \emph {et~al.}(1953)\citenamefont
  {Metropolis}, \citenamefont {Rosenbluth}, \citenamefont {Rosenbluth},
  \citenamefont {Teller},\ and\ \citenamefont {Teller}}]{Metropolis1953}%
  \BibitemOpen
  \bibfield  {author} {\bibinfo {author} {\bibfnamefont {N.}~\bibnamefont
  {Metropolis}}, \bibinfo {author} {\bibfnamefont {A.~W.}\ \bibnamefont
  {Rosenbluth}}, \bibinfo {author} {\bibfnamefont {M.~N.}\ \bibnamefont
  {Rosenbluth}}, \bibinfo {author} {\bibfnamefont {A.~H.}\ \bibnamefont
  {Teller}}, \ and\ \bibinfo {author} {\bibfnamefont {E.}~\bibnamefont
  {Teller}},\ }\href {\doibase 10.1063/1.1699114} {\bibfield  {journal}
  {\bibinfo  {journal} {J. Chem. Phys.}\ }\textbf {\bibinfo {volume} {21}},\
  \bibinfo {pages} {1087} (\bibinfo {year} {1953})}\BibitemShut {NoStop}%
\bibitem [{\citenamefont {Torrie}\ and\ \citenamefont
  {Valleau}(1977)}]{Torrie1977}%
  \BibitemOpen
  \bibfield  {author} {\bibinfo {author} {\bibfnamefont {G.}~\bibnamefont
  {Torrie}}\ and\ \bibinfo {author} {\bibfnamefont {J.}~\bibnamefont
  {Valleau}},\ }\href {\doibase 10.1016/0021-9991(77)90121-8} {\bibfield
  {journal} {\bibinfo  {journal} {J. Comput. Phys.}\ }\textbf {\bibinfo
  {volume} {23}},\ \bibinfo {pages} {187} (\bibinfo {year} {1977})}\BibitemShut
  {NoStop}%
\end{thebibliography}%

\appendix
\section{Generalization to arbitrary dimension}\label{app:nd_hamiltonian}
We discretize the Hamiltonian \preequ\eqref{equ:surface_model} on a $d$-dimensional grid with  grid spacing $(\Delta x)_i$ in direction $i$. We arrive at
\begin{equation}\label{equ:hamilton-disc-d}
    \begin{aligned}
        &\mathcal{H} = \frac{\gamma}{2} \left[\prod_{k=1}^{d-1} (\Delta x)_k\right] \sum_{j_1}^{n_1-1} \cdots \sum_{j_{d-1}}^{n_{d-1}-1}\left\{ \vphantom{\sum_{i=1}^{d-1} \frac{\left(h_{j_1,\ldots,j_i+1,\ldots,j_{d-1}}-h_{j_1,\ldots,j_i,\ldots,j_{d-1}}\right)^2}{(\Delta x)_i^2}}\right. \\
        & \qquad \left. \sum_{i=1}^{d-1} \frac{\left(h_{j_1,\ldots,j_i+1,\ldots,j_{d-1}}-h_{j_1,\ldots,j_i,\ldots,j_{d-1}}\right)^2}{(\Delta x)_i^2}\right\},
    \end{aligned}
\end{equation}
where $h_{j_1,\ldots,j_{d-1}} = h_{\{j\}} = h(j_1 (\Delta x)_1,\ldots,j_{d-1} (\Delta x)_{d-1})$ and $n_i$ and $(\Delta x)_i$ are the number of nodes and the discretization step size in dimension $i$, respectively. Calculating the derivative with respect to $h_{\{j\}}$ yields the force
\begin{equation}\label{equ:force_d_dims}
    \begin{aligned}
        F_{\{j\}} &= \gamma \left[\prod_{k=1}^{d-1} (\Delta x)_k\right] \\
                  &  \quad \times \sum_{i=1}^{d-1} \frac{h_{\ldots,j_i+1,\ldots} - 2h_{\ldots,j_i,\ldots} + h_{\ldots,j_i-1,\ldots}}{(\Delta x)_i^2}.
    \end{aligned}
\end{equation}
and the equation of motion
\begin{equation}\label{equ:dynamics-subst-d}
    \dot{h}_{\{j\}} = \beta \gamma \Pi D_0 \sum_{i=1}^{d-1} \delta_i^2 h_{\{j\}} + \sqrt{2D_0} \eta_{\{j\}}.
\end{equation}
Here we have introduced the abbreviations
\begin{equation}
    \Pi = \left[\prod_{k=1}^{d-1} (\Delta x)_k\right]
\end{equation}
and
\begin{equation}
    \delta_i^2 h_{\{j\}} = \frac{h_{\ldots,j_i+1,\ldots} - 2h_{\ldots,j_i,\ldots} + h_{\ldots,j_i-1,\ldots}}{(\Delta x)_i^2}.
\end{equation}

\section{Contact distance density distributions}\label{app:contact_distance_dists}
In \presec\ref{sec:pde_solution} we have derived an approximate solution for the distribution of contact distances between two interfaces, $u(l)$, given a rate function $k(\ell)$:
\begin{equation}\label{equ:pi_solution_final_app}
  u(\ell) \approx \frac{ k(\ell)^{3/4} e^{-\tilde{S}_0(\ell)}}{2\Dint^{1/2} k(\ell_0)^{1/4} Z}.
\end{equation}
In this section we explicitly calculate $u(\ell)$ for exponential and Gaussian rate functions and test our results by comparing $u(l)$ to distributions obtained from a simple Gaussian random walk model. Random walkers are started at $\ell_0$ and their positions are updated according to
\begin{equation}
  \ell_{i+1} = \ell_i + \sqrt{2 D \Delta t}\xi
\end{equation}
where $\xi$ is a random number chosen from a Gaussian distribution with zero mean and unit variance. Each timestep the walker reacts with probability
\begin{equation}
  P_\text{r} = 1 - e^{-k(\ell_i) \Delta t}.
\end{equation}
If a reaction occurs the position is saved and added to a histogram. The results of these simulations are shown in \prefigs\ref{fig:comparison_numeric_pi_by_D_exp} and \ref{fig:comparison_numeric_pi_by_D}.

\subsection{Exponential rate functions}
\begin{figure}[tb]
    \centering
    \includegraphics[width=1.0\linewidth]{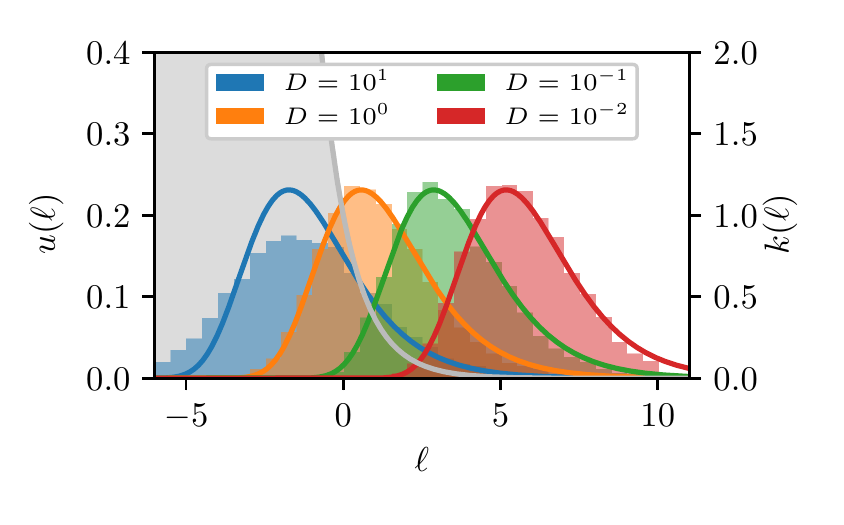}
    \caption{Comparison of \preequ\eqref{equ:pi_exp_u_solution} (lines) to results obtained from random walk simulations (histograms) for an exponential rate function $k(\ell)$ and different values of the diffusion coefficient $\Dint$. The other parameters are: $\nu = 1$, $\Lambda = 1$, $\ell_0 = 10$ and $\Delta t = 0.1$. The gray curve indicates the values of the rate function $k(\ell)$.}
    \label{fig:comparison_numeric_pi_by_D_exp}
\end{figure}
We first calculate $u(l)$ for the exponential rate function
\begin{equation}\label{equ:app_exponential_k}
  k(\ell) = \nu e^{-\ell/\Lambda},
\end{equation}
where $\ell_0 > \ell$. Substituting the ``action'' $\tilde{S}_0$, given by
\begin{equation}
  \begin{aligned}
  \tilde{S}_0(\ell) &= \sqrt{\frac{\nu}{D}} \int_{\ell_0}^{\ell} \intv{\ell'} ~ e^{-\ell'/(2\Lambda)} \\
  &= \sqrt{\frac{4 \Lambda^2 \nu}{D}} \left[e^{-\ell/(2\Lambda)}-e^{-\ell_0/(2\Lambda)}\right],
  \end{aligned}
\end{equation}
into \preequ\eqref{equ:pi_solution_final_app} yields the distribution
\begin{equation}\label{equ:pi_exp_u_solution}
  u(\ell) = \frac{1}{Z} k(\ell)^{3/4} \exp\left[-\sqrt{\frac{4 \Lambda^2 \nu}{D}} e^{-\ell/(2\Lambda)}\right].
\end{equation}
This result becomes exact in the limit $D \to 0$ and, indeed, it is in excellent agreement with the data obtained from random walk simulations as $D$ becomes small. The results are shown in \prefig\ref{fig:comparison_numeric_pi_by_D_exp}. Note that the constants $Z$ have been obtained by numerically normalizing the distribution $u(\ell)$.

We can also predict the most likely contact distance $\ell^\ast$ by setting the derivative of $\log u(\ell)$ to zero and solving the resulting equation, which yields
\begin{equation}\label{equ:l_star_exp}
  \ell^\ast = \Lambda \log\left[\frac{16}{9}\frac{\Lambda^2 \nu}{D}\right].
\end{equation}

\subsection{Gaussian rate functions}\label{sec:gauss_solution}
The rate function $k(\ell)$ is now given by a Gaussian,
\begin{equation}\label{equ:app_gaussian_k}
    k(\ell) = \nu\exp\left[-\frac{\ell^2}{2 w^2}\right],
\end{equation}
where we assume that $\ell_0 \gg w > 0$, $\nu$ is a positive prefactor, and $\ell < \ell_0$. Substituting this rate function into \preequ\eqref{equ:action_s0} yields
\begin{equation}
    \tilde{S}_0 = \sqrt{\frac{\nu}{\Dint}} \int_{\ell_0}^{\ell} \text{d}\ell' ~ \exp\left[-\frac{\ell^2}{4 w^2}\right],
\end{equation}
which can be expressed in terms of an error function:
\begin{equation}
    \tilde{S}_0 = \sqrt{\frac{\nu\pi w^2}{\Dint}} \left[\mathrm{erf}\left(\frac{\ell_0}{2w}\right) -\mathrm{erf}\left(\frac{\ell}{2w}\right)\right]
\end{equation}
Substituting into \preequ\eqref{equ:pi_solution_final_app} we arrive at
\begin{equation}\label{equ:gauss_prob_dist}
  u(\ell) = \frac{1}{Z} k(\ell)^{3/4}\exp\left[\sqrt{\frac{\nu\pi w^2}{\Dint}} ~ \mathrm{erf}\left(\frac{\ell}{2 w}\right)\right].
\end{equation}
The maximum of this $u(\ell)$ can again be calculated, yielding
\begin{equation}
    (\ell^\ast)^2 = 2 w^2 W\left(\frac{8}{9}\frac{w^2 \nu}{D}\right),
\end{equation}
where $W$ is the Lambert-W function. In the limit of large $L$ we can approximate $W(x) \sim \log(x)$, yielding
\begin{equation}
    \ell^\ast = w \sqrt{2 \log\left(\frac{8}{9}\frac{w^2 \nu}{D}\right)}.
\end{equation}

\begin{figure}[tb]
    \centering
    \includegraphics[width=1.0\linewidth]{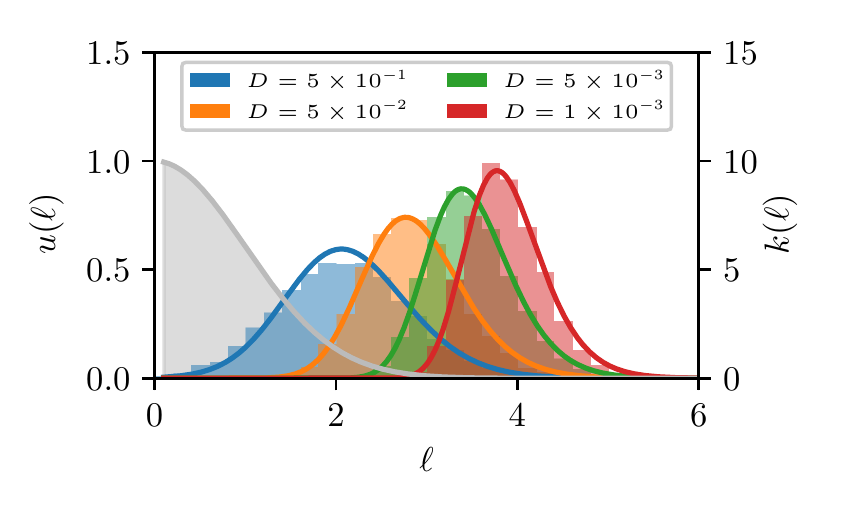}
    \caption{Comparison of \preequ\eqref{equ:gauss_prob_dist} (lines) to results obtained from random walk simulations (histograms) for different values of the diffusion coefficient $\Dint$. The other parameters are: $\nu = 10$, $w = 1$, $\ell_0 = 10$, and $\Delta t = 0.1$. The gray curve indicates the values of the rate function $k(\ell)$.}\label{fig:comparison_numeric_pi_by_D}
\end{figure}

\Prefig\ref{fig:comparison_numeric_pi_by_D} shows a comparison of this solution to the simulation results where the distribution $u(\ell)$ has been normalized numerically. Also here the agreement between the two is excellent in the limit $\Dint \to 0$.

\section{Mean first passage time calculation}\label{app:MFPT_calculation}
In \presec\ref{sec:rate_calc} we have outlined two different methods of calculating the mean first-passage times (MFPTs): the \emph{direct} method and what we termed the \emph{Poisson} method. In this appendix we provide the parameters used to carry out these calculations as well as the algorithmic details of the direct method.

\begin{figure}[tb]
    \centering
    \includegraphics[width=1.0\linewidth]{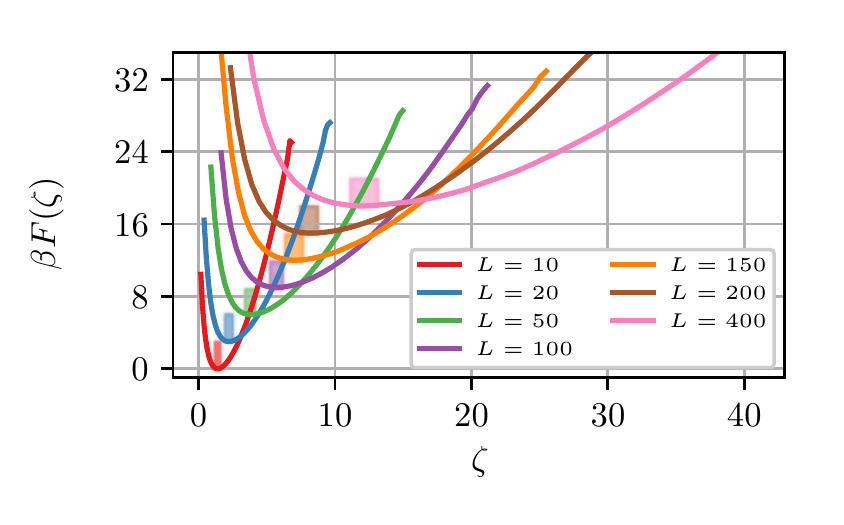}
    \includegraphics[width=1.0\linewidth]{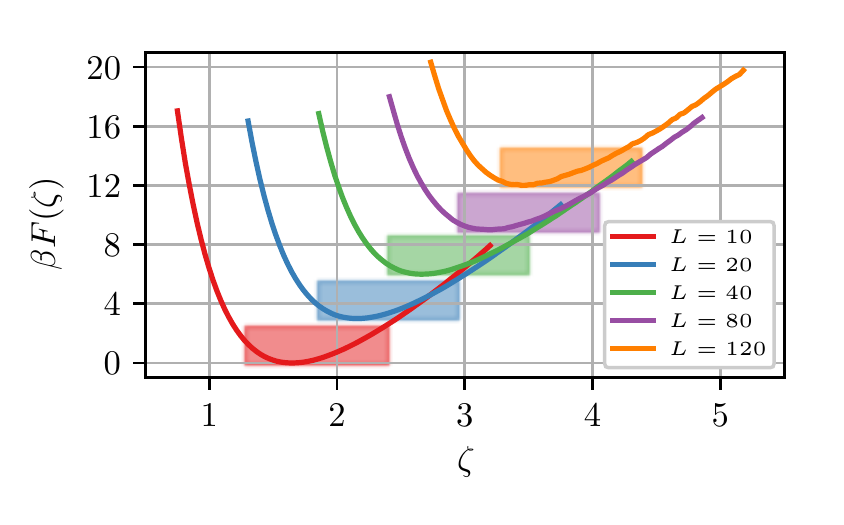}
    \caption{Free energies $\beta F(\zeta) = -\log P(\zeta)$ as a function of the maximum relative height, $\zeta$, for the (1+1)- (top) and (2+1)-dimensional (bottom) Edward Wilkinson interface. For clarity, the curves have been shifted for different system sizes. The shaded boxes indicate the equilibrium regions $\mathcal{E}$ that were used in the MFPT calculations. These results have been obtained using umbrella sampling simulations with harmonic biases that were subsequently matched using the WHAM method\precite\cite{Ferrenberg1989,Kumar1992,Grossfield2013}. For details on these simulations see \preapp\ref{app:mc_simulations}.}\label{fig:free_energies}
\end{figure}

Both methods require a choice of the equilibrium region $\mathcal{E}$ which are gathered in \pretabs\ref{tab:mfpt_calc_pars_1d} and \ref{tab:mfpt_calc_pars_2d}. These regions were chosen based on a previous calculation of the free energy $\beta F(\zeta) = -\log P(\zeta)$, where $P(\zeta)$ is the equilibrium probability density as a function of the maximum relative height, $\zeta$. The results of this calculation are shown in \prefig\ref{fig:free_energies} together with the chosen regions that are centered around the minimum of the respective free energy.
\begin{table}[tb]
    \centering
    \caption{Simulation parameters used to calculate MFPT data for the (1+1)-dimensional EW interface. Shown are the length of individual simulations $\mathcal{T}$, the extent of the equilibrium region $\mathcal{E}$, and the number trajectories used, $n_\text{R}$, as a function of linear system size $L$. For $\zeta_\text{low}$ and $\zeta_\text{high}$ the corresponding value of $\tilde{\zeta}$ is given in brackets.}\label{tab:mfpt_calc_pars_1d}
    \begin{tabular}{|r|r|r|r|r|}
      \hline
        \multicolumn{1}{|c|}{$L$}     & \multicolumn{1}{c|}{$\mathcal{T}$} & \multicolumn{2}{c|}{$\mathcal{E}$} & $n_\text{R}$ \\
      \hline
         &                 & \multicolumn{1}{c|}{$\zeta_\text{low}$}          & \multicolumn{1}{c|}{$\zeta_\text{high}$} & \\
      \hline
      10 &  $1.5 \times 10^7$  &               1.298 (0.410) &        1.614 (0.510) & 16\\
      20 &  $1.5 \times 10^7$  &               2.000 (0.447) &        2.447 (0.547) & 16\\
      50 &  $1.5 \times 10^7$  &               3.423 (0.484) &        4.130 (0.584) & 12\\
     100 &  $1.5 \times 10^7$  &               5.209 (0.521) &        6.209 (0.621) & 12\\
     150 &  $1.5 \times 10^7$  &               6.380 (0.521) &        7.605 (0.621) & 12\\
     200 &  $1.5 \times 10^7$  &               7.367 (0.521) &        8.781 (0.621) & 12\\
     400 &  $1.5 \times 10^7$  &              11.155 (0.558) &       13.160 (0.658) & 10\\
     500 &  $1.5 \times 10^7$  &              12.477 (0.558) &       14.713 (0.658) & 10\\
    1000 &  $1.5 \times 10^7$  &              17.646 (0.558) &       20.808 (0.658) &  8\\
    2500 &  $1.5 \times 10^7$  &              27.900 (0.558) &       32.900 (0.658) & 10\\
    5000 &  $1.5 \times 10^7$  &              39.457 (0.558) &       46.528 (0.658) &  9\\
     \hline
    \end{tabular}
    \caption{Simulation parameters used to calculate MFPT data for the (2+1)-dimensional EW interface. Shown are the length of individual simulations $\mathcal{T}$, the number of trajectories used, $n_\text{R}$, and the extent of the equilibrium region $\mathcal{E}$ as a function of linear system size $L$. For $\zeta_\text{low}$ and $\zeta_\text{high}$ the corresponding value of $\tilde{\zeta}$ is given in brackets.}\label{tab:mfpt_calc_pars_2d}
    \begin{tabular}{|r|r|r|r|r|r|}
      \hline
      \multicolumn{1}{|c|}{$L$}     & \multicolumn{1}{c|}{$\mathcal{T}$} & \multicolumn{2}{c|}{$\mathcal{E}$} & \multicolumn{1}{c|}{$n_\text{R}$}\\
      \hline
         &                 & \multicolumn{1}{c|}{$\zeta_\text{low}$}          & \multicolumn{1}{c|}{$\zeta_\text{high}$} & \\
      \hline
      10 &   $3 \times 10^7$ & 1.287 (-0.55) & 2.388 (0.55) &    11\\
      20 &   $3 \times 10^7$ & 1.840 (-0.55) & 2.941 (0.55) &    11\\
      40 &   $3 \times 10^7$ & 2.393 (-0.55) & 3.494 (0.55) &    11\\
      80 &   $3 \times 10^7$ & 2.946 (-0.55) & 4.047 (0.55) &    10\\
     120 &   $3 \times 10^7$ & 3.269 (-0.55) & 4.371 (0.55) &     9\\
     200 &$2.14 \times 10^7$ & 3.677 (-0.55) & 4.778 (0.55) &     6\\
     400 &$2.63 \times 10^7$ & 4.230 (-0.55) & 5.331 (0.55) &     7\\
     800 &$1.22 \times 10^7$ & 4.783 (-0.55) & 5.884 (0.55) &     4\\
     \hline
    \end{tabular}
\end{table}

In the direct method we average the waiting times between configurations in $\mathcal{E}$ and the system reaching a checkpoint $\zeta_i$ the next time (see \prefig\ref{fig:mfpt_calc_sketch}). Here, we write $T_j^{(i)}$ for the times a crossing of checkpoint $\zeta_i$ occurs and $t_{jk}^{(i)}$ for the times when the system is found in region $\mathcal{E}$ in between the hits at time $T_{j-1}^{(i)}$ and $T_j^{(i)}$, i.e. all the times where the next crossing occurs at $T_j^{(i)}$. $N_j^{(i)}$ is the number of times $t_{jk}^{(i)}$ observed. The MFPT for checkpoint $\zeta_i$ is then given by
\begin{equation}
  \tau(\zeta_i) = \frac{1}{\sum_{j} N_j^{(i)}} \sum_{j} \sum_{k=1}^{N_j^{(i)}}\left(T_j^{(i)} - t^{(i)}_{jk}\right),
\end{equation}
where the $\sum_j$ runs over all checkpoint crossings $j$. This can be rewritten in the computationally more convenient form of
\begin{equation}
  \tau(\zeta_i) = \frac{1}{\sum_{j} N_j^{(i)}} \sum_{j} \left(N_j^{(i)} T_j^{(i)} - \sum_{k=1}^{N_j^{(i)}} t^{(i)}_{jk}\right).
\end{equation}
Hence, to calculate the MFPTs directly, two sums have to be kept and updated each time the interface $\zeta_i$ is crossed: $\sum_j N_j^{(i)}$, $\sum_{j} (N_j^{(i)} T_j^{(i)} - \sum_{k=1}^{N_j^{(i)}} t^{(i)}_{jk})$.

\section{Direct- vs. Poisson method in the (1+1)-dimensional EW model}\label{app:additional_results}
In \presec\ref{sec:rate_calc} we present mean first-passage times (MFPTs) calculated using two different methods: (i) a direct average of waiting times between visiting the $\mathcal{E}$ region (see \preapp\ref{app:MFPT_calculation}) and crossing a given checkpoint $\zeta_i$, and (ii) an evaluation that assumes that the statistics of these crossing are Poissonian statistics. In this appendix we present an analysis of the differences between the results obtained using these two methods and their dependence on system size $L$.

\begin{figure}[htpb]
    \centering
    \includegraphics[scale=1.0]{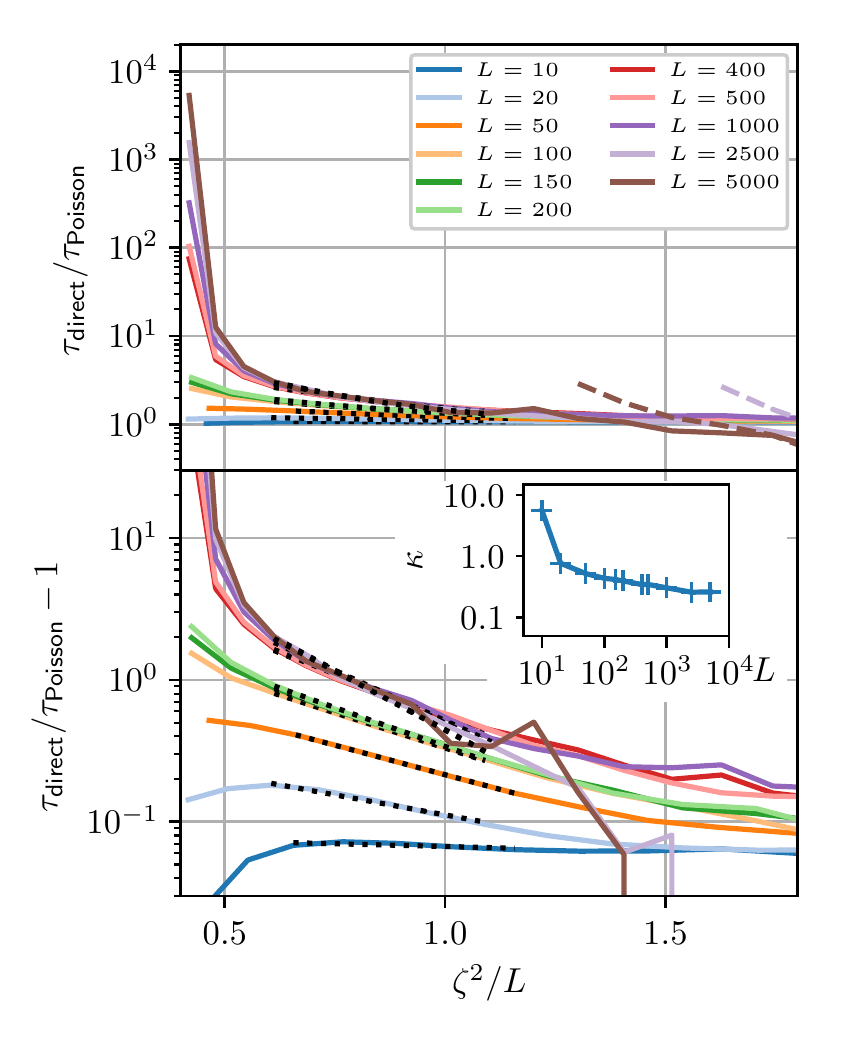}
    \caption{\figtop Ratios of mean first-passage times for (1+1)-dimensional EW interfaces calculated using the direct method to the ones obtained using the Poisson method, $q = \tau_\text{direct} / \tau_\text{Poisson}$. The dashed lines represent the expected values for large $\zeta - S(l)$, $(\mathcal{T}/4)/ \tau_\text{Poisson}$, that come about due to the finite length of the trajectories used to calculate $\tau_\text{direct}$. The dotted black lines are fits to the function $q = 1 + A \exp(-\tilde{\zeta}/\kappa)$. \figbottom $q-1$ plotted on a logarithmic scale. \figinset fitted parameters $\kappa$ as a function of inverse system size $1/L$.The data suggests that $\lim_{L\to\infty} \kappa(L) = const.$.}
    \label{fig:mfpt_histo_analysis_1d_mfpt_ratio}
\end{figure}

\begin{figure}[tb]
    \centering
    \includegraphics[width=1.0\linewidth]{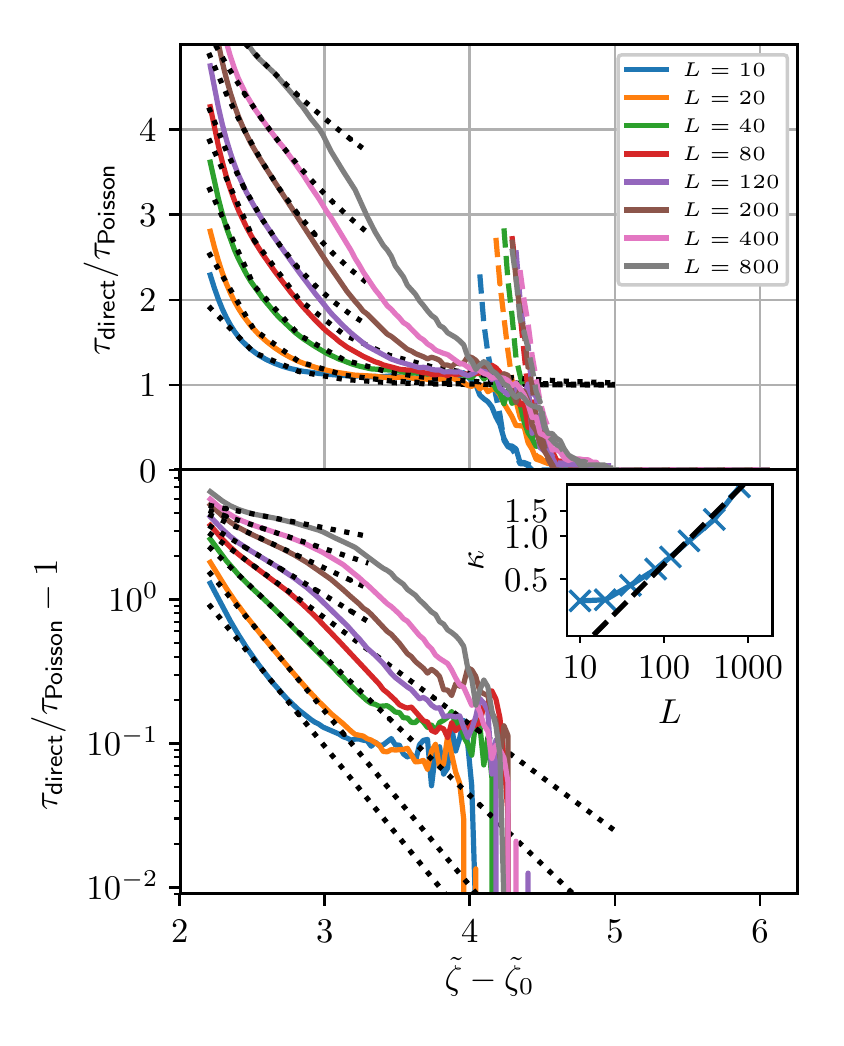}
    \caption{Same analysis as in \prefig\ref{fig:mfpt_histo_analysis_1d_mfpt_ratio} for the (2+1)-dimensional interface. \figinset $\kappa$ as a function of $L$. The dashed line indicates a fit of $\kappa(L) = A L^{B}$, where $A \approx 0.03$ and $B \approx 0.67$. In contrast to the (1+1)-dimensional system, $\kappa$ grows with system size.}
    \label{fig:mfpt_histo_analysis_2d_mfpt_ratio}
\end{figure}

\Prefigs\ref{fig:mfpt_histo_analysis_1d_mfpt_ratio} and\prefigNoWord\ref{fig:mfpt_histo_analysis_2d_mfpt_ratio} show the ratios of the MFPTs calculated using direct- and the Poisson method, $q(\zeta) = \tau_\text{direct}/\tau_\text{Poisson}$, for the (1+1)- and the (2+1)-dimensional EW interface, respectively.

As was noted in the main text, the direct method is limited by the length of trajectories used because the a priori probability of observing a given waiting time is not the same for all waiting times $t_\text{w}$. We assume the distribution of waiting times, $\rho_\text{w}(t_\text{w})$, to be time-independent and that subsequent checkpoint crossings are independent of each other. Consider a set of simulation runs of length $\mathcal{T}$ over which we calculate the average waiting time of such events.

We first calculate the waiting time for a single fixed starting time of the waiting period, $0 \le t < \mathcal{T}$. The expected observed average waiting time starting from $t$ $\left< t_\text{w}\right>_t$ is given by
\begin{equation}
  \begin{aligned}
    \left< t_\text{w} \right>_t &= \frac{1}{Z(t)}\int_t^{\mathcal{T}} \intv{t''} \left(t''-t\right) \rho_\text{w}(t''-t) \\
    &= \frac{1}{Z(t)}\int_0^{\mathcal{T}-t} \intv{t'} t' \rho_\text{w}(t'),
  \end{aligned}
\end{equation}
where in the second line we have substituted $t' = (t'' - t)$ and $Z(t) = \int_0^{\mathcal{T}-t} \intv{t'} \rho_\text{w}(t')$ normalizes the distribution of events observed in the limited time $\mathcal{T}-t$. Now we can average over the starting points $t$ to arrive at the overall expectation value for the observed waiting times:
\begin{equation}\label{equ:wait_time}
  \left< t_\text{w} \right> = \frac{1}{\mathcal{T}} \int_0^\mathcal{T} \intv{t} \frac{1}{Z(t)} \int_0^{\mathcal{T}-t} \intv{t'} t' \rho_\text{w}(t')
\end{equation}
As an illustrative example, consider waiting times that are distributed according to a Poisson distribution: $\rho_\text{w}(t_\text{w}) = \tau^{-1} e^{-t_\text{w} / \tau}$. A calculation of $\left<t_\text{w}\right>_t$ yields
\begin{equation}
  \left<t_\text{w}\right>_t = \tau + \frac{\mathcal{T} - t}{1 - e^{(\mathcal{T}-t) / \tau}}.
\end{equation}
As expected, the second term vanishes in the limit $\mathcal{T} \to \infty$ to recover the well known result for Poisson distributions.

The result of the second integration in \preequ\eqref{equ:wait_time} can not be expressed in terms of basic functions, however, in order to assess the effect of limited simulation time we now consider the regime where $\mathcal{T} - t \ll \tau$. Expanding $\left<t_\text{w}\right>_t$ to first order in $(\mathcal{T}-t)/\tau$ yields
\begin{equation}
  \left<t_\text{w}\right>_t \approx \frac{1}{2}\left(\mathcal{T} -t\right).
\end{equation}
With this approximation carrying out the second integral in \preequ\eqref{equ:wait_time} then gives
\begin{equation}
  \left<t_\text{w}\right> \approx \frac{\mathcal{T}}{4},
\end{equation}
which is solely determined by the length of the trajectories. The same result can be achieved by simply assuming that $\rho_\text{w}(t_\text{w})$ is a constant over the range $[0,\mathcal{T}]$. This result is represented by dashed lines in \prefigs\ref{fig:mfpt_histo_analysis_1d_mfpt_ratio} and\prefigNoWord\ref{fig:mfpt_histo_analysis_2d_mfpt_ratio} and is in excellent agreement with the observed decline of $q(\zeta)$ at large $\zeta$.

For both dimensionalities we observe an exponential decay of $q(\tilde{\zeta})$ with $\tilde{\zeta}$ before the deviations due to finite $\mathcal{T}$ drop $q$ to zero. The indicated fits of $q(\tilde{\zeta}) = 1 +  A \exp[-(\tilde{\zeta}-\tilde{\zeta}_0) / \kappa(L)]$ to the data reveal a qualitative difference between the (1+1)- and the (2+1)-dimensional system: while the parameter $\kappa(L)$ approaches a constant value as $L$ increases for (1+1)-dimensional interfaces, in the (2+1) dimensional case $\kappa(L)$ grows with system size.

\section{Scaling of MFPTs with system size}\label{app:scaling}
\begin{figure}[tb]
    \centering
    \includegraphics[width=1.0\linewidth]{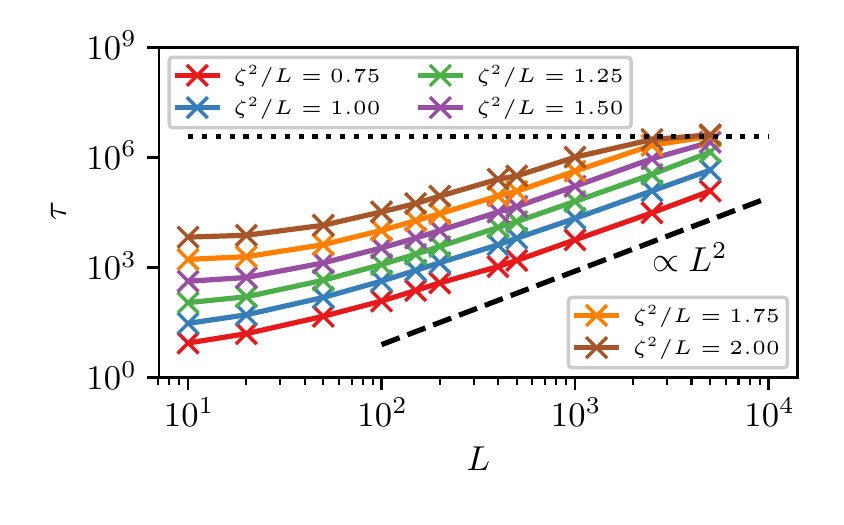}
    \caption{Mean first passage times $\tau$ as a function of linear system size $L$ at fixed values of $\tilde{\zeta} = \zeta/L^2$ for (1+1) dimensional EW interfaces. The dotted line indicates a value of $\tau = \mathcal{T}/4$, where $\mathcal{T}$ is the length of the simulations used to calculate the MFPTs.}
    \label{fig:mfpt_histo_analysis_1d_mfpt_scaling}
\end{figure}

\begin{figure}[htpb]
    \centering
    \includegraphics[scale=1.0]{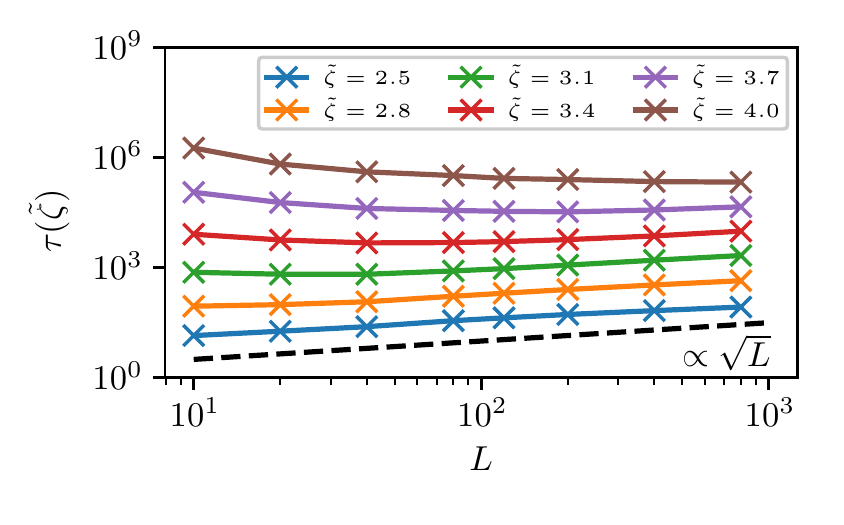}
    \caption{Mean first passage times $\tau$ as a function of linear system size $L$ at fixed values of $\tilde{\zeta} = \zeta - S(L)$ for (2+1) dimensional EW interfaces.}
    \label{fig:mfpt_histo_analysis_2d_mfpt_scaling}
\end{figure}

In \prefigs\ref{fig:mfpt_histo_analysis_1d_mfpt} and\prefigNoWord\ref{fig:mfpt_histo_analysis_2d_mfpt} in the main part, we have shown examples of the scaling behavior of $\tau(L)$ at fixed values of $\tilde{\zeta}$. \Prefigs\ref{fig:mfpt_histo_analysis_1d_mfpt_scaling} and\prefigNoWord\ref{fig:mfpt_histo_analysis_2d_mfpt_scaling} show the functional form of $\tau(L)$ for a broader range of $\tilde{\zeta}$ values.

\section{Detection of cluster geometry in the Ising model}\label{sec:geometry_detection}
We present an algorithm, that detects the geometry of a given cluster in simulations of the 2- and 3-dimensional Ising model with periodic boundary conditions. This includes the detection of holes in slabs. As a first step, we identify the largest cluster of neighboring spins that point in a particular direction within the system, the so called geometric cluster\precite\cite{Schmitz2013,Binder2016}. The neighbors of a given spin are all spins that can be reached by making a step of $\pm 1$ in each direction separately. Diagonally displaced spins are not considered neighbors.

\begin{figure}[htpb]
    \centering
    \includegraphics[scale=1.0]{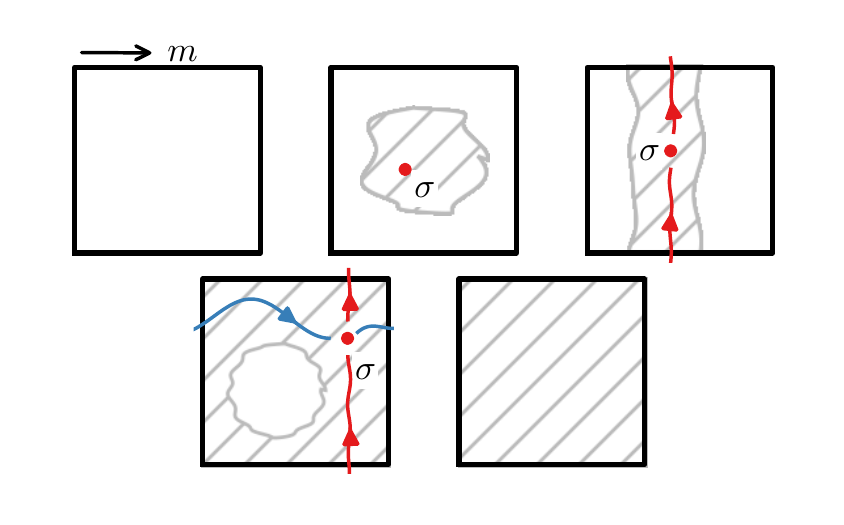}
    \caption{Sketch of 5 different states that are observed as a function of increasing magnetization in the 2d Ising model. In order from top left to bottom right they are referred to as: homogeneous-down, disk-up, slab, disk-down, and homogeneous-up. The red dots are randomly chosen spins and the lines indicate paths that connect those spins to one of their periodic images without leaving the cluster. In the disk-down configuration two paths are found. One leading to a copy of the spin that is offset by the $\vec{s}_y$ box vector (red) and one that leads to a spin that is offset by the $\vec{s}_x$ vector (blue).}
    \label{fig:geometry_plot}
\end{figure}

The number of distinct geometries that are thermodynamically stable states if one fixes the magnetization of the system, depends on the number of dimensions. Consider clusters in two dimensions that consist of spins that point up. There are three stable geometries in order of increasing magnetization\precite\cite{Leung1990}: a disk of spins pointing up, a slab that points up next to a slab that points down, and a disk that points down (see Fig. \ref{fig:geometry_plot}). In the following we will refer to these states as disk, slab, and opp-disk, using the \emph{opp-} prefix to identify configurations in which the largest cluster of spins that point in the opposite direction has the given geometry.

To distinguish between these geometries, we introduce an algorithm that counts the directions in which a cluster is connected to itself. Figure~\ref{fig:geometry_plot} shows a sketch of the principle used to identify these geometries. We pick a random spin $\sigma$ from our cluster and ask the question: how many periodic copies of $\sigma$, that are found in adjacent copies of the simulation box, are connected to $\sigma$ via a path that does not leave the cluster? Additionally, we require that these copies are distinct, in the sense that they are not related to each other by an inversion around $\sigma$.
\begin{table}[htpb]
    \centering
    \caption{Table of possible geometries in two dimensions, the number of closest periodic images of a spin that can be reached without leaving the cluster, $n_\text{c}$, and the value for the largest cluster in the opposite direction $\bar{n}_\text{c}$.}
    \label{tab:geometry_2d}
    \begin{tabular}{|c|c|l|l|}
        \hline
        $n_\text{c}$ & $\bar{n}_\text{c}$ & Geometry & \\
        \hline
        0   &   2   &   Disk & non-spanning\\
        \hline
        0   &   1   &   Disk\footnotemark[1] & non-spanning\\
        \hline
        0   &   0   &   Disk\footnotemark[1] & non-spanning\\
        \hline
        1   &   1   &   Slab & spanning\\
        \hline
        1   &   0   &   Opp-Disk\footnotemark[1] & spanning\\
        \hline
        2   &   0   &   Opp-Disk & spanning\\
        \hline
    \end{tabular}
    \footnotetext[1]{In these cases the disks are separated from their periodic images by two spins that are offset by one diagonal step.}
\end{table}
\begin{table}[htpb]
    \centering
    \caption{Table of possible geometries in three dimensions, the number of closest periodic images of a spin that can be reached without leaving the cluster, $n_\text{c}$, and the value for the largest cluster in the opposite direction $\bar{n}_\text{c}$.}
    \label{tab:geometry_3d}
    \begin{tabular}{|c|c|l|l|}
        \hline
        $n_\text{c}$ & $\bar{n}_\text{c}$ & Geometry & \\
        \hline
        0   &   3   &   Sphere & non-spanning\\
        \hline
        1   &   3   &   Cylinder & spanning \\
        \hline
        2   &   2   &   Slab & spanning \\
        \hline
        2   &   3   &   Hole & spanning \\
        \hline
        3   &   3   &   Hole-Opp-Hole & spanning \\
        \hline
        3   &   2   &   Opp-Hole & spanning \\
        \hline
        3   &   1   &   Opp-Cylinder & spanning \\
        \hline
        3   &   0   &   Opp-Sphere & spanning \\
        \hline
    \end{tabular}
\end{table}

In the case of a disk, the answer is 0; there is no path to any periodic copy of a given spin. If the cluster is a slab the answer is $1$ since we can find a path along the $y$-direction to a spin offset one box-length up and another one, one box-length down. However, these two spins are related to each other by an inversion around $\sigma$, and, hence, we discard one of them. The last example is the opp-disk; here we can reach copies in $y$-direction and copies in $x$-direction so the answer is $2$.

In order to also handle unusual cluster geometries like slabs that are aligned along the diagonals of the simulation box or clusters that span multiple periodic boxes before they are connected to themselves, we can generalize this concept by looking for the ``closest'' copies of our spin in terms of a \emph{box-distance}, $\Delta$. To define $\Delta$ we first note that the offset of each periodic copy of $\sigma$, can be written as
\begin{equation}
    \Delta \vec{r} = \sum_i b_i \vec{s}_i,
\end{equation}
where the $b_i$ are integer coefficients and the vectors $\vec{s}_i ~ (i=x,y,\ldots)$ span the simulation box. The box-distance $\Delta$ is then given by
\begin{equation}
    \Delta = \sqrt{\sum_i b_i^2}.
\end{equation}
By looking for the copies with smallest $\Delta$, we make sure that we start searching for copies in the neighboring boxes first.

The algorithm that distinguishes between different geometries then proceeds as follows:
\begin{enumerate}
    \item\label{bul:first} Pick a random spin $\sigma$ from the largest cluster in the system that points in the direction of interest.
    \item For $d$ dimensions, up to 2$d$ of the periodic images of $\sigma$ that can be reached via a path that does not leave the cluster, will have the minimum value of $\Delta$, $\Delta_\text{min}$. Find all periodic images with $\Delta_\text{min}$.
    \item Pairs of these images will be related by an inversion around $\sigma$. Keep only one copy of each of these pairs.
    \item\label{bul:last} Count the number of these periodic images found, $n_\text{c}$.
    \item Repeat steps \ref{bul:first} to \ref{bul:last} for the largest cluster pointing in the opposite direction. The resulting count is called $\bar{n}_\text{c}$.
    \item To determine the geometry refer to tables \ref{tab:geometry_2d} and \ref{tab:geometry_3d} that summarize the possible geometries in two and three dimensions and the associated values of $n_\text{c}$ and $\bar{n}_\text{c}$,.
\end{enumerate}

The additional geometries encountered in 3 dimensions shown in Tab. \ref{tab:geometry_3d}, are cylinders, slabs with a hole (i.e.~slabs where there is a connected path through the slab that consists of spins that point in the opposite direction), and the corresponding opp-geometries. Also, in small systems configurations that contain an up-slab and a down-slab, both with holes, can be found\footnote{Think of two slabs next to each other. A hole is formed in one of them by flipping a path of spins that connect the opposite cluster to itself. Afterwards the same can be done for the opposite cluster as well, since those two paths do not necessarily cross} (Hole-Opp-Hole).

\section{MC simulations}\label{app:mc_simulations}
In the case of a (1+1)-dimensional interface the equations of motion~\eqref{equ:dynamics-subst} are derived from the Hamiltonian
\begin{equation}\label{equ:hamilton-disc-1}
  \mathcal{H} = \frac{\gamma}{2} \sum_{i=0}^{N-1} \Delta x \left(\frac{x_{i+1} - x_i}{\Delta x}\right)^2.
\end{equation}
The generalization to higher dimensions is straightforward and given in \preapp\ref{app:nd_hamiltonian}.

Free energy landscapes as a function of $\zeta$, $F(\zeta)$, are obtained using Metropolis Monte Carlo\precite\cite{Metropolis1953} umbrella sampling\precite\cite{Torrie1977} simulations. Configurations are biased using a set of harmonic spring potentials and the resulting histograms are subsequently joined using the Weighted-Histogram-Analysis-Method\precite\cite{Ferrenberg1989,Kumar1992,Grossfield2013} (WHAM)\footnote{Code available at \texttt{http://membrane.urmc.rochester.edu/}, Version 2.0.9.}.

\section{Calculation of cluster size from the magnetization}\label{sec:slab_width_from_magn}
If a configuration from an Ising model simulation consists of a single large cluster, the size of the cluster can be estimated from the total magnetization. This estimate will be accurate, if the average magnetizations inside and outside of the cluster are close to the bulk values $m_\text{u} = -m_\text{d} = \left< m \right>$. This is expected to be true in the bulk of the cluster, if the cluster is a large enough slab, so that the width of the interfaces is small relative to the width of the cluster. Due to the symmetry of the Ising model with respect to a flip of all the spins, effects that stem from the surfaces of the cluster average out. Hence, we can write
\begin{equation}
	N m \approx n \left< m \right> - (N-n) \left< m \right>,
\end{equation}
where $m$ is the magnetization of configuration, $N$ is the total number of spins and $n$ is the number of spins in the cluster of interest. The cluster size can then be estimated by simply rearranging to
\begin{equation}
	n \approx \frac{N}{2}\left( \frac{m}{\left<m\right>} + 1 \right).
\end{equation}
\end{document}